# *Is Triboelectricity Confusing, Confused or Complex?*

Karl P. Olson and Laurence D. Marks*
Department of Materials Science and Engineering
Northwestern University, Evanston IL 60201
*Corresponding author: Laurence.marks@gmail.com

**Abstract**

In this report, we look at the fundamental physics of triboelectricity, charge transfer due to contact and sliding. While much of the report focuses upon recent advances such as the incorporation of flexoelectric contributions, we also include older work, some from centuries ago, which can only now be fully understood. Basic concepts and theories ranging from elements of tribology and contact mechanics through semiconductor built-in potentials, electromechanical terms, mechanochemistry and trap states are briefly described, linking to established surface science and interface physics. We then overview the main models that have been proposed, showing that they all fall within conventional electrostatics combined with other established science. We conclude with some suggestions for the future. Based upon this overview, our conclusion is that triboelectricity is a slightly complex combination of classic tribology and standard electrostatic phenomena that can be understood using the generalized Ampère's law connecting the electric displacement field with both Coulomb and polarization contributions, and the free carrier density, that is $\mathbf{\nabla} \cdot \mathbf{D} = \rho_f$. Triboelectricity may be confusing, it is not really confused if care is taken, but it is complex.

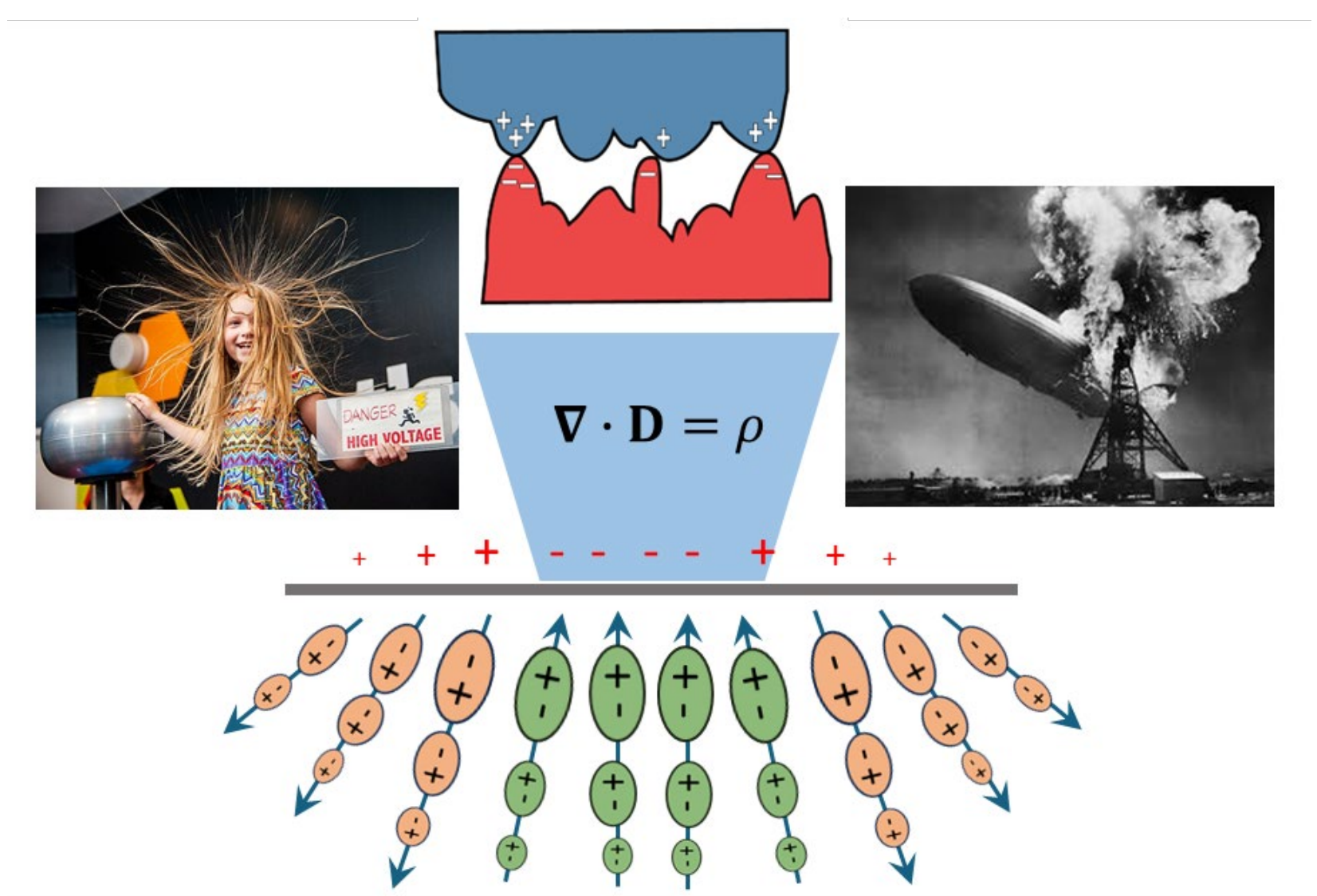

## 1. Introduction

From teaching high school students to industry, the generation of static electricity by charge transfer when two materials contact or are rubbed against each other is well known, a phenomenon called either contact electrification, if there is no sliding, or triboelectricity in the more general case. It was recorded by Thales of Miletus around 585 BC [1] after rubbing amber with fur, although it may have been known earlier [2] since there is extensive evidence for amber jewelry being used before this [3, 4], for instance in the late Neolithic period [5] as shown in figure 1. The word triboelectricity has roots in the Greek words *tribo*, to rub, and *ēlektron* for amber, so had the original meaning of *"rubbing amber"*, which is how it was first observed millennia ago. There are several other terms which have the same meaning, in particular *tribocharge* for the charge transferred, and *frictional electrification, triboelectrification,* and *the triboelectric effect* for the general process.

Static electricity is a true cross-disciplinary phenomenon, with consequences far beyond the simple ones of receiving an electric shock after stroking a cat on a dry day. It can range from a nuisance as in a 2024 TikTok on defrizzing hair [6], to fundamental science such as how millimeter size grains aggregate during planetary formation [7], quality control problems during the processing of pharmaceutical powders [8], as a component of how butterflies collect pollen [9] and can be exploited in the refinement of inorganic powders [10] or wheat bran [11], for energy harvesting [12, 13] or in the recycling industry for separating granular waste [14] – an incomplete list that continues to grow. While most commonly found for solid-solid contacts, triboelectric charging can occur in many other cases, for instance with sliding water droplets are described by Faraday in 1843 [15], the classic waterfall effect analyzed by Nobel laureate Lenard in 1892 [16], and current safety issues with pneumatic transport systems [17]. Static electricity is to blame for at least two major fires or explosions ***each day*** in manufacturing plants worldwide [18].

How is it possible that a phenomenon that has been known for at least as long as recorded history remains one of some confusion, with to date still significant debate in the literature? We believe that there are three main reasons for this:

1. While many of the key components such as the contact potential between dissimilar materials, band bending at junctions, and the role of asperities in contacts are well documented and can be found in the older triboelectric literature, they have not always been included in analyses.
2. One of the key terms, namely the flexoelectric effect, was not established until the 1960s and only became a topic of interest to the scientific community (outside of some work involving liquid crystals) from about 2001 onwards, and only considered as a component of triboelectricity starting from 2019 [19]. Many results which are otherwise confusing follow directly when it is included; indeed, we suspect that if the flexoelectric effect had been discovered a century earlier, triboelectricity would be settled science.

3. It can be highly irreproducible. We believe this is because many experiments have not controlled key parameters; similar to the earliest semiconductor devices, impurities and contaminants can have large effects, for which as we will see there is already evidence. We believe that this irreproducibility can be controlled if care is taken, and we will make some suggestions later.

To expand briefly on irreproducibility, the sensitivity of insulators and semiconductors to inadequate control is well known; it is not so well known that tribology (friction and wear) can also be sensitive. A classic example is Leonardo da Vinci's seminal work on friction. In his notebook pages [20] one of the three observations he made was that friction coefficients had a consistent value of 0.25. Da Vinci's experiments were recently reproduced by Pitenis, Dowson and Sawyer [21], who concluded that his results were a consequence of rough preparation and uncontrolled surfaces, and probably due to carbon layers which are sometimes called tribopolymers or friction polymers [22, 23].

The purpose of this report is to provide an overview of what we believe is a more complete and *ab-initio* description of triboelectricity and contact electrification for solids. While there are some similarities with tribocharge between solids and fluids or between two fluids, we will limit our scope. We use the term "*ab-initio*" here to indicate that we are not using any empirical models, and everything can be traced back to fundamental solid-state quantum mechanics. We will attempt to be somewhat comprehensive, paying attention to both older work as well as newer work; beyond acknowledging precedence, sometimes the older work is more thoughtful even though it may be harder to find and they had less sophisticated experimental tools and theoretical models. We will take a reserved approach to the use of neologisms when there are older and established terms; we strongly believe it is always important to properly credit early work rather than reinventions. As relevant we will mention ambiguities and uncertainties, with more on these in the discussion. If certain prevalent statements in the literature violate accepted science we will, hopefully politely, mention that there are issues.

Because of our self-limitation about neologisms, there are certain terms which we believe are really all part of triboelectricity and should not be separated, for instance the so-called *tribovoltaic effect* [24, 25] or *tribotronics* [26]. Hence, we will not include any specific discussion on these. Since we are focusing here on the fundamentals, we will not discuss in detail the large body of work on exploiting tribocharge to harvest energy using what are called Triboelectric Nanogenerators (TENGs), except where they shed light on the fundamentals.

The structure of this report is as follows. In the next section we will define the problem, introducing some of the known experimental constraints that models of triboelectricity have to explain, as well as a framework of **Drivers**, **Triggers**, **Mechanisms** and **Dependencies** [27] that we will use to clarify the different processes (scientific sub-topics). We will also provide a decision tree to guide how we believe the problem needs to be approached. Following this, in section 3 we go into brief details about the main contributors. These include aspects of the tribology of asperity contacts as well as electromechanical terms such as flexoelectricity [19], standard interface and surface physics phenomena such as band bending and the built-in potential for contacts [28, 29], as well as others contributors which are known but are not always considered in the triboelectric literature such as the role of interface dipoles, chemisorption and trap states. How these all fit together as various approaches that have been considered in the literature is the focus of section 4, in some cases with a few minor extensions of textbook results, for instance solving for the semiconductor-semiconductor band bending with charge separation, and pointing out the importance of effective medium approaches to model inhomogeneities of surfaces and material. It is not a case of "one model is right and the rest are wrong", instead in many cases the tribocharge is due to a combination of processes involving the transfer of electrons, ions or both. Following this, we will briefly discuss the key step of disengagement and how charge return, rectification, irreversibility and dissipation play a role. Finally, we end with a discussion, and some suggestions for the future. We acknowledge that we cannot cover all of these in great detail, and many of the sections merit, and in some cases have review articles of their own; we apologize in advance for any omissions of citations.

Triboelectricity can be confusing because it spans many sub-fields of science, sometimes confused because different names are used for the same process, and it is certainly complex. Our overall conclusion is that the general form of Ampère's law used to understand much of electrostatics has a leading role in triboelectricity, combining the electric field, band bending and polarization contributions to the electric displacement field together with free and bound charges. These must be combined with classic tribology. The electric field dependence has been known for more than a century as the Volta-Helmholtz hypothesis, band bending has been invoked but not always treated quantitatively, while inclusion of electrical polarization via electromechanical terms is relatively new and fills in many gaps.

## 2. Defining the problem

Triboelectricity must be consistent with established solid-state physics. Most of the important contributing terms are not new, and can be found in work more than fifty years ago such as the reviews of Vick [30], Harper [31, 32], and Montgomery [33], the book by Loeb [34], and the slightly more recent review of Lowell and Rose-Innes [35]. Other established terms include processes such as quantum tunnelling [36] and Schottky barrier effects, as well as modern approaches to asperity contacts based upon the approach of Bowden and Tabor [37, 38]. For instance, in 1958, Levy, Wakelin, Kauzmann, and Dillon [39] indicated that asperity-asperity contacts played a role in triboelectric charging. The term which was not known until recently is flexoelectricity. Flexoelectricity was only documented in the 1960s [40-43], when it was sometimes called “bending piezoelectricity” [44, 45], but remained largely unnoticed for inorganic solids until Ma and Cross demonstrated in 2001-2002 that it could be large in relaxor ferroelectrics [46-48]. The first consideration of flexoelectricity as a contributor to triboelectricity was in our earlier work in 2019 [19].

We will argue that all models and explanations must also be consistent with the large body of experimental data, some of which is quite old. For instance, the first major scientific study was by Gilbert in 1600 [49], who mentioned in his seminal work that humidity played a role. Other constraints have also been known for more than a century, for instance the 1867 observation of Harris [50] that charge transfer can reverse with force. Sometimes the literature may appear to have internal contradictions. For instance in 1923 Richards observed that collisions of a solid insulator and a metal led to charge transfer of the opposite sign to sliding (which he called frictional effect) [51]. As will be detailed later, this is not a contradiction, it connects to competition between the contact potential and electromechanical effects, particularly flexoelectricity.

For terminology it is important to distinguish between the different potential terms that are often used. In the electrochemical literature the difference in the electron chemical potential between two materials in vacuum is the “contact potential” or “Volta potential difference” [52]. The “Galvani potential” is the chemical potential when the presence of polar molecules, dipoles and any Fermi level pinning is included, and is the sum of the Volta potential and what is called the surface potential [52]. However, in the surface science and triboelectric literature, the dipoles and similar are considered as modifying the work function. Consistent with the latter we will use the contact potential as a term that is not simply a material constant, but changes with phenomena such as dipoles and Fermi level pinning. We note that in the semiconductor literature the term “built-in potential” is used, replacing contact potential in the same sense of including the additional contributions. There is also what is called the “effective contact potential” which is some weighted mean over inhomogeneities such as different orientations.

Experimental constraints that must be explained include:

1. The role of the work functions, including the Helmholtz-Volta model of contact potential. This works well to explain simple metal contacts [53, 54] but often less so for other types of contacts.

2. The influence of band structure [30, 32, 35, 55, 56] and interface or trap states [30, 57-59]. It is well documented that these matter, but not all models include them qualitatively and very rarely quantitatively.
3. The role of dielectric constants, including the empirical Coehn's law that materials with lower dielectric constants tend to charge negative versus positive for those with a higher dielectric constant [51, 60-62].
4. The effect of the surface texture [63-65] and particle size [66, 67] of contacting bodies on the charging. For instance, Varghese *et al.* demonstrated that surface texture involving micropyramid arrays can increase the tribocharging by two orders of magnitude compared to flat surfaces [64].
5. The role of the environment around the contact, including both the temperature [68-70] and humidity [49, 71-75]. As we will see later, humidity can have more than one role.
6. How the mode of contact [58, 76-79] (impact, normal contact, rolling, or sliding) and the speed of sliding [24, 80-84] matters, including the angle of particle contacts [78, 79, 85-88]. For instance rolling typically leads to less charge than sliding, and Nordhage and Backstrom even reported a change in sign between them for Mg on NaCl [58].
7. Influence of the force of contact [80, 81, 84, 89-93] or speed of impact [51, 89, 94-98], including why the tribocurrent scales as a reduced power of the force. For instance in 1977 Nordhage and Backstrom reported a power-law dependence of 0.36, which is close to 1/3 [58], which they note is what would be expected for a Hertzian contact.
8. Why charge transfer can change sign on contact versus pull-off [19, 99], and can change sign from soft to hard contacts [50, 100, 101] and also with strain [102-104] and rubbing versus just contact [51]. For instance, Sun *et al.* [100] observed local surface charging negative with a Kelvin tip force of 115 nM, which reversed sign when the force was increased to 1150 nN.
9. The influence of surface acidity [105-107], conductivity [108, 109], or surface chemical species [110]. As we will discuss later, there are connections here to chemisorption induced changes in work function.
10. Why charge transfer can be inhomogeneous [33, 111-114], often a mosaic with regions of positive and negative charges for contact between nominally identical materials.
11. How there can be charge transfer between nominally identical materials with different curvatures [101, 115-117]. This was first observed in 1910 by Jamieson [115], whose work remained largely uncited until recently; this is a textbook example of the flexoelectric effect.

12. Why charge transfer depends on the surface roughness [118-120], macroscopic strain [91, 121], and curvature [115, 118, 122] of the contacting bodies. For instance Xie *et al*. [91] suggest that the in-plane stresses due to contacting surfaces may be a key process for charge transfer between chemically identical surfaces.

The literature spans more than two centuries, with rediscoveries or reclassifications not uncommon. To better organize the analysis, we believe that it is important to delineate the interpretations that have appeared into what we call **Drivers**, **Triggers**, **Mechanisms** and **Dependencies** [27]:

A **Driver** is the relevant free energy term that controls the thermodynamics of charge transfer. By definition, a **Driver** must correspond to a reduction in the free energy of the system, either in a closed thermodynamics sense or open thermodynamics where the total system is considered including the environment. One of these is the contact potential difference.

A **Trigger** is the local process which leads to charge transfer, typically elastic and/or plastic contacts.

The **Mechanism** is the atomistic or nanoscale process that leads to the charge transfer. It frequently is associated with a specific structural **Trigger** and has to be in response to some **Driver**. One mechanism would be hydrolytic bond breaking.

The **Dependencies** are factors, such as details of contact or the environment, which influence any of the above, for instance the humidity.

As we describe both the background in the next section and the models and explanations in the one after that, we will indicate to which of these various contributions belong. To illustrate how these fit together, figure 2 is a schematic decision tree which we argue contains the key contributors to understanding triboelectricity. The schematic traces how from differences in work function of the bulk material, the nature of the surface and interface (including states) needs to be included with adjustments from these and chemisorption. Then the issue of doping and Fermi level pinning must be included if relevant. Note that here we will not make a major differentiation between semiconductors and insulators since both have carriers and electromechanical terms; in many respects the difference between the two is more a question of the application of interest than fundamental physics. After this, the force applied and to what extent electromechanical or mechanochemical contributions matter needs to be added. Finally, after charge transfer occurs, there is the question of irreversibility which can lead to either trapping of charge carriers, conduction away or diode behavior preventing backflow of charge.

The decision tree is slightly complex, but it must be in order to accommodate the very wide range of situations where triboelectric charge transfer occurs. Fortunately, each of the different components has a sound scientific foundation, although there are still details which may need further exploration both theoretically and experimentally; we will return to these much later.

# 3. Background concepts and theory

## 3.1 Elements of tribology and contact mechanics

Whenever two bodies contact or slide against each other this involves asperities as illustrated in figure 3. What we see as two flat surfaces at the macroscale are really asperity contacts at the micro to nanoscale, each of which can be approximated by a spherical asperity on a flat surface. Depending upon the roughness of the surface, the asperity contacts can be anywhere from a few nanometers to a few microns in size; typically, there is a distribution of sizes. The macroscopic laws of friction that are taught in high school are a consequence of changes in the number of asperity contacts with load, as first detailed by Bowden and Tabor [37, 38]; they are not a fundamental property of the contact or sliding process. More recent details for rough surfaces can be found in the work of Persson [123-126] and others [127-129], an incomplete list. At each of these asperity contacts there can be charge transfer, and we can simplify the problem to considering a sum of parallel single contact charge transfers. The precise form of the elastic deformation depends upon whether there are only normal forces, the nominal case for pure contact, or also sideways forces that lead to shear, particularly when there is relative motion.

The simplest, and most useful model for a single contact was first described by Hertz [130]. It is for the case of a sphere contacting an elastic surface, and to produce an analytical solution he made some approximations. Figure 4 shows the dimensionally scaled stress and strain in the normal direction for a Hertzian contact. One of the fundamental terms is the contact radius which scales as

$$a \propto (FR/Y)^{1/3} \tag{1}$$

with *F* the force of contact, *R* the radius of the sphere, and *Y* an effective modulus of the contact. The maximum stress is under the tip and changes sign outside the contact region so that the largest strain gradients are near the edge of the contact area. The strain gradients scale inversely with the asperity contact radius,

$$\varepsilon' \propto 1/R \tag{2}$$

so are large at the nanoscale which will become important later. Related forms are known for other shapes and play a role in triboelectric analyses as shown in Table 1 and discussed by Olson and Marks [131]. There is good evidence that charge transfer correlates with Hertzian contact mechanics, for instance the force to the one-third power dependence has been measured in detail by Escobar, Chakravary and Putterman [81, 90] and is shown in figure 5. We note that the dependence also matches that for a flexoelectric solution during sliding [132], which we will return to much later.

Table 1. Scaling of Hertz solutions for the contact radius *a*, stress *σ*, flexoelectric polarization $P_{FXE}$, total flexoelectric bound surface charge $Q$, and sliding tribocurrent *J* for various rigid asperity shapes contacting an elastic flat surface. These are for 3D shapes of a spherical contact, the circular disc of a cylinder, and a circular cone, as well as 2D shapes extended in the surface plane of a roller, flat punch and a triangular prism. The scaling depends on the contact force *F*, the asperity radius *R* or half-angle *α*, the elastic modulus *Y*, and an effective, weighted mean flexoelectric coefficient *μ*.

| Asperity Shape | $a \propto$ | $\sigma \propto$ | $P_{\mathrm{FXE}} \propto \mu\sigma/aY$ | $Q \propto a^2 P_{\mathrm{FXE}}$ | $J \propto aP_{\mathrm{FXE}}$ |
|---|---|---|---|---|---|
| Sphere | $\left({FR}/{Y}\right)^{1/3}$ | $\left({FY^2}/{R^2}\right)^{1/3}$ | ${\mu}/{R}$ | $\mu\left({F^2}/{RY^2}\right)^{1/3}$ | $\mu\left({F}/{R^2 Y}\right)^{1/3}$ |
| Cylinder | $R$ | ${F}/{R^2}$ | ${\mu F}/{R^3 Y}$ | ${\mu F}/{RY}$ | ${\mu F}/{R^2 Y}$ |
| Cone | $\left({F\tan\alpha}/{Y}\right)^{1/2}$ | ${Y}/{\tan\alpha}$ | $\mu\left({Y}/{F\tan^3\alpha}\right)^{1/2}$ | $\mu\left({F}/{Y\tan\alpha}\right)^{1/2}$ | $\mu/\tan\alpha$ |
| Roller | $\left({FR}/{Y}\right)^{1/2}$ | $\left({FY}/{R}\right)^{1/2}$ | ${\mu}/{R}$ | ${\mu F}/{Y}$ | $\mu\left({F}/{RY}\right)^{1/2}$ |
| Punch | $R$ | ${F}/{R}$ | ${\mu F}/{R^2 Y}$ | ${\mu F}/{Y}$ | ${\mu F}/{RY}$ |
| Triangular Prism | ${F\tan\alpha}/{Y}$ | ${Y}/{\tan\alpha}$ | ${\mu Y}/{F\tan^2\alpha}$ | ${\mu F}/{Y}$ | ${\mu}/{\tan\alpha}$ |

For completeness, we note that the gradients in the Hertz solution diverge on the circle at the edge of the contact. This can cause numerical issues without further analysis, for example by including higher order elasticity terms or using numerical solutions which break the assumed symmetries of the Hertzian model.

The strict Hertz solution is for a purely normal force. There are other, similar approaches where shear forces are also included [130, 133-135]. As illustrated for the radial stress in figure 6, shear reduces the symmetry of the stresses and strains and leads to a significant dipolar character; further details can be found in the supplemental material to Olson and Marks [132]. Because of this asymmetry there can be a current when shear is present that is not found in the symmetric normal load case, as also discussed in Olson and Marks [132]. Values for how these scale with shape, the "sliding current" are also included in Table 1. Note that shear can also be important for problems involving powder collisions, as there will always be some tangential forces except for the rare case of a head-on collision. For example, the dependence of tribocharge on the incidence angle has been investigated for colliding particles [85-87] where both the normal and tangential components have been shown to be important; shear effects are discussed in further detail in section 4.11.

While there have been many efforts to improve upon the Hertz solution [127, 136, 137], it remains a cornerstone of much of contact mechanics because it works very well for most properties. To go beyond requires complex analytical analysis or numerical techniques, the latter now being widely available in codes such as finite element methods. These can be important; for instance, the approximations that Hertz made forces symmetry on the solution for contacts of identical materials which would lead to no charge transfer due to flexoelectric contributions. In numerical calculations there is tribocharging for different curvatures of the same material as shown by Mizzi and Marks [101].

Perhaps the largest weakness of the Hertzian approach is the neglect of adhesion, which can have a significant contribution, particularly for separation of contacts and soft materials such as polymers. For these cases adhesion needs to be included, for which there are well established models [138, 139]. With adhesion there are some differences around the edges of the contact, and also tension during separation; neither of these are present otherwise. There can also be additional terms when shear is involved. As first analyzed in 1942 by Bowden, Moore and Tabor [140] there can be ploughing of material at the front of a sliding asperity, a process which often leads to wear (material removal or transfer) as has been extensively investigated in the tribology literature [141, 142].

Any form of tribological contact involves asperities of different sizes and their elastic and plastic deformation as well as adhesion [124, 126]. The inelastic components of these deformations as well as processes such as the nucleation of dislocations [143] or the creation of new grain boundaries are what, macroscopically, is observed as friction. In many cases there is also transfer from one material to the other as what are called “transfer layers” [144], and there can also be what are called “third bodies” (additional compounds or microstructures) formed during sliding [145]. Both of these will become relevant later when we discuss different models that exist in the literature, particularly those related to material transfer. There is starting to be some inclusion of their consequences for triboelectricity, for instance the experimental observation by Zhang *et al*. that transfer films can lead to an inversion of the contact potential [146].

Some care is needed when generalizing from a single contact to a number of asperities [37, 38]. The macroscopic laws of friction are a consequence of statistical changes in the number and area of contact. A common approach (e.g. [123]) is to consider that the asperities are just at the plastic yield strength, and that with a higher stress they deform. The total contact area is then the applied load divided by the yield strength, often a small fraction of the macroscopic contact area. This means that the conversion from single asperity to multiple contacts is complex. In 2004, Chang *et al*. [147] observed that the tribocharge between several self-mating metals varies inversely with the true contact area, and a recent analysis by Zhai *et al*. [120] indicates that the charge transfer scales with the force divided by the true contact area, with more of the charge transfer on the larger microcontacts. As shown in the experimental results of figure 7, for different stiffnesses and forces the charge varies, tending to be larger for the smaller roughness amplitude. Note that smaller amplitude roughness is consistent with smaller radius of curvature (larger curvature) asperities. There is scaling with both the contact stiffness and contact area, which is comparable to that for a single asperity but more complex. There is starting to be some analysis of the true contact areas, for instance the recent work by Wang *et al*. imaging the interface using astigmatic optical microscopy [148]. We feel this area is still developing and will make a few further comments later in section 4.12 about effective medium approaches and also in the discussion.

Tribological contacts involve both **Triggers** and **Dependencies**; contact alone is not a **Driver**, but two asperity contacts can influence each other, and the strain often leads to polarization, to which we turn next. They are triggers because contacts are needed for charge transfer.

## 3.2 Electromechanical contributions

Deformation of an insulator can create polarization, and this in turn leads to surface bound charges which play a central role in triboelectricity [149]. It is useful to separate this physics into two parts, a thermodynamic energy for the coupling of deformation and polarization, and the Maxwell equations. Both are continuum approaches, although there are direct connections to the quantum formulations as will be mentioned later. The thermodynamic energy density combining both the strain and polarization terms can be written as (after reducing the number of parameters) [150, 151]:

$$W = (1/2)\chi_{ij}^{-1}P_iP_j + (1/2)C_{ijkl}e_{ij}e_{kl} - d_{ijk}P_ie_{jk} - \chi_{ij}^{-1}\mu_{jklm}(P_i\,\partial e_{kl}/\partial x_m - e_{kl}\,\partial P_i/\partial x_m) - P_iE_i - e_{ij}\sigma_{ij} \tag{3}$$

Here $P_i$ is the polarization density vector, $\chi_{ij}$ the susceptibility, $C_{ijkl}$ the elastic stiffness tensor, $e_{ij}$ the strain, $d_{ijk}$ the third-order piezoelectric tensor, $\mu_{ijkl}$ the fourth-order flexoelectric tensor, $E_i$ an external electric field vector, and $\sigma_{ij}$ the applied stress, all of which are implicitly functions of the position vector $x_m$.

We will caution readers that the literature is inconsistent about the notation used for the flexoelectric tensor. Our convention is "left to right and alphabetical" with implied summations. The first index connects to the direction of the polarization (and the susceptibility when that is relevant). The second and third indices are used for the strain, with the last being for the spatial derivative. This convention matches that of early work [40, 152, 153], but almost every other permutation can be found in the literature.

The external electric field will correlate with the contact potential (or built-in potential for semiconductors) between two materials as well as that from external charges, and the stresses will be those due to the contact forces between asperities or an asperity and a flat surface. We note that there can also be existing polarization in electrets, a point we will return to later. In addition, we will comment here that the effects of an external electric field term connects to some measurements for granular materials, see the work by Hu *et al.* [154] who observed changes with the sign of the field, and also the references therein.

As indicated by the indices in equation (3), (electro)mechanical and dielectric properties are not necessarily isotropic. In much of the triboelectric literature, including work herein, this is ignored for simplicity. However, anisotropy is likely key in some cases; for example, the sign of the flexoelectric coefficients can vary in different directions (see Table 4 later) and often elastic moduli are anisotropic. Many of the terms are also frequency dependent, particularly electrostatic ones.

The thermodynamic energy density is a functional (integrated over position). At equilibrium we obtain the relationship for the polarization:

$$P_i = \chi_{ij}E_j + \chi_{ij}d_{jkl}e_{kl} + \mu_{ijkl}\,\partial e_{jk}/\partial x_l \qquad (4)$$

and stress

$$\sigma_{kl} = C_{ijkl}e_{ij} - d_{ikl}P_i + \chi_{ij}^{-1}\mu_{jklm}\ \partial P_i/\partial x_m \qquad (5)$$

Note that this represents an "effective stress" that can have a role in plastic deformation or for boundary conditions at interfaces and surfaces.

The flexoelectric terms will play a significant role later, so a little expansion is relevant. As illustrated in figure 8, for a material with a centrosymmetric unit cell as at the top, with a homogeneous strain there is no net polarization, the inversion center is preserved. If the unit cell is asymmetric there is polarization with homogeneous strain – the piezoelectric effect, shown in the middle. However, only a limited number of materials have asymmetric unit cells and display a piezoelectric effect. With inhomogeneous strain there is always polarization in semiconductors and insulators because inversion symmetry is destroyed by strain gradients, as shown at the bottom. This is the flexoelectric effect, whose presence is independent of the unit cell symmetry [150, 151, 155-157].

The flexoelectric tensor $\mu_{ijkl}$ describes the polarization due to a strain gradient as

$$P_i = \mu_{ijkl}\, \partial e_{jk}/\partial x_l \quad (6)$$

Often, effective flexoelectric coefficients that more conveniently relate to experimental geometries are used by combing different components and taking account of the elastic modulus tensor [158]. These can differ from the single component microscopic coefficients; for example, the calculated microscopic coefficient for (100) $SrTiO_3$ is 36.9 nC/m [159] while experiments bending $SrTiO_3$ in the (100) direction find effective coefficients of 12.4 nC/m [160].

It is important to emphasize that the flexoelectric term includes a number of different terms, not just those due to the movement of (screened) ions as sketched in figure 8. These correspond to changes in the energy levels of the electrons, what is called the deformation potentials, as well as variation with strain of the work function, the mean-inner potential and also surface dipoles. As pointed out in 2014 by Stengel [161], surface terminations matter. These additional terms are variations of Coulomb potential terms, there is duplicate terminology here, and we will describe them further in the next section 3.3.

For completeness, we will note that in conventional electrostatics a metal does not have a flexoelectric (or piezoelectric) contribution. There is recent work on what are called “polar metals” [162], and investigations of flexoelectric-like contributions in some cases [163-165], as well as experimental observation of converse flexoelectricity in topological semimetals [166]. We will take a conservative approach here and assume that these terms can be neglected for all experimental cases of tribocharging, but admit that future work may prove this to be an invalid assumption.

In most cases elastic strains scale with the size of the features which lead to a deformation, here the size of the asperity contact discussed in section 3.1. Integrating the total contribution of the terms linear in the strains leads to energy contributions which scale as the cube of the contact radius. In contrast, the presence of a derivative in the flexoelectric contributions leads to a scaling as the square of the size. Hence the ratio of the flexoelectric energy contribution to the linear energy terms (e.g. piezoelectricity) scales inversely with size, so flexoelectricity becomes very significant at the nanoscale as has been mentioned by several authors [150, 156, 167-170].

While polarization is fundamental, for triboelectricity charges are more important. To convert to these some extra steps are required as we have discussed recently [149]. First, there is a standard decomposition of the polarization density $\mathbf{P}(\mathbf{r})$ as a function of position $\mathbf{r}$ into what are called “bound charges” [171]. This takes the form:

$$\mathbf{P}\cdot\mathbf{n} = \rho_b^{s\prime}\,;\ -\nabla\cdot\mathbf{P} = \rho_b \tag{7}$$

where the first equation on the left is only evaluated at the surface with $\mathbf{n}$ being the surface normal, and the second is the bulk bound charge [171]. The term “bound” is used since these charges are not free to move around but are associated with either the surface termination or changes in the polarization. For most purposes, it is easier to consider other terms such as the potential associated with the polarization and the overall charge balance in terms of these bound charges.

The second term relates to Maxwell’s equations, in particular the electric displacement field $\mathbf{D}$ (see [171-173]) via

$$\mathbf{D} = \varepsilon_0\mathbf{E} + \mathbf{P} \tag{8}$$

with $\varepsilon_0$ the vacuum permittivity and $\mathbf{E}$ the electric field from earlier. The general form of Ampère’s law is then

$$\nabla\cdot\mathbf{D} = \varepsilon_0\nabla\cdot\mathbf{E} + \nabla\cdot\mathbf{P} = \rho_f \tag{9}$$

where $\rho_f$ is the free charge density, for instance electrons, holes, or mobile ions. Note that by combining equation (7) and (9) we now have both bound and free charges. Applying the condition that the displacement field is zero outside a solid gives a boundary condition evaluated at the surface:

$$\mathbf{D}\cdot\mathbf{n} = \varepsilon_0\mathbf{E}\cdot\mathbf{n} + \mathbf{P}\cdot\mathbf{n} = \rho_f^S \tag{10}$$

with $\rho_f^S$ the surface free charge density. This contains an effective bound surface charge $\varepsilon_0\mathbf{E}\cdot\mathbf{n}$ from the electric field and is an important charge balance equation. Note that the surface here may be the interface between two materials. The equations above are standard, and the interpretation of polarization in terms of a bound charges can be traced back to the original work of Maxwell [174], see the 1951 analysis of Fisher [173] for more specifics.

There is a clear connection between these and both triboelectricity and electrostatic induction. This boundary condition is used in both the Belincourt method for measuring piezoelectric method and flexoelectric coefficient measurements. As illustrated in figure 9, piezoelectric measurements use a parallel plate configuration [175-179] while flexoelectric use either bending [151, 158, 160, 180] or a pyramid [151, 168, 181, 182] – note that the pyramid is effectively an asperity contact. In both cases the strain is oscillated to remove the contribution from the electric field $\varepsilon_0\mathbf{E}\cdot\mathbf{n}$, and the charge (effectively a tribocharge) is measured as a function of the applied strain. The potential connection between piezoelectricity and triboelectricity is well documented, but it was shown by Peterson in 1949 [183] and Harper in 1955 [184] that it was not a dominant factor. Flexoelectricity was not known until much later, so the connection to triboelectricity was not investigated until recently [19]. (In hindsight, the title of the 2006 paper by Cross [182] "*Flexoelectric effects: Charge separation in insulating solids subjected to elastic strain gradients"* was farsighted.)

This boundary condition is also sometimes used for triboelectric energy harvesting, although as we have pointed out previously [149] it has been misquoted. The mistake is that a term of $\mathbf{P}_s$ for a "surface polarization" is added [185-187], but this is inappropriate as it already exists in the boundary condition (10), as documented in electrostatics textbooks [171, 172].

The net effect of the polarization due to the inhomogeneous strain from a contact is analogous to a capacitor, as illustrated in figure 10 where we are using the physics convention with the arrow tip at the positive pole of the bulk polarization. For a parallel plate capacitor as shown in figure 10a there are bound charges inside the dielectric which are compensated by free carriers on the metal plates on either side. At an asperity contact as shown in figure 10b there will be spatially decaying polarization and associated bulk and surface bound charges. (Note the reversal of the polarization in figure 10b outside of the contact. )The bound charge far below the contact will decay to zero so there is no need for a lower contact. For a metal/insulator contact, the surface bound charges at the top will be balanced by interface charges on the metal. (What happens at the rest of the surface is more complex, and we will return to this later.) The polarization decays on a scale of about three times the contact radius, so to a first approximation contributions from adjacent asperities can be ignored.

The above is using the classical approach to polarization described in terms of microscopic dipoles, but there is also what is called the "modern theory of polarization" [188, 189] which is more general and rigorous. In this, the quantum mechanical electron density redistribution is included and polarization is described at a fundamental level in terms of the Berry phase [190], thereby avoiding some ambiguities. As analyzed in detail by Vanderbilt and King-Smith [191], the connection between the surface bound charge and the normal component of the polarization can be validated at the quantum mechanical level in terms of the Berry phase [190], and similarly for an interface. Hence we can use the continuum bound charge approach so long as we are not working at a sub-Angstrom scale).

There are also other non-linear terms such as contributions from gradient elasticity which can be relevant. An important clarification is that by "gradient elasticity", we mean here the non-linear terms for single crystals, not those which correspond to microstructure contributions as originally discussed by Mindlin [192], see for instance the discussion by Askes *et al.* [193]. The gradient scale parameter used, a length, corresponds to the size scale at which gradient terms become comparable to the conventional elastic contributions, and should be of the order of unit cell sizes for perfect crystals [192, 193]. Our analyses indicate that these regularize the numerical solutions, removing some artificial abrupt changes in charges and potentials [194] where we used an existing estimate of the gradient scale parameter for $SrTiO_3$ of $\ell$=4 Å by Stengel [195], which is a little larger than published estimates for metals [196].

These additional terms are also sometimes used in the mechanics literature where slightly different forms are used [197, 198], for instance for problems such as flexoelectric beams [197, 199] or cracks [200]. These may include higher-order terms [197-199, 201, 202] or coupled stress approaches [203]; hopefully there will be more work in the future using these approaches for triboelectric problems. Another approach which has been used in the liquid crystal literature is based upon the 1969 work of Meyer [43], which Petrov modelled as changes in curvature for membranes [204]. (The name "flexoelectricity" comes from the analysis of de Gennes and Prost [205] for liquid crystals, and was later adopted by the community for other materials.) With liquid crystals there can be much larger effects as the direction of the molecules can align; see for instance the articles in the book edited by Buker and Eber for further details [206].

Adding additional complexity to the problem is that plastic deformation can affect the electromechanical contribution, for example high dislocation densities suppressing piezoelectricity by over 50% in $K_{0.5}Na_{0.5}NbO_3$ [207] or can enhance piezoelectric responses in $BaTiO_3$ by a factor of 19 [208]. There are also contributions to flexoelectricity, for instance twin boundaries have comparable contributions to that of the bulk in $LaAlO_3$ [209]. Motion of ferroelectric domain boundaries can also have a major effect, for instance increasing the flexoelectric response of relaxor ferroelectrics by an order of magnitude [210]. There are also grain size effects on permittivity, and therefore flexoelectricity. In $SrTiO_3$ these cause changes in the flexoelectric coefficients by a factor of 3 for grain sizes from 0.39 to 10.1 μm [211]. The above is only a very brief survey, the relevant literature here (particularly for flexoelectric materials) is quite extensive.

For completeness, it should be mentioned that the different terms for the variation of the energy and polarization in a material have been considered in a very simplified "electron cloud overlap" model, where strains are considered to deform the orbitals around the atoms [212]. This is a nice pictorial approach. However, it misses the established physics of Berry phase polarization, long range screening, exchange-correlation in a multielectron system, electron delocalization, as well as relativistic and spin contributions for electrons [213] and other well-documented solid-state physics [214]. From a view of scientific rigor, this cloud model is a reinvention of established electromechanical and mechanochemical science. We will therefore not consider it further.

The electromechanical terms are **Drivers**; charge transfer due to the polarization from strain or strain gradients is frequently going to be an important process.

## 3.3 Coulomb potential terms

There are a number of terms which are connected to the electromechanical contributions but are often considered slightly differently as alluded to in the previous section, so we will use a different section for them.

The most apparent one is the Coulomb term connecting the polarization and electric fields with any local charges; this plays a role in electrostatic induction which we will discuss later in section 4.9. These terms are **Drivers**.

The second is a number of corrections to the bulk flexoelectric terms for a finite crystal as mentioned in section 3.2. It is important to clarify terminology and usage. One approach is to consider the variations with strain of work functions, surface dipoles and related terms as independent, microscopic contributions. An alternative is to bundle all these in with the bulk response, and treat electromechanical responses as macroscopic observables via the tensor coefficients in equations (3-6). We choose here the second approach which is consistent with our bundling of dipole and similar terms into the work functions. (There is nothing wrong with the other approach, it is just not the convention we are using here.)

These corrections were not known in the early flexoelectric literature when theory and experimental measurements often differed in both magnitude and sign [151, 158, 159]. While they were initially disputed [215], it is now established that the corrections are comparable in magnitude to the bulk flexoelectric response [161, 216, 217]. With strain the energy levels of electrons change, and this is described by what are called deformation potentials [218-220], whereas the average potential is called the mean-inner potential. (In the flexoelectric literature the latter has been called a "surface contribution" because it is only present with finite samples [221], but we prefer not to use this terminology [217].) Using an approximation based upon the electron scattering factors for atoms proposed by Ibers [222], the mean inner potential term can be approximately calculated for any material [217].

To illustrate these further, Table 2 shows some of the values for the flexocoupling voltage (coefficient divided by the dielectric constant) calculated by Mizzi and Marks [217]. The bulk ionic and mean inner potential terms depend upon orientation, and the surface dipole term on the termination. For instance, for silicon (100) with a 2×1 surface reconstruction, the contributions to the flexocoupling voltage (flexoelectric coefficient divided by the dielectric response) of the mean inner potential is 5.4V, whereas the contribution from the breaking of symmetry (figure 8) is 0.8V [217]. Note that silicon has a bulk flexoelectric coefficient even though there are no ions. When there is a charge carrier population, the deformation potential leads to an electron redistribution, and therefore a polarization. This is dependent on the deformation potential of the relevant band and is implicitly included in the bulk and MIP terms of Table 2. The magnitude of deformation potentials are in the eV range: silicon and germanium valence band dilatation deformation potentials are about 2 eV (per unit of dilatation) [223] and the conduction and valence band deformation potentials in $SrTiO_3$ have similar magnitudes [224].

Table 2. DFT calculated flexocoupling voltages for several surfaces with different terminations. The mean-inner potential contribution ($f_{MIP}$) is the sum of the bulk ($f_{MIP}^{bulk}$) and surface ($f_{MIP}^{surf}$) components. The total flexoelectric response ($f_{total}$) is the sum of $f_{MIP}$ and the bulk flexocoupling voltage calculated from values in [159] and included as ($f_{FeX}$). All values correspond to beam bending.

| | | $f_{FeX}$(V) | $f_{MIP}^{bulk}$ (V) | $f_{MIP}^{surf}$ (V) | $f_{MIP}$ (V) | $f_{total}$ (V) |
|---|---|---|---|---|---|---|
| $SrTiO_3$ (100) | SrO | -7.7 | 8.1 | -1.9 | 6.2 | -1.5 |
| | TiO2 | | | 1.8 | 9.9 | 2.2 |
| | c(4x2) | | | 0.0 | 8.1 | 0.4 |
| | (2x1) | | | 0.7 | 8.8 | 1.1 |
| MgO (100) | Bulk | -2.4 | 6.1 | -0.4 | 5.7 | 3.3 |
| MgO (111) | Mg-oct | -4.7 | 8.9 | -2 | 6.9 | 2.2 |
| | O-oct | | | 0.5 | 9.4 | 4.7 |
| | (1x1)H | | 8.8 | -0.3 | 8.5 | 3.8 |
| Si (100) | (1x1) | 0.8 | 5.5 | 0.0 | 5.5 | 6.3 |
| | (2x1) | | 5.5 | -0.1 | 5.4 | 6.2 |

Beyond these terms there are electronic terms present due to the difference in the energy of the electrons between dissimilar materials. In the triboelectric literature, these are referred to as contact potentials. For metals, the contact potential is the difference between the work functions. Note that here we are defining the work function as including dipole and crystallographic orientation effects discussed in the next sections; see also section 4.3 and figures 20-22 later. In a triboelectric contact, the net effect of contact potential differences is to create an electric field across the interface of contacting materials, thereby contributing to the $P_iE_i$ term in equation (3).

Until now, our discussion of electronic contributions has considered only those present in every contact (including between neutrally charged bodies). However, in most environments and especially in those where triboelectricity is of interest, the contacting bodies will be charged. This might be a smoothly-distributed surface charge on a metal particle, or a mosaic distribution of charge densities on a surface with lower conductivity [114]. Electric fields from charges already extant on contact surfaces also cause changes also due to the $P_i E_i$ term of equation (3). This is demonstrated in figure 11a, where Matsusaka *et al.* [225] showed, as others also have [94, 226, 227], that charge transferred during particle impacts depended on their initial charge. As the figure indicates, there is more positive charge transfer the more negative the initial particle is. This is an illustration of electrostatic induction which is an important **Driver** that we will return to later in section 4.9.

As early as 1920, Richards [89] and others [95, 225, 228-230] showed that there is typically a maximum saturation charge for multiple contacts; figure 11b shows an example of this also from the work of Matsusaka *et al.* which in this case is about 17 nC. We infer that this number matches when the electron-electron Coulomb repulsion of the charges on the sphere balances contact potential or similar **Drivers**.

The terms discussed in this section directly contribute to the electric potential in contacting materials and, as a result, the thermodynamic energy density, and are therefore **Drivers** of charge transfer.

## 3.4 Interface/surface structure, chemisorption and work functions

There are local phenomena at the interface or surface where chemisorption and/or changes in chemistry take place which can have a major influence on the work functions and therefore contact potentials. While these are standard in surface science, it does not appear that there is the same level of recognition in the community working on contact electrification and triboelectricity, although it was discussed as early as 1953 by Vick [30] and Peterson in 1954 [231].

The contact potential between two materials is not a material parameter; it is sensitive to the surface structure and crystallographic orientation of the material. This has a long history, for instance in 1928 Farnsworth and Rose demonstrated that there was a contact potential of 0.463(2) V between a Cu (111) oriented surface and a (100) oriented one [232]. As shown early by Bardeen [233] and then more completely by Lang and Kohn [234], the electrons in a material leak out into the vacuum leading to a dipole at the surface, For simple metals this increases the work function. Depending upon the planar density, this leakage can be larger or smaller, leading to the well documented variation of work functions with orientation. For reference, an extensive list of calculated values for elemental crystals has been published by Tran *et al.* [235]. There is also extensive evidence for local variations of work functions in polymers due to fluctuations in either or both dopant concentrations and polymer configuration and microstructure. For instance, Semenikhin *et al*. [236, 237] report variations of about 45 mV, Barisci *et al*. [238] 100 mV and O'Neil *et al*. [239] 200 mV.

As an illustration of another way the work function can vary, Table 3 shows calculated values for the work function for different surface reconstructions of $SrTiO_3$ (001); see reference [217] for more details. The strontium-oxide terminated surface has a low work function, while the commonly observed 2x1 reconstruction [240] has a much higher work function.

Table 3: The calculated work function (WFn) for different surface terminations of $SrTiO_3$ (001) in volts. Depending upon how the surface has been treated all of these are possible and would lead to quite different contact potentials. See [217, 241] for full descriptions of the surface structures.

| Surface | SrO | $TiO_2$ | (1x1)DL | (2x2)A | (2x2)C | C(4x2) | (2x1) |
|---|---|---|---|---|---|---|---|
| WFn (V) | 4.0 | 5.6 | 6.0 | 6.7 | 7.4 | 7.2 | 8.2 |

There can also be significant changes due to chemisorption, see for instance the calculations of Leung *et al*. [242]. These are not monotonic, for instance Leung *et al.* found that a monolayer of oxygen on the (100) surface of tungsten decreases the work function by 0.54 V, whereas on the (110) surface it increases the work function by 1.1 V. On most oxide surfaces (and sometimes other materials) water can dissociatively adsorb leading to hydroxide with the more positive hydrogen further from the surface. This hydroxide dipole will reduce the work function. As an illustration of this, figure 12 plots the work function of a MgO (111) surface with a hydroxide termination where some of the hydroxides have been replaced by chloride [27]. Since chloride is a negatively charged adsorbent without a major dipole, the work function increases from 2.6 V with just hydroxide, to 5.6 V with saturated chloride. These results are consistent with observations by Lowell and Brown in 1988 [243] of changes in the tribocharging of different metals with polyvinyl alcohol treated with halogens to replace hydroxide groups, as well as similar results of Tanoue *et al.* [244]. Lowell and Brown observed that there can be a change in the sign of the charge transfer, which is also consistent with the magnitude of the work function changes above. This is also consistent with work where a correlation was found with surface acidity [105-107], since surface acidity and electron acceptance are linked to surface dipoles.

Similar is the observation of a linear relationship with thiol chain length of the surface dipole for monolayers of alkyl thiols adsorbed on gold by Evans and Ulman [245], and comparable results for alkanethiols by Alloway *et al*. [246] and Lee *et al*. [247]. How large the effects of chemisorbants are depends in part upon the polarizability of the surface; as reviewed by Zojer *et al*. [248], cases where the surface is more polarizable show larger work function changes on chemisorption. (Note the connection here to the empirical Coehn's law that materials with lower dielectric constants tend to charge negative versus positive for a higher dielectric constant [60-62].) Comparable changes are known for semiconductors, for instance changes in the work function of GaAs by about 1 V after adsorption of bezoic acid derivatives measured by Bastide *et al*.; the more electronegative is the tail of the acid outside the surface, the larger is the work function [249].

We note that there are effects going the other way, that is electric fields (which could be electromechanical as in section 3.3 or Coulomb as in section 3.4) change chemisorption – the contributions are coupled. This has been studied in electrochemistry, for instance the work by Mei *et al.* [250] who show significant changes in the adsorption energy of several species with fields of the order of 10 V $nm^{-1}$. Flexoelectric fields, for example, can reach this magnitude [19].

The same type of effect can occur at interfaces, and there is extensive literature on this. There can be penetration of the electron states in the metal into an adjacent insulator leading to what are called *metal induced gap states* [251, 252]. For instance, for a Pt/Ir tip on $SrTiO_3$ [56], and for other metal/oxide contacts [253], the change in the effective work function due to dipoles at an interface can be from 0.1 V to 1 V. As one example, figure 13 shows a plot of the electrostatic potential as a function of position for an interface between either nickel or aluminum and a thin layer of nickel oxide taken from reference [254]. For the Ni/NiO system the interface dipole points into the metal, whereas it is in the other direction for Al/NiO. There is also other evidence for interfacial dipoles [255-258]. This is not limited to inorganic systems; it is also known that there are often large interface dipoles between organic semiconductors or polymers and metals [248, 259-263].

Impurities at interfaces can also have a large effect. For instance the work function of thin MgO film can be changed by Mg atom incorporation at the MgO/Ag(001) interfaces[264, 265] while in other research calculations suggested that the work function of MgO/Ag (001) could be changed by up to 1.3 V by oxygen impurities [266]. The effect of different metal supports on the work functions and adsorption energies of oxide surfaces have been found in multiple metal/oxide systems [265-280].

The experimental evidence for the role of chemisorbants and surface structure in triboelectric charging has a long history. For instance, in his 1905 textbook Hadley mentions that there is a change in the sign of glass rubbed with silk if they are both first dried in a sand-oven (150-250 °C) versus heating the glass with a Bunsen burner (>1000 °C) [281]. At that time tools were not available to work out what this was due to; a modern interpretation is that a sand-oven will drive off weakly chemisorbed species such as adsorbed water, whereas the much higher temperature of a Bunsen burner will desorb strongly chemisorbed species such as dissociated water (hydroxides at the surface). This interpretation is consistent with the recent work of Groshean and Waitukaitis who showed that cleaning and baking samples was of critical importance in obtaining reproducible results for single-collision experiments [282]. There is also literature such as the analysis by Debeau in 1944 [283] who showed that charge transfer between quartz and sodium chloride on nickel could be fit by a classic Langmuir isotherm [284], and that chemisorbed water was a "poison" for his system. There is also work where the changes of work function with adsorption has been studied for industrial purposes, for instance to facilitate choice of materials for hair combs to reduce static electricity [109], and the work by Lowell and Brown in 1988 mentioned above [243]. Quite recently Byun *et al.* have exploited these processes to improve triboelectrification using electron donating or accepting groups [285]. More discussion will be given later in section 4.7 on humidity and section 4.8 on the environment.

These terms are a mixture of **Drivers** and **Dependencies**. Some are drivers as the contact potential between two bodies depends upon the difference between their work functions, so changing those will change both the magnitude and sign of thermodynamically downhill charge transfer. Others are dependencies which are external factors during experiments such as humidity, which have indirect effects on the contact potential. Incomplete experimental control of impurities and surface adsorbents may well be a significant source of irreproducibility in triboelectric experiments.

## 3.5 Selvedge layer electronic structure (band bending)

The selvedge layer is a region which can range from a nanometer to even hundreds of nanometers below the surface of a material over which there are changes in properties; we differentiate this from the atomic layer immediately at the surface. The most important properties here are how the energy levels for contact of two dissimilar materials relate to the chemical potential of the electrons, the Fermi level, and change as a function of depth within the materials. This is referred to in the semiconductor literature as band bending. Over a region which is of the order of the Debye length there is a change in the number of occupied dopant levels as well as holes in the valence band and electrons in the conduction band. The width of this region can range from 10 nm for materials with high dopant levels, to microns for low doping. Measurements of the penetration depth of tribocharge are of a comparable scale to the Debye length [286-289] of 10-600 nm, although smaller values of 3nm have been reported for fluoropolymers [290]; see also the early theoretical estimates by Chowdry and Westgate [291] of 100 nm.

The most widely known example of band bending is the p-n junction, where the alignment of Fermi levels during contact of an n-type and p-type semiconductor creates a rectifying diode. A band diagram of this process is shown in figure 14a, and in figure 14b for the similar case of equal types but different doping levels. Similar band bending occurs in any contact involving two different materials with at least one non-metal. Beyond a p-n junction, this can result in the formation of (an incomplete list) Schottky diodes (figure 14c), a metal-insulator-semiconductor contact (figure 14d), or Ohmic contacts (figure 14e). In all cases, the details depend upon the work functions, Fermi energies, the doping level and possible pinning states. In addition, there can be diode barriers to charge flow. Which of these is experimentally relevant is not always documented in triboelectric publications.

While band bending was first analyzed for problems such as metal contacts on doped silicon, it is a general phenomenon and has recently been well documented for polymer-metal contacts, see for instance the reviews by Opitz [263] and Zojer *et al*. [248].

In some cases, such as in Schottky contacts, charge carriers can induce charges on the opposite body and be influenced by the resulting potential. This is known as the image potential or image-force. The image charge of opposite charge in the metal produces the same potential and attracts the charge carrier, reducing the electronic barrier. Image effects are typically small in magnitude compared to electromechanical terms, and our calculations indicate that they can shift the barrier slightly away from the interface [56].

A phenomenon of some importance is Fermi-level pinning, as this complicates the contact potential and may well be a source of some of the confusing results in the literature, particularly for inorganic insulators and polymers. If there are a significant number of available states at (or near) the interface, these can dominate so the band bending between the two materials is determined by them, not the intrinsic work function differences of the bulk – see figure 14c-d. (We will return to this in section 4.3 and figure 21.) These states may be intrinsic to the material or come from chemisorption. Often raw samples left in the air for some time will have defects and chemisorbants at the surface. In surface science one of the standard techniques is to heat samples to remove chemisorbants and reduce the density of defects at surfaces. It is worth mentioning that changes in tribocharging on annealing surfaces can be traced back over a century to the detailed study in 1915 by Shaw [292]. There is also more recent work indicating that annealing can significantly change flexoelectric coefficients, an alternative process where there are (presumably) changes in the concentration and nature of defects in the selvedge region altering the flexoelectric response [293].

We note that the total charge transferred from (or to) the depletion zone equates to the free carriers needed to compensate it via the generalized Ampère equation and the electric divergence field at the surface, as discussed in section 3.2.

Band bending near the surface of materials in contact results in potential gradients (or equivalently bound charges or polarization) that act as **Drivers** of charge transfer. In cases involving rectification, such as p-n or Schottky junctions, they also provide a **Mechanism** for the asymmetry of charge transfer during a contact versus detachment as briefly described later in section 5.2.

## 3.6 Mechanochemistry and tribochemistry

The term "mechanochemistry" has become popular in the 21st century and refers to chemical transformations initiated or sustained by mechanical force [294, 295]. A similar term that is more common in the tribology community is "tribochemistry" [296, 297]. The key component in both cases is the formation of some species due to an applied mechanical action (stress) as illustrated in figure 15. In the case shown the state on the right becomes lower in energy after the application of stress, and the activation energy barrier is reduced. The example shown would lead to a pressure-induced chemical reaction or phase change, for which an example is applying pressure to ice to melt it. Changes in activation energy barriers are also common, for instance in the transformations of carbon containing materials to graphitic carbons – see Hoffman and Marks [22] and references therein, and also Wang *et al.* [23].

We will only discuss mechanically enhanced or mechanically driven chemical reactions involving surface sites of the rubbed materials and/or species adsorbed on their surfaces. We will also limit our discussion to reactions that might result in charge transfer, as opposed to reactions that are enhanced or made possible by triboelectric charging, which is a separate area of study [298-300]. Finally, for the sake of completeness it should be noted that there is a long history of interpretations in tribology in terms of stress-assisted processes dating back to the early work of Prandtl [301] and Eyring [302] as reviewed by Spikes [303].

One major problem with some of the literature needs to be clearly stated – duplication of terms. In the classic physics literature variations in the positions of energy bands as a function of pressure is described in terms of deformation potentials as introduced in 1950 by Bardeen and Shockley [218], see also the analyses by Van de Walle and Martin [219] as well as Franceschetti *et al.* [220]. These terms are included in the electromechanical contributions described in section 3.2-3.5. The chemistry literature instead describes how stress and strain modify reactions as mechanochemistry. To expand on this, in terms of density functional theory there are two ways to calculate the effect of strain and strain gradients:

1. For any given strain, solve for the orbitals and energies for a given set of atomic positions. The effective potentials based upon the ion positions, Coulomb contributions and exchange-correlation terms are calculated self-consistently for each set of positions.
2. Determine the change in the potential as a function of the atomic positions due to the ions, Coulomb contributions and exchange-correlation terms. Then treat this as a perturbation when calculating for different positions.

Of these, approach 1 is mechanochemical, while approach 2 is electromechanical. Both approaches are "right"; they are just alternate ways to describe the same phenomenon assuming that all terms are fully included. As a caveat, with finite computational resources and pragmatic approximations the mechanochemical approach may be better for local changes, and the potential perturbation better for long-range contributions.

There is also potentially confusing duplication in what the charge carriers are. In a chemical approach it is common to consider ions, which are localized charges of either sign on atoms. Often a more physics approach starts from a delocalized model where the charge (of either sign) is spread over a number of sites. However, these can also be local charges, in which case they would be described more as polarons [304, 305].

It is important to mention that external mechanical forces, which are typically considered in mechanochemistry, and the electric field from long-range electromechanical contributions can be strongly synergistic. As an example, figure 16 shows calculations from Scheele and Neudecker of the effect of the two for breaking the sulphur-sulphur bond of a linear disulphide molecule [306]. There is a close to quadratic reduction with electric field of the force required to break the bond, and for fields larger than 0.025 atomic units (12.9 V $nm^{-1}$) the dimer is unstable. Fields of this magnitude are reasonable both for contact and electromechanical potentials .

Within a chemical context these can be considered as **Drivers** of chemical reactions that may lead to charge transfer, particularly if the reactions are heterolytic (involving unequal charge partitioning).

## 3.7 Trap states and irreversibility

Charge transfer due to contact or sliding is an irreversible process, which requires something more than equilibrium thermodynamics. As an example, if we consider infinitesimally slow contact and separation of two electrically insulated metals with different work functions, there will be charge transfer on contact due to the contact potential and any Coulomb coupling between the two metals. As they separate, the Coulomb coupling drops, so thermodynamically some of the charge can transfer back; the importance of back transfer has been known for some time [77, 307]. If charge could freely tunnel between the two then most of it would transfer back on separation; when this stops or there are other processes that prevents this is important. One possible reason for the irreversibility is that after the transfer, charge moves to some other state in the material, different in energy or location. If the back flow of charge from that state is forbidden or hindered then it would be a charge trap, leading to irreversibility.

As illustrated in figure 17, empty trap states will play a significant role in leading to an irreversible change. Here the concept is that on contact, local electron (16a) or hole (16b) states can become energetically accessible due to band bending, and then participate in charge transfer with a reduction in the total energy. The charge carriers may not readily transfer back as the contact is released due to either or both the energy barrier or because they are spatially located away from the interface. The trap states may be at the surface, such as Tamm states [157, 308], or they could be in the bulk and related to point defects [309]. For the case of two identical materials with different curvatures, the flexoelectric contributions and band bending on the two sides will be different as discussed by Mizzi and Marks [101]. Consequently, there will be a preference for the occupation of trap states on one of the two materials, consistent with inhomogeneous charging connected to local asperities.

There has been significant literature where a role for trap states in static charging has been invoked dating back to the early work of Vick [30], Fukada and Fowler [57], as well as Nordhage and Backstrom [58] and others [291, 310-312]. There is also work by Lowell who observed that surface states formed on exposure to the atmosphere mattered for some polymers [313]. A recent review has been given by Molinié [59]. It is worth noting that in the catalysis and oxide surface science literature, it is known that water is a pernicious chemisorbant and often will passivate lower co-ordination or reduced atoms at surfaces. Since such surface defects are potential electronic trap states there is a strong circumstantial case for a connection between humidity and trap states. There is some newer literature, for instance the recent work by Xu *et al.*, which has made the important step of using modern semiconductor characterization methods on the traps and aspects of the transport, see for instance figure 18 which shows how charge can enter or leave trap states, as well as experimentally measured trap state energies and densities [314].

The presence of trap states provides a **Mechanism** for the irreversibility of charge transfer and can also be considered as a **Trigger**.

# 4. Proposed Drivers, Triggers, Mechanisms and Dependencies

Many different factors and models have been proposed over the years (centuries) to explain triboelectricity. Some of these are empirical, which at least in the early days was appropriate. Quite a few are not very rigorous, and if they do not base their arguments on accepted fundamental science we are not going to consider them further. Our view is that any legitimate model must involve a proper thermodynamic path with a **Driver** for the system to change to a lower free energy where the different contributions can be (or have been) independently measured. More recently, some authors have been attempting to tackle the problem at a more fundamental level, for instance by involving specific calculations of work function changes or trap states using density functional theory (DFT) approaches [315-322]. These are starting to make some headway, although as yet most have been for models which are much smaller than real systems due to the computational expense.

In the next sections we will discuss the main approaches.

## 4.1 Frictional electrification

No overview of triboelectricity would be complete without mentioning "frictional electrification," which occurs throughout the literature dating back to at least Peclet's 1834 work *"Memoir on Electricity produced by Friction"* [323]. Sometimes the term is used in the spirit of "charge transfer when there is also friction", i.e., referring to sliding versus contact electrification. What we need to be clear about is that friction is a consequence of a range of dissipative terms such as the excitation of phonons, the creation of crystallographic defects such as dislocations and grain boundaries by plastic deformation, as well as other dissipative processes. Friction is not a driver or a fundamental force, so we will argue that the term *frictional electrification* should be used with care, if at all.

Related, to what extent charge transfer contributes to friction has been debated in the literature. Some have argued that it is significant [324, 325], and others that it is small [39, 184]. Two interesting recent papers claim the fabrication of superlubricious (very low friction coefficient) triboelectric nanogenerators (TENG) [326, 327]. In such a device, the retarding frictional force from the charge transfer must be small. As a caveat, a detailed analysis by Guerret-Peicourt, Bec and Treheux suggested that the connection to friction is via trapped charges, not the full tribocharge, which may explain some discrepancies [328] particularly for open circuit versus closed circuit operation.

Independent of the extent charge transfer contributes to friction, in some cases there may be a connection, for instance the work of Nakayama [329], who found that the surface potential and friction coefficients had a positive correlation. There is good evidence that when there is charge transfer between two bodies this leads to increased adhesion, which in turn will increase the dissipation from friction, and also the deformation at asperities and the true contact area. For instance, Horn and Smith [330] measured additions to the work of adhesion of between 6 and 9 J/m$^2$ for non-sliding contacts between smooth insulators, with similar values from other authors [331, 332]. This adhesion will change the contact pressure, hence the effective friction and consequently the electromechanical contributions. In addition, when there is more friction there will also be more shear deformation of contacts and hence electromechanical contributions during sliding.

Charge transfer does not occur because of the triboelectric effect, although this statement does appear in the literature. The triboelectric effect (or frictional electrification) is a ***name*** for charge transfer due to sliding and/or contact; it is not a **Driver**.

## 4.2 Triboelectric Series

A commonly used approach is the concept of a triboelectric series, first proposed by Wicke in his 1757 PhD Thesis [333]. He arranged different materials into a series where, depending upon their positions, the earlier ones in the series would charge negative when in contact with those later in the series. Interestingly, this work predates that of Rev. A. Bennet in 1789 [334] who started to formalized how different materials led to tribocharge, see also the analysis by Elliot [335]. Bennet is generally credited with the first work on what he termed "adhesive electricity", which Volta in 1800 called "contact electricity" [336], see also the translation of Volta's 1800 letter in the book by Dibner [337]. From a tribological viewpoint, Bennet's original term has relevance; his model would map to both Hertz and JKR theories [138, 139] whereas Volta's approach is a Hertz model [127, 130], really more for batteries and electrochemical potentials. However, Volta's terminology has become the norm.

Triboelectric series have some scientific justification, but there are also major problems. The justification for them is that there is some charge transfer (for gentle contacts) which depends upon the contact potential between two materials; a necessary condition for equilibrium is that the chemical potential of the electrons (Fermi levels) is the same. We leave further discussion of the contact potential to the next section. One key weakness is that many of what were considered as "materials" when Wicke wrote his thesis are composites for which the concept of a single electronic property is known to be an inadequate description. For instance, wood is a composite so the work function can at best be considered as some weighted mean of the different constituents. Another is that the concept of a triboelectric series that just depends upon what the two materials are ignores the well-established role of crystallography, interfaces and band-bending described in sections 3.2-3.5 and below in section 4.3.

It has also been documented for more than a century that the triboelectric series frequently fail to predict results and cannot explain established phenomena such as charging due to contacts of two pieces of the same material. There are many cases of "cyclic triboelectric series", and an example is shown in figure 19, adapted from Pan and Zhang [338]. Around the circle every material will have a positive charge with respect to the material clockwise from it, which contradicts the concept of a triboelectric series.

Unfortunately, triboelectric series are often taught in high schools and even continue to be used in some publications. This is an example of the "lie-to-children" phenomenon discussed by Jack Cohen and Ian Stewart in their 1994 book [339]; we simplify so children understand by being inaccurate.

The triboelectric series can be thought of as an empirical **Driver** which connects to the contact potential. However, it is not *ab-initio* science so it is not an approach which explains tribocharging and we argue should be used with extreme care.

## 4.3 Contact potential

The contact potential, sometimes called the Volta-Helmholtz model, is a fully rigorous concept that has been known to play a role for more than half a century, see for instance the 1977 review of earlier work by Gallo and Ahuja [340]. In all materials, the equilibrium energy of the electrons, also called their chemical potential or Fermi level, depends upon the chemical composition and the arrangement of the atoms. The difference in the electron chemical potential between two materials is the contact potential, including polar molecules, dipoles and any Fermi level pinning. Electrons, holes or other charge carriers will be transferred until the chemical potentials are the same as illustrated in figure 20. As mentioned above, this is a partial rationalization for the use of a triboelectric series parameterization.

Careful experiments with metals have confirmed this, as well as an implicit connection between location of the metals in a triboelectric series and the sign of charge transfer [35, 341]. We should point out that for a metal-insulator-metal system, Fermi level pinning in the insulator will not play a role in changing the contact potential if the insulator is thin and the pinning is similar at both interfaces. Similarly, if there are chemisorbants at the interface between two metals there may be significant cancellation of adsorbate effects. Hence, it is reasonable that even without completely clean metal to metal contacts, a simple interpretation holds and the contact potential $V_c$ will primarily be the difference between the two work functions $V_c \approx \Delta\phi$.

While for metal on metal the evidence for a key role for the contact potential is clear, even early it was unclear whether this worked for dielectrics. As illustrated in figure 21, it is still correct to consider the contact potential as the difference in work functions far into the bulk. If the Fermi level is not pinned, this is what will govern the band bending. However, if the Fermi level is pinned, as illustrated by the red states at the interface, then $V_c = V_D - V_{CB}$ which corresponds to the change in energy to bring the defect states to the metal Fermi level. In the semiconductor literature, this is referred to as the built-in potential, although it would map to the term contact potential in the triboelectric field.

These factors are almost certainly a major contributor to why experimental measurements of charge transfer involving insulators or conductors often do not line up with the differences in work functions measured using bulk materials. We also note that in any experimental measurement of the work function of an insulator or conductor in isolation the native band bending of the free surface will need to be accounted for as this is not present at an interface.

It is useful to repeat here the standard abrupt pn junction approximation for the depletion width that can be found in textbooks [28, 29] – noting that the metal can be approximated as a semiconductor with a very high density of states, and the above definitions of the built-in potential which is not always the same as the contact potential. The approach involves matching the electric displacement field at the boundary which is how similar problems are handled elsewhere in this manuscript. In the simplest model, all the carriers are assumed to be transferred over a certain width (there are more complex models). The total width of the depletion zone is:

$$W = x_p + x_n = \sqrt{(\frac{2\epsilon}{e})V_c(1/N_p + 1/N_n)} \tag{11}$$

where $x_p$ and $x_n$ are the p-type (larger work function) and n-type (smaller work function) widths, with a dielectric constant $\epsilon$, the electron charge $e$, the built-in potential $V_c$ and number densities of charge carriers in the two sides $N_p$ and $N_n$; see figure 22 for the standard representation of the potential V and electric displacement field component normal to the interface D. For a pn contact with the same host but different dopants, the built-in potential is the difference between the conduction (or valence) bands of the two, so it is not a simple material property. For an inhomogeneous semiconductor the values are a little more complex, but the problem is established semiconductor physics [28, 29].

Being specific, the two widths are

$$x_p = W\frac{N_n}{N_p+N_n}; x_n = W\frac{N_p}{N_p+N_n} \tag{12}$$

with the total charge transferred per unit area of the contact of:

$$Q = ex_pN_p = ex_nN_n \tag{13}$$

For a metal with a high number of charge carriers the depletion width becomes very small, so does not really exist in any meaningful way. However, the total charge transferred which is an upper bound for a single contact is still approximately correct. If we take the metal as the n-type material (smaller electron affinity), and use as limit of $N_n \rightarrow \infty$ these reduce to

$$W = x_p = \sqrt{\left(\frac{2\epsilon}{e}\right) V_c / N_p} \, ; Q = \sqrt{2e\epsilon V_c N_p} \tag{14}$$

The depletion region can act as a capacitor, which has one noteworthy consequence. There is recent data indicating that doping alters apparent flexoelectric coefficients quite dramatically [342-344] which are believed to be due to Maxwell-Wagner polarization [345], also called the barrier-layer effect [346, 347]. (Remember that our convention is to bundle all microscopic contributors into the macroscopic electromechanical response, in which case it is correct to consider an enhanced flexoelectric response.) The higher the doping, the smaller the width so the larger the capacitance. This is believed to lead to apparent enhancements of the flexoelectric coefficient via a surface piezoelectric contribution. We speculate that there may be comparable changes in the triboelectric behavior, fertile grounds for future work as we mention later in section 6.

When two materials are separated by some distance d, as shown in figure 22, continuity of the electric displacement field across the divide leads to a linear potential variation (electric displacement field is constant in the gap), similar to a capacitor. The various terms for the electrostatics in equations (11) to (14) only change by replacing the built-in potential $V_c$ by $V_c - Qd/\epsilon$. For instance, the limit for the charge with an n-type metal is then, substituting into equation (14) and solving:

$$Q = eN_p\left(\sqrt{d^2 + \epsilon V_c / eN_p} - d\right) \tag{15}$$

The dependence here is consistent with Maxwell's equations: more charge transfer with more available carriers in the dielectric because the modulus of the electric displacement field at the interface is larger, plus an increase with the work function difference. It is similar to the simple capacitor model (see section 5.1), with corrections for the carrier variations. We note that more accurate models based upon experimental measurements of the band bending are available in a few cases, for instance for $SrTiO_3$ [348, 349] or by *ab-initio* methods [350, 351]. Also, while the solution for a metal is a reasonable approximation, it does neglect contributions such as the Friedel oscillations of the electron density [352, 353] as well as multilayer relaxations of the atomic positions [354, 355] which will alter slightly the electrostatics near the surface.

It merits noting that while the above is standard, textbook semiconductor physics, it is rarely described in the triboelectric literature; we think this may well contribute to the confusion. It can be found in the work of Zhang *et al.* [356] who unfortunately do not emphasize the connection to standard semiconductor devices. He *et al.* do include that they are using a conventional interface model [357]. We believe this is an area where much more experimental work should be done, particularly connections to Fermi level pinning and measurements of the band bending coupled with triboelectric measurements.

The more complicated behavior of the contact potential and depletion region also connects to the early work in 1898 of Coehn [60] relating charge transfer and the dielectric constant in what became known as “Coehn’s law”, see also the work by Richard in 1923 [51] as well as more recent work [55, 358, 359]. The dielectric constant has a natural role in screening of the potential, so will alter the effect of any interface dipole. There are other consequences of the dielectric constant which we will return to when we discuss the electromechanical contributions in section 4.11.

Note that many of the materials that have been used are polycrystalline or composites, which can have different work functions in different areas, as mentioned in section 3.3. Sometimes effective contact potentials are assigned to non-metallic bodies to preserve the model. “Bodies”, and not “materials”, is used because this effective potential depends on many factors. Assigning contact potentials, or even effective contact potentials, to *materials* fails to capture same material tribocharging and the dependence on pre-contact strain. As discussed earlier in sections 3.3-3.5, there is well established science that indicates that contact potentials are not simple terms but have strong variability. It would be more rigorous to use some form of weighted effective work function similar to the proposal of Tran *et al.* [235], or for dielectrics a weighted effective built-in potential; see also section 4.12 later. We note that there is triboelectric evidence for this in the results of Shio in 2001 who observed charge transfer between polycrystalline and monocrystalline ice [360].

There are cases where there are clear experimental connections to the earlier discussion of the effect of surface chemisorption in section 3.4, for instance a dependence of tribocharge upon the surface polarity of polymers [361, 362]; a more complete list of some of the different cases where surface chemisorbants affect tribocharging can be found in the recent reviews by Lyu and Ciampi [363] and by Malekar *et al*. [364].

For completeness, it should be mentioned that in some cases effectively the same arguments as those for a bulk band-structure are used, but with a more molecular-based description in terms of the highest occupied molecular orbitals (HOMO) and lowest unoccupied molecular orbitals (LUMO). These approaches are in principle identical, the description of the electrons is just using more localized atomic-like orbitals instead of extended plane waves or other types of bases.

To summarize, the contact potential (or built-in potential) should be generalized and written in a form

$$V_c \approx \Delta\phi + \mathbf{P}_{\boldsymbol{Dipole}} \cdot \mathbf{d}_{\boldsymbol{Int}} + \Delta\phi_{Bending} \tag{16}$$

with the bulk work function difference $\Delta\phi$, $\mathbf{d}_{\boldsymbol{Int}}$ an interface separation associated with the polarization due to an interface dipole $\mathbf{P}_{\boldsymbol{Dipole}}$ and $\Delta\phi_{Bending}$, a possible band bending correction due to Fermi level pinning. Associated with this there will be some charge transfer, which, from the generalized Ampère equation from section 3.2 (or the divergence theorem), connects to the normal component of the electric displacement field from the combination of work function differences, polarization and the depletion zone. Note that the connection between the charge and contact potential is therefore more complex for a dielectric.

The contact potential is a **Driver**, and we will argue that it should always be included as a factor. It justifies in terms of modern solid-state physics the Helmholtz-Volta hypothesis, the first of the experimental constraints in section 2. However, we need to take account of the variability due to chemisorption, band bending and other effects, and not take the contact potential as an invariant materials property.

### 4.4 Local heating and electronic excitation

One suggestion in the literature is that electron transfer is a result of elevated temperatures due to local heating at the contact interface. As far back as 1941, Frenkel suggested [365] that hotspots could heat by 1000 K on one side of a semiconductor-semiconductor contact, and that this would excite carriers to higher energy levels which could then drop into lower energy states in the other material; here thermal energy is the **Driver**. This was taken further by Henry in 1953 who hypothesized that in asymmetric rubbing, temperature gradients were responsible for tribocharging between identical materials [112]. Some measurements by Bowles in 1961 [366] partially supported this for similar polyethylene surfaces, although their results indicate that only 10-20% of the total charge transfer could be attributed to temperature differences. Opposing this in 1985, Lowell and Truscott [367] observed that their charge transfer was not affected by sliding speed, which they argued was a counterindication. Note also that in 1960 Harper challenged Frenkel's calculation, arguing that with a corrected time scale the temperature rise would be negligible [32].

More recently, Shin *et al*. [368] proposed a governing law based on thermoelectric potentials at the interface. Other works have indicated that temperature is important because thermionic emission and tunneling are involved in triboelectricity [56, 369], or that the Seebeck effect matters [370], but that these are only a component of a complete theory. Correlations have been observed between thermoelectric coefficients and triboelectric output, but little further detail has been shown [371, 372]. Zhang *et al*. [24] suggest that effects from frictional heating can in fact be separated from a purely triboelectric (or “tribovoltaic”) contribution, further supporting that local heating is important but does not explain everything.

There is little doubt that frictional heating occurs; one can use friction with sticks to produce a fire [373], although it will generally take longer than a minute. For typical loads and velocities of bearing materials, the local contact temperature rises have been extensively analyzed [374-379], see also the review by Kennedy [380] and the recent comparison by Umar *et al*. [381]. The maximum steady-state temperature rise is of the order of

$$\Delta T_{max} = \frac{qa}{k} = \frac{\mu P v a}{k} \tag{17}$$

with $q$ the heat flux, $a$ the contact radius, $k$ the thermal conductivity, $\mu$ the friction coefficient, $P$ the applied pressure and $v$ the velocity. (This assumes that all the frictional energy becomes heat.) With a reasonable speed of 1 $m \cdot s^{-1}$ for copper, silicon, quartz and PVC as test examples, the change in temperature is less than 1 °C for a 1 μm contact. It is not unreasonable that local flash temperature rises could be a factor of 10-100 higher than these estimates, but these are still modest; there is experimental evidence for flash temperature rises of 100-200 °C from tribological measurements of polymers [377] and Raman spectroscopy for a steel pin on sapphire [382].

The above experiments are for machines with relatively large loads and in some cases high friction coefficients (dry sliding). However, triboelectric experiments are expected to have much smaller temperature rises, consistent with recent measurements by Lee *et al*. of a few degrees [371], and less than one degree by Kim *et al*. [383]. For completeness, we also note the report of tribocharge for a superlubricious contact where the friction coefficient is very low so heating should be minimal [326, 327].

In addition, heating as a source of charge transfer does not match with many of the experimental constraints in section 2 ranging from charge transfer from contacts without sliding to charge transfer between identical materials or curvature effects. Certainly, the temperature will act as a **Dependency**, influencing thermionic emission over electronic barriers, the conductivity, and the population of trap states, among other electronic effects, which have implications for other **Drivers** and **Mechanisms**, but temperature as **Driver** by itself seems unlikely.

A related model that has sometimes been invoked in the literature is electronic excitations driven by friction [80, 81, 384-390]. For instance, Xu *et al.* [81] argued that for a sliding contact involving a semiconductor pn junction friction produces electron/hole pairs which can migrate due to the contact potential thus leading to charge transfer. A different proposition by Zheng *et al*. [387] invokes the frictional energy release as the source for electronic excitations. This would be a proposed **Mechanism**, but the key problem with it is what is the **Driver**? Appropriate **Drivers** would be the electromechanical, Coulomb and mechanochemical terms detailed in sections 3.2-3.6. As mentioned in section 4.1, friction is a consequence of energy dissipative processes, not a fundamental force. Our interpretation here is consistent with the interpretation for a sliding Schottky nanocontact of Liu *et al*. [385]. We are not aware of any definite evidence that would support a hypothesis of electronic excitations as a **Driver** for sliding or normal contacts.

## 4.5 Mechanochemistry and tribocharge

Mechanochemistry was first connected to triboelectricity by Henniker [391], who suggested that mechanically produced free radicals were the source of transferred electrons. Related work has suggested that during contacts ions are produced [392-394], or that protons are transferred for polymers [395]. There is newer data which supports a role for ion movement due to stress as a component of mechanochemical ionization and charge transfer, for instance DFT observations of heterolytic ion formation by Sukhomlinov *et al*. [317].

As mentioned in section 3.6, some care is needed due to overlapping terminologies. Localized charges as in ions and delocalized charges (electrons or holes) as excitons in semiconductors are not that different. Both can be part of the **Mechanism** for charge transfer, which should be distinguished from the **Drivers**. We will argue that the presence of ions does not prove mechanochemistry is dominant, just as their absence does not disprove a role for it.

There are also ambiguities in terms of which processes are referred to as mechanochemistry. For instance both Balestrin *et al*. [396] and Shin *et al*. [397] refer to charge separation associated with fracture as being part of mechanochemistry, as against the established description as fractocharging which dates back to Faraday in 1839 [398], see for instance the work of Kornfeld in 1978 [399]. We note that there is strong evidence for flexoelectricity playing a role, since the strain gradients at a crack tip are very large [400-402], and may even play a role in bone repair [403, 404].

A recent example of mechanochemistry interacting with triboelectricity has been given by Fatti *et al*. [320]. In this work (see figure 23) the authors used pseudopotential calculations with the PBE functional [405] to apply a compression to a flat, planar silica/gold contact (hence without a strain gradient, but with piezoelectric contributions implicitly included), see figure 23a, terminating the silica with hydrogen atoms to preserve valence and charge neutrality. They used models of fused hydrogen terminated silica and quartz, as well as a boron doped material. They calculated the enthalpy for transfer of a hydrogen atom from the silica to the gold as a function of the applied pressure, finding a reduction with pressure (figure 23b,d) for the transfer energy, and also charge (electron) transfer (figure 23c,e). Their results points towards a hydrolytic charge transfer of the hydrogens used to passivate the silica, effectively an ion transfer. In the supplemental material the authors also include additional information for variations in the arrangement of the hydrogenated silica atoms at the interface, showing that the different dipoles will lead to different charge transfer consistent with the earlier discussion in section 3.4.

Mechanochemistry is clearly important in their example, but it contributes indirectly because the chemical reactions do not provide the potentials that are needed to drive charge transfer, but instead facilitate the transfer by changing the interface dipole. As mentioned earlier, an equivalent description of this would be in terms of the deformation potentials for different orbitals (energy bands) and interface dipoles.

Changes in the positions of the energy levels, whether it is described as an electromechanical term or as mechanochemistry/tribochemistry is certainly a **Driver**.

## 4.6 Material transfer

Material transfer has been evoked as part of tribocharging [406, 407]. This may be atoms or ions [392, 393], or fragments of molecules fragments [408, 409], and is a topic that has been debated [409, 410]. It may also involve preexisting ions in polymers [392]. While often tribocharge corresponds to the transfer of electrons, this is not the only **Mechanism** and ions are quite possible, for instance there are many materials where the charge carrying species are protons [411-414]. There is also extensive evidence of heterolytic cleavage (or heterolysis) where bond breaking leads to positive and negative ion pairs [415, 416], and also ion emission from surfaces during cleavage [417-419]. There is also literature correlating material transfer and possible heterolytic cleavage with tribocharging [420].

One caveat about material transfer during sliding is that it is established to occur completely independent of any charge transfer [145, 421, 422]; it is a standard tribological phenomenon. Often the softer of two materials in contact will transfer to the surface of the harder one, for instance as evidenced by in-situ observations of transfer of $MoS_2$ layers [423]. As mentioned earlier, graphitic carbon (or friction polymer) coatings or transfer layers are common in many systems [22, 23], and may well have been present and unsuspected in many experiments. Figure 24 illustrates that various different carbon species can be formed at contacts, with also the formation of what is called a tribo-film and a scar where some of the material has worn away; for triboelectricity, replace the counterpart ball with a second material. There have been some experiments [146] that indicating that here are changes in the tribocharging after the formation of a transfer layer. However, there is no evidence as yet to clarify whether the charge transfer only occurs with material transfer being the **Mechanism**, as against both processes occurring in parallel.

It is possible that in some cases material transfer of charged species is part of the **Mechanism** where an existing electric field due to the electromechanical and contact potential terms assists in a heterolytic cleavage. However, there is no indication that it is a **Driver**, these are either related to electrostatic potential differences or changes in the relative levels of electrons with strain as discussed in section 3. In this we are in agreement with some of the existing literature [424]. Note that this does not negate the possibility of ion transfer in response to electrostatic potentials, as against electrons – both will respond to a potential difference.

## 4.7 Humidity

That water and humidity influence triboelectricity is well-known, dating back to at least the seminal work of Gilbert in 1600 [49]. The chemisorption of water, particularly onto oxides, is one of the most extensively studied topics in surface science [425]. To understand how it can influence tribocharging, we need to include aspects of the science from a wide variety of fields and subfields; A full review of humidity effects would by itself be a significant undertaking. Herein we will point towards the main processes by which humidity is a **Dependency**; one or more of these may be active, and they could also compete so some care is needed. Figure 25 summarizes the main **Mechanisms** whereby water can play a role. While it may be critical for some combinations of materials, at the same time in other cases triboelectric charging does not require the presence of water or similar media, as indicated by tribocharging in vacuum [426] and under oil (i.e., a nonpolar liquid) [427]. We will also include in this section a model which to our knowledge has not appeared in the literature, but has extensive circumstantial evidence for it, namely the effect of humidity and water chemisorption on friction and hence shear.

### *4.7.1 Work function modification*

Humidity and other chemisorbed species can change the contact potential (a driver discussed in section 4.3) [428-430]. This is not a small effect; existing data in the surface science literature indicates that the changes can be 1-2 V, which is comparable to the contact potential. In most cases, the hydrogen will be on top of the outermost atoms, but there are also cases where it resides slightly below the outer layer [431].

### *4.7.2 Passivation of trap states.*

Hydrogen (from water) is known to be a possible dopant leading to trap states [432], and both water and hydrogen can passivate traps, particularly at surface defects [433-435]. In the kinetic cases of charge saturation, humidity is also important, as higher humidities can cause more ions to be adsorbed at surfaces [436].

#### *4.7.3 Conduction mechanism*

Mechanisms for ionic transfer which involving water have been proposed [393], for example the transfer of ions via water bridges between contacting materials. There is also strong evidence for electron conduction via water bridges from STM experiments [437, 438]. As mentioned earlier, conduction via the movement of protons is also well documented [412-414]. There is also compelling evidence for rapid diffusion of hydrogen on oxide surfaces with typical activation energy barriers of 1eV or less [425, 439]. In careful measurements, Arridge [440] demonstrated that the diffusion rate of charge carriers on nylon depended upon the humidity, and there is other evidence such as the measurements of surface conductivity by Awakuni and Calderwood [441]. A related process is described by Gil and Lacks where hydrolytically dissociated water, as H+ and OH- ions, may accelerate conduction [442].

#### *4.7.4 Charge dissipation*

Humidity directly affects electrical breakdown [443], and thus charge saturation in the breakdown-limited cases. For instance, Amiour, Ziaria, and Sahli found that higher humidities caused increased surface potential decay rates on Kapton HN polyimide films by a factor of five between 35% and 95% humidity, correlating with the formation of additional shallow, lower energy trap states in humid conditions, see figure 26 [444].

#### *4.7.5 Changes in shear due to friction changes*

There is extensive literature indicating that humidity can have a large effect upon friction coefficients. For instance, in 1948 Savage pointed out that the lubricious properties of graphite were not intrinsic to the structure, but depend upon chemisorbed species such as water molecules [445]. Many other cases are documented in the tribology literature as discussed in the 1990 review by Lancaster [446] and more recently by Chen *et al*. [447]. Any change in the friction coefficient will have a corresponding change in the shear stresses and strains. As will be discussed more in section 4.10, it is known that sliding leads to significantly more charge transfer, and we argue that this is due to the shear components of the contact.

Humidity is clearly a **Dependency** through multiple paths. In some respects, humidity could be considered as an “**anti-Trigger**” or “**anti-Driver”** if the passivation of surface traps prevents irreversible charge transfer.

### 4.8 Environment beyond humidity

Other environmental factors can also act as **Dependencies**. For example, atmospheric pressure and composition, like humidity, modifies electrical breakdown, causing changes in the triboelectric charge saturation of up to a factor of 3 from atmospheric to vacuum pressures [230]. Adsorption of various species, beyond those derived from water mentioned above, can change the magnitude and sign of triboelectric charge, see for instance the overview by Nguyen *et al*. [448] and the analysis by Biegaj *et al*. of glass beads with different surface functionalization contacting stainless steel [449]. The authors of the latter found that more hydrophobic terminations have more charge transfer by about a factor of 3, which is also consistent with surface dipole shifts of the work function. For instance, they find significantly more transfer for fluorine terminated surfaces compared to hydroxide; fluorine at the surface has a different sign of surface dipole than hydroxide, similar to chlorine as discussed in section 3.4.

It should be mentioned that air contains carbon dioxide, which can also be chemisorbed on many surfaces, both metals [450-452] and oxides [452, 453]. While this is not as well studied as water, adsorbed species derived from carbon dioxide will similarly cause a surface dipole induced change in the work function. For instance there can be a work function change of 0.4V for $CO_2$ adsorption on a Ni film [451], and work function changes have been used in $CO_2$ detection [454].

Temperature is another environmental **Dependency**. Temperature is a modifier of drivers in many ways, for example changing the Fermi-Dirac distributions of electrons, as well as altering thermally assisted emission which can be important for charge transfer in Schottky contacts [56, 69, 349]. Temperature also has direct effects on breakdown and other charge dissipation mechanisms, and therefore on charge saturation mechanisms [455].

## 4.9 Electrostatic Induction

Electrostatic induction is the long-range coupling of both the polarization and electric field, the electrostatic potential (including polarization contributions), and any charges for two bodies which are not in contact. It dates back to the earliest days of electrostatics, being first described by John Canton in 1753 [456]. A typical case is shown in figure 27 for a gold leaf electrostat, based upon an 1881 image from Sylvanus Thompson [457]. While it is often considered as just charge redistribution, herein we will include the polarization component induced by external electric fields as well. Note that terms such as the image charges and potential mentioned in Section 3.3 are also part of electrostatic induction and should not be double-counted.

One important clarification about terminology is necessary. It is common to see descriptions with TENGs of using the "coupling of triboelectrification and electrostatic induction". Both of these are charge redistribution processes, with nominally tribocharge involving contacts whereas induction does not. In any triboelectric contact and separation there will be electrostatic induction *after* the initial charge contact, as nicely demonstrated experimentally by Musa *et al.* [458] Note that neither triboelectricity or induction are fundamental process, they are consequences of electrostatic **Drivers**.

Devices using electrostatic induction are well established; miniature microphones as in webcams use electrostatic forces to convert sound to current using a condenser setup with either an external bias [459, 460] or an electret [461, 462]. Beyond its role in charge transfer, induction is also often used as a charge collection [463] or measurement [393] mechanism in triboelectric experiments. Electrostatic induction is one of the oldest methods of energy harvesting, dating back to at least the 1978 work of Jefimenko and Walker [464] who used disk electrets with a permanent polarization; further details can be found in the review by Beeby *et al*. [465].

The physics is described by the combination of the electromechanical contributions of section 3.2 as well as band-bending in sections 3.3-3.6, including the non-contact case described in section 4.3. If both sides of two materials separated by a gap are charge neutral without polarization there are no coupling terms (except for Casimir forces which we will neglect as too small to be important, and in principle a Van der Waals like induced bulk polarization which will also be small). After charge redistribution there will be coupling terms, for instance charge on the surface of a metal will couple with polarization of a dielectric. (Note that the two should be calculated self-consistently.) As we have discussed previously [194] these image potential terms can have a significant role.

There is another term which we should draw attention to which is called **Electrostatic Adhesion** (or electroadhesion). This is a topic which has been studied recently in the context of human fingers and touchscreens, for instance references [466-468]. Persson has analyzed the Coulomb component of this in a general sense [469, 470], while Xu *et al*. [471] have explored some of the consequences for contact electrification. As a caveat, to our knowledge the full electromechanical form including flexoelectric contributions has not to date been used, so there is more work to be done.

## 4.10 Sliding versus contact

It is established that sliding produces more charge than contact or rolling [76-79, 88], including results with particle contacts [78, 79, 85-88]. The first theory to try and explain this, attributed to Volta [54, 472], was that sliding increased the number of contacts per second, leading to more charge transfer. This matches the general experimental trend, which shows more charge transfer with faster sliding [24, 80-84]. However, it does not explain the differences between sliding and rolling, where the number of asperity contacts are similar, but the tangential forces are not.

Various other explanations have since been offered, including a larger contact area [473], more material transfer [114, 474] (see section 4.6), increased local heating [24] (see section 4.4), the movement of space charge regions [326], or based on conversion of frictional work into electronic excitations as discussed in section 4.4. To date these do not qualitatively agree with experimental trends, as pointed out by others [424, 475], none have shown quantitative agreement, and some include assumptions which are not backed up by external measurements and theory (see the Supplemental notes to Olson and Marks [132] for further discussion).

One model published in 2020 by Alicki and Jenkins [475] that has potential relevance for sliding treats triboelectricity as an irreversible process. These authors argued that the electrons in two materials that slide against each other have different velocities, and quantum effects cause this imbalance to pump electrons from one material to the other. Electrons are pumped in both directions, but small differences in the potential energy landscapes for the two surfaces can cause net charge transfer. This is an interesting concept which would be a **Driver** founded in established quantum mechanics. What is not so clear as yet is how this would incorporate phenomena such as charging of identical materials, the normal force dependence, or other established features such as curvature effects since this model requires relative motion. As the authors state, more careful experimental testing is needed to test its importance.

We have recently proposed a different approach which is based upon fact that asymmetric deformation of asperities occurs during sliding. In particular, asperity contact involves shear during sliding, whereas rolling and normal contact do not involve. The asymmetry of the strains due to that shear force (see figure 5, and also figure 29 later) and the resultant polarization cause increased and continuous charge transfer [132]. For a normal force case with symmetry, an infinitesimal displacement in any direction will lead to cancelling changes in the interface bound charge. In contrast, when the symmetry is broken by shear there will be a difference in the change of the interface bound charge at the leading and trailing parts of an asperity in contact with the substrate which we argue will lead to extra charge transfer. We have included this in the next section which deals in a more general fashion with the electromechanical contributions.

Additionally, sliding could lead to asperities bending due to friction with a flat surface or contact with asperities on the second body. This bending would cause flexoelectric fields with magnitudes relevant for charge transfer. To our knowledge, this has not been explored in detail as yet.

### 4.11 Electromechanical, particularly flexoelectric

Electromechanics, the coupling of electronic and mechanical behaviors of materials, can be associated with experimental triboelectricity results that date back to the work of Jamieson [115] and Shaw [68], who showed that charge transfer depended on macroscopic bending and surface strain. (The connection to electromechanical phenomena has not always been explicitly stated.) The most widely known electromechanical phenomenon is piezoelectricity, the electric polarization produced by strain. Piezoelectricity may explain a small part of triboelectricity, as suggested in 1941 by Martin, who measured the tribocharging of wool and hair [476] and also both Peterson [183] and Harper [184] who measured the small contribution it had to tribocharging of quartz crystals. However, piezoelectricity only occurs in non-centrosymmetric materials, and therefore cannot explain triboelectricity in centrosymmetric crystals. Instead, flexoelectricity, the coupling of electric polarization with strain gradients, is present in all non-metals and can explain many experimental trends [19]. This has been analyzed in detail in recent articles at Northwestern [101, 131, 132, 149, 194] and by others [477-482], and we will provide a brief overview in this section. All electromechanical phenomena lead to polarization so are **Drivers**.

Recapping what was mentioned earlier, when two materials rub together, the actual contact typically occurs between surface asperities over a much smaller area than the apparent contact [37]. This results in nanoscale contact where the strain gradients are large (for metal/oxide contacts, ~$10^7$ $m^{-1}$) and produce flexoelectric potentials (~1-10 V) large enough to drive charge transfer [19]. Not only is bulk flexoelectricity important, but terms that are often called "surface flexoelectricity" are also significant as mentioned in Section 3.3.

Combining contact electromechanics and of the contact potential, many experimental results in the triboelectric literature can be qualitatively explained, and in a few cases semi-quantitatively. Our models of force-dependent current-bias curves in nanoscale Schottky diodes [56], charge transfer between impacting bodies [194], the triboelectric current between sliding bodies [132] as well as the shape of contacting asperities [131, 483] quantitatively agree with experiments. There is also other work in the recent literature which support flexoelectricity playing a role, for instance analysis of contact between identical materials [477, 484], in multilayer graphene [478], contact electrification between a nanoscale tip and flat polymers [480], the formation of oxygen radicals in MgO nanocube ensembles [485], and the performance of n-Si Schottky nanogenerators [481]. There are also other papers where a general role for flexoelectric contributions has been discussed [81, 397, 486-488] and modelled in simplified cases [489]. We note that recent work that shows large increases in apparent flexoelectricity in doped samples [342-344, 490-492] due to the barrier-layer effect [345-347] is likely connected to the large changes in triboelectric currents with doping levels of semiconductors [132, 386, 387, 481], although this is not as yet unconditionally proven.

From the earlier details in section 3.1, the magnitude of the flexoelectric contribution will scale as the flexoelectric coefficient divided by the Youngs modulus to some power. As an indication of this, some values are shown in Table 4 adapted from the Supplemental material of Mizzi, Lin and Marks [19] with Youngs moduli from the literature. The largest values occur for the relaxor ferroelectric PIN-PMN-PT and $BaTiO_3$, the latter being associated with the ferroelectric distortion.

Table 4: Values for the mean, effective flexoelectric coefficient, flexovoltage, Youngs modulus and a scaling ratio for a sphere for several different materials. A printed sign means that this was measured; when it is absent the values are magnitudes.

| Material | Orientation | μ (nC/m) | f (V) | Y(GPa) | $\mu Y^{-1/3}$ | Reference |
|---|---|---|---|---|---|---|
| $SrTiO_3$ | [001] | +6.1 | +2.3 | 238 | +0.98 | [158] |
| $SrTiO_3$ | [101] | -5.1 | -1.9 | | -0.82 | [158] |
| $SrTiO_3$ | [111] | -2.4 | -0.9 | | -0.39 | [158] |
| $BaTiO_3$ | [001] | +200 | +22 | 67 | +49.24 | [493] |
| $BaTiO_3$ | [110] | -50 | -6 | | -12.31 | [493] |
| $BaTiO_3$ | [111] | -10 | -2 | | -2.46 | [493] |
| $DyScO_3$ | [110] | -8.4 | -42 | 214 | -1.40 | [494] |
| $TiO_2$ | [001] | +2 | +1.3 | 230 | +0.33 | [342] |
| PIN-PMN-PT | [001] | ~ 4e4 | ~ 1400 | 134 | ~7816 | [495] |
| PVDF | | 13 | 160 | 3 | 9.01 | [496] |
| Oriented PET | | 9.9 | 289 | 4 | 6.24 | [496] |
| Polyethylene | | 5.8 | 273 | 0.25 | 9.21 | [496] |
| Epoxy | | 2.9 | 84 | 3.1 | 1.99 | [496] |
| P(VDF-TrFE-CTFE) | | 3.5 | 10 | 0.2 | 5.98 | [497] |
| Jeffamine | | 0.1 | 4.5 | 1.0 | 0.10 | [181] |
| Rubber | | 0.5 | 20 | 0.01 | 2.32 | [181] |

The key results of an electromechanical approach that includes contact potentials are:

1. It is not independent of other physics, terms such as the contact potential, band bending and surface charges are part of the overall electromechanics [19, 56, 101, 131, 132, 149, 194], specifically the electric displacement field.
2. It connects seamlessly with the established physics of polarization, bound charges and surface charges [149].
3. It is not dependent on the material system; contact potentials occur in all materials and flexoelectricity in all non-metals.

4. It is an *ab-initio* model, as all the terms can be independently measured or calculated, although the database of flexoelectric coefficients is currently somewhat small and calculations are not simple [159, 195, 224].
5. With polarization and electrostatics part of the formulation, the Coehn's law dependence upon the dielectric constant is automatically included [51, 55, 60, 358, 359]. We note that the magnitude of the flexoelectric coefficients divided by the dielectric constant (called the flexocoupling voltage) was estimated by Kogan [40] to be roughly in the range of 1-10 V, hence large dielectric constants typically lead to larger flexoelectric charging.
6. The general Landau expansion with polarization of equation (3) indicates that the surface bound charge and hence the tribocharge will depend upon the intrinsic polarization of an insulator. This was observed in 1980 by Robins, Lowell and Rose-Innes [498].
7. The original use of the "flex" in flexoelectricity was related to bending/curvature [42]. Hence there are potential differences based upon differences in local curvature [101], which explains phenomena such as charge disproportionation between particles of different sizes [117] as well as same material charging, including the classic early result of Jamieson [115].
8. Differences in charging with strain also come naturally from the model, since there will be a strain gradient of approximately $\epsilon/R$ for a applied strain of $\epsilon$ with $R$ the asperity radius; this is comparable to the misfit strain in quantum dots, see reference [499] for specifics. (Remember that the normal stress on an external free surface is zero, but will tend to that of an applied bulk macroscopic tension or compression with depth.) The presence of polar structures due to flexoelectricity that relate to this has recently been experimentally demonstrated by Shang *et al.* [500].
9. Changes in the sign of charge transfer with pressure, which has been known for more than a century (Harris observed the phenomenon in 1867 [50] as did Shaw in 1917 [68]), drop out of the analysis of the band bending, for example by Zener tunneling [101]. This is shown in figure 28 for contact between a silicon sphere (orange) on a strontium titanate flat. At high loads, on the right, the conduction band of the silicon dips below the top of the valence band of the strontium titanate. This is consistent with the experimental results of Sun *et al*. who observed a change in the sign of the transfer using Kelvin probe microscope with different forces [100], which Qiao *et al.* interpreted as related to flexoelectricity [479]. It is also consistent with our measurements of band bending due to flexoelectricity with a conductive AFM tip [56], and see also other data for band bending in the review of Xia *et al.* [157].
10. It connects to work where increased stress reduces the charge transfer in some materials, as observed by Zhang and Shao [501], opposite to most other work. There is no reason why the flexoelectric charge transfer should have the same sign as the contact charge transfer, they can compete as mentioned by Qiao *et al*. [479], which follows directly from the general form of Ampère's law of equations (7-10).

11. The prediction of a change in the sign of the polarization outside a contact to that inside (see figures 10b, 29) is consistent with observations by Terris *et al*. in 1989 [502], who observed bipolar surface potential regions of comparable size. We note that further evidence that this can be due to polarization (as against free carrier transfer) was provided by Cunningham [503].
12. It connects to related experimental work on phenomena such as fractoemission, fractoelectrification, triboluminescence, and tribo-excited exoelectron or x-ray emission [417, 504-507]. At crack tips, there are large flexoelectric contributions [401-403, 508] which could lead to the formation of energetic electrons.
13. Analysis yields variations in the tribocharge as a function of asperity shape which are in both qualitative and semi-quantitative agreement with experiment [131].
14. The approach yields numbers for the charge transfer which are in quite good agreement with experiment [19, 101, 131, 194].
15. When applied to a sliding contact, it indicates that there will be significantly more charge, a result which has been known experimentally for more than a century but never well explained. As shown in figure 29, shear breaks the symmetry of the polarization field so the induced charge at the leading and trailing edges of the contact are different [132]. There is a simple solution for the (maximum) current $I$ for a sliding velocity of $\boldsymbol{v}$ involving the line integral around the contact $\boldsymbol{\ell}$ of a triple product with the electric displacement field $\boldsymbol{D}$

    $$I = \oint_{\partial A} \boldsymbol{v} \cdot (\boldsymbol{\ell} \times \boldsymbol{D})\, d\boldsymbol{\ell} \tag{18}$$

    Note that this yields the force to the one-third power dependence of the tribocurrent shown earlier in figure 5. The enhanced tribocharging with shear is consistent with a range of experimental results as described previously [132] as well as others such as the dependence of tribocharge on the incidence angle for colliding particles [85-87] and that large charge transfer is associated with sliding (when the relative velocity is finite) as against stick (when it is zero) in a stick-slip contact [509, 510].

This evidence suggests that electromechanical terms are key **Drivers** in triboelectricity, along with the contact potential. They do not explain everything by themselves, but we believe are an important component that was missed until recently.

### 4.12 Inhomogeneity and an effective medium approach

Dating back to at least the work of Hull in 1948 [111] and Henry in 1953 [112] there is evidence for a mosaic of inhomogeneous charging, particularly for experiments involving nominally two identical materials; early work is discussed by Montgomery [33]. There have also been more recent experimental results at higher resolutions [113], particularly using scanning probe methods [114, 511].

There are two main processes which can explain this. Firstly, as discussed in section 3.4 for crystalline solids, whether they are metals or insulators one contributor will be orientation, The work function is not a material property, it can have a strong crystallographic dependence [232, 235] so will vary with grain orientation in a polycrystalline material. For polymers there is also good evidence for inhomogeneous variations in the work function [236-239].

Secondly, as discussed in section 3.1, it is established that contacts between two surfaces involve asperities of different sizes, with the net friction and contact area statistical consequences of how these contact vary with applied pressure as first analyzed by Bowden and Tabor [37, 38, 123-129], For any contact, for example the illustrations in figure 3, in some cases the curvature will be larger on the top surface, in others on the bottom. Thus different asperity contacts will lead to different local flexoelectric contributions which in turn leads to a charge mosaic as analyzed by Persson [477].

Both of these, and also inhomogeneity of dopants and trap states, will lead to variations in the tribocharge. For both tribocharging of identical materials and non-identical materials, it is probably more appropriate to consider effective medium theories (e.g., [512-519]). These reduce the behavior of a composite with a range of properties to an equivalent effective response employing different models for the reduction. For these the magnitude of the variation with position of the relevant parameters matters, herein the work function and electromechanical terms, this variation being called the "heterogeneity contrast". If the heterogeneity contrast is small, then a rule of mixtures or similar may be applicable, for instance shared load bearing in a mechanical system. If it is large, the effective medium may be dominated by the strongest component, as in many fiber-reinforced composites, or by the weakest as in fracture at grain boundaries. The use of an effective work function [307, 520-522] is common, and an effective medium approach has been used by Chen *et al.* [523] to improve the treatment of the dielectric properties of a composite material used for triboelectric energy harvesting. We suspect that the full charge transfer should also be treated in a similar fashion, and there may be similarities to the behavior of nanowire arrays [524, 525] or effective medium approaches to induced polarization [526]. The analysis by Zai *et al.* [120] suggests that a rule of mixtures may not be appropriate. However, there is as yet insufficient evidence to generalize so this is a topic for further work.

## 5. Disengagement

So far, we have described processes where charge is transferred from one material to another on contact and/or sliding. The **Drivers** for this are a combination of the contact potential, Coulomb terms and electromechanical contributions, modified by various dependencies. However, this is not everything, as we need to consider what happens when the materials are separated, the stage at the end of the Decision Tree (figure 2).

There is extensive evidence that irreversibility plays a significant role. As demonstrated as early in 1920 by Richards [89], for repeated contacts without sliding the total charge increases with the number of contacts to some saturation value, see also figure 11b from earlier. As shown by Peclet in 1834 [323], there is a similar saturation of charge with the duration of sliding contact. Both of these indicate that charge can flow from one material to another irreversibly, accumulating with multiple events. A nice illustration of this is hysteretic behavior of the contact force response observed by Seol *et al*. [527] and shown in figure 30. The authors observed that the open circuit voltage as the contact force was reduced did not follow the same path as when it was being increased. The accumulation and hysteresis are analogous to conventional ratchetting – what goes up does not always go down. As one final example before being more specific, we note that it is common in triboelectric experiments to use a Hertzian contact model for the area of, for instance, the force dependence of particle contacts on substrates. The area used is the *maximum value*, which drops as the contact separates, so this implicitly assumes irreversibility.

As one point of clarification, nominally electromechanical terms such as piezoelectric and flexoelectric contributions are reversible. That is, the polarization decreases back to zero as the strain is reduced so long as there is no permanent plastic deformation. However, just because the polarization is reversible does not mean that the movement of free carriers is reversible; in triboelectricity we measure macroscopically how these or charged molecular fragments have changed locations.

There are a number of established models that reveal **Mechanisms** for irreversibility which we will discuss in this section. While engagement is a well-defined problem where most of the parameters have been independently studied in other areas, how charge will behave on disengagement is much more specific to triboelectric charging, so there is less connected science, more ambiguity, and ample room for further research.

### 5.1 Capacitor model

If the two bodies are metals the contact is completely ohmic. At the interface there will be a dipole due to free carriers as discussed in section 4.3 and shown in figures 10 and 13, and this will balance the contact potential. This dipole will be very localized, equivalent to a polar bond, with perhaps minor Friedel oscillations [352] going into the bulk of the two materials.

In this case there are no flexoelectric contributions, although there could be effects due to differences in the deformation potentials of the two materials and foreign atoms. As the metals separate the charge can partially reverse as the contact force is decreased. For the open circuit case the charge left after disengagement would correlate with the energy of a capacitor created between the two materials with a balance of the electrostatic interaction across the opening gap and the potential energy associated with the contact potential difference. The idealized problem has been analyzed some time ago, for instance by Harper [528], Lowell [341], Lowell and Rose-Innes [35], and reviewed by Castle [307]. The standard approach is to assume that below some separation distance $d$ tunnelling can occur [36], and use this to determine the charge transfer. (We note that thermionic emission return currents may also play a role as discussed in 1977 by Ruckdeschel and Hunter [529], Lowell and Rose-Innes in 1980 [35] and more recently be others [70, 530, 531]. Water menisci may also be important for current flow from retracting asperities, similar to STM tips where there is experimental evidence [437, 438].) This gives for the charge $Q$ per unit area (that corresponds to the free carrier density at the interface used earlier) and total charge $Q^{tot}$ a form similar to

$$Q^{tot} = A_c Q = \epsilon_0 A_c V_c / d \ = C_o V_c \tag{19}$$

for a contact area $A_c$ (that is often approximated using a Hertzian model), a contact potential of $V_c$ and an effective capacitance of $C_o$. Typically, the separation *d* is about 1 nm and the effective capacitance is introduced to account for the fact that the contact is between multiple asperities. This form appears to be very appropriate for discrete particle collisions with metals because the charge cannot escape except for that which tunnels back during separation, this is an open circuit condition.

When one of the two (or both) is an insulator, the problem is more complex. In the limit of a very soft contact without significant adhesion, the electromechanical terms can be ignored in which case only the band bending matters. The band bending, the types of materials in contact and their electronic properties such as the doping level, determine the relevant transport mechanisms. If there are no traps and the contact is completely ohmic with negligible adhesion and also open circuit, the limit will be similar to the metal-metal case above and still well described by equation (15) with

$$Q^{tot} = eA_c N_p \left( \sqrt{d^2 + \epsilon V_c / e N_p} - d \right) \tag{20}$$

since the forces at electromechanical contributions will go to zero as the contact disengages. Note that this scales as 1/d for large separations, similar to the condenser model.

There may be a thin insulating layer of an oxide or a chemisorbed layer on the metals, in which case it will be a metal-insulator-metal contact, which behaves like a capacitor across the insulating layer (which may be polarizable). So long as the layer allows for charge transfer by either simple or thermally assisted tunnelling the results should be similar.

## 5.2 Diode irreversibility

In systems with diode behavior, such as Schottky metal-insulator contacts or p-n junctions, the charge flow across the interface is limited by an electronic barrier rather than a tunneling gap. For metal-semiconductor contacts an empirical "effective work function" has been used to describe the semiconductor (or insulator) as early as 1954 by Peterson [520], and can be found in other publications such as the 1997 overview by Castle [307]. Unconnected to triboelectricity, the term "effective work function" has an accepted definition in the semiconductor interface literature [521, 522] to account for Fermi level pinning and the dipole contributions described earlier; see also the earlier discussion in section 4.12 about effective medium theory. In many cases it appears to be rigorous to use this interface literature definition for soft contacts, although to date there seems to be no testing whether the numbers from the triboelectric and interface fields are comparable or not.

Consider the case of the Schottky diode shown earlier in Figure 21. In the limit of a soft contact there will be current flow from the metal to the semiconductor when the contact is formed, this is the forward direction of the diode where current can readily flow. However, as the contact disengages current would flow back from the semiconductor to the metal, which is the reverse direction and hence will be limited. We note the importance of rectifying contacts is not new, and can be found for instance in the work of John *et al.* [226], who observed for pre-charged sodium chloride particles impacting on a titanium plate, different charge transfer depending upon the sign of the pre-charge, shown in Figure 31. This is consistent with pn junction rectification being formed as John *et al.* state.

If we assume that the Fermi level pinning in the semiconductor is the same when they are separated as it is when joined, the problem is similar to a metal-metal contact but now with the contact potential the difference in energy between the metal Fermi level and the energy of the surface pinning states. Charge would flow both on approach and on contact, but back current would be limited by the diode characteristics.

For a contact with more force, the flexoelectric and piezoelectric contributions will need to be included. The flexoelectric contributions will modify the behavior of the diode via band bending as has been established experimentally in the literature [157, 532-534] and which we have previously modelled for the Schottky case [56]. Image force lowering can also play a role in Schottky diodes, although our calculations imply that this is small compared to flexoelectric effects for typical asperity contacts [194].

Note that the polarization contribution does not need to be of the same sign as the contact potential contribution, and the flexoelectric coefficient can change sign with both material and orientation as mentioned in Table 4. Hence the flexoelectric and contact potential terms can compete as previously mentioned by Qiao *et al.* [479]. This is consistent with results such as those of Zhang and Shao [501], who for their materials found a decrease in the tribocharge with increasing contact stress, opposite to most other work [80, 81, 84, 90-93].

In addition, if there is adhesion then, using Johnson-Kendall-Roberts [535, 536] contact models (not Hertzian), the flexoelectric charge transfer will change sign during disengagement as we pointed out previously [19]. It is possible that in some cases it is the charge transfer during adhesive separation that dominates.

Along with the capacitor model discussed in section 5.1, the formation of a diode at a contact and its modification by electromechanical effects form **Mechanisms** for irreversibility which do not necessarily require impurities, defects, or other imperfections.

## 5.3 Trap states and irreversibility

For a closed circuit case the charge may escape to a large reservoir which would provide irreversibility; this is thermodynamically downhill due to the entropy of mixing of the carriers in the reservoir. This is the classic method used in the early electrostatic generators called influence machines [537, 538]. When there are traps, the problem will be similar to a diode, with the reduction in the energy of carriers when they fall into the traps playing the same role as the barrier in the diode. There has been extensive discussion of trap states [57, 314, 539, 540], which have also been referred to as "backflow stuck charges" [316]. Trap states are another **Mechanism** for irreversible charge transfer.

One caveat; with a finite size contact there is also the question of what takes place at the surface away from the immediate contact, as we have very recently discussed [194]. For instance, in specific calculations, if conduction across the surface is fast the flexoelectric contributions approximately cancel and the contact potential terms dominate [194]. It is possible that surface conduction can alter the diode behavior, particularly chemisorbed water from humidity.

## 5.4 Dissipation

If the surface charge is large enough then it can force a plasma discharge [511, 541, 542], and the latter can depend upon the environment [543]. In that case, the saturation charge is influenced by factors that affect atmospheric breakdown rather than those that directly influence drivers of triboelectricity. An alternative is that the potential difference due to the charge can be larger than the band gap in which case Zener tunnelling can occur [544-546], as experimentally analyzed via in-situ measurements in a transmission electron microscope [494].

We also need to consider dissipation of triboelectric charge away from the surface, see also the review by Molinie on charge decay [59]. Mechanisms for charge dissipation can involve conduction across the surface or into the bulk of a material, transfer to a gas or liquid (e.g., by evaporation of charged water molecules [547]), and electron emission or tunnelling [307, 529], among others [548]. In addition, occupation of trap states by transferred charge can limit the rate and amount of charge dissipated [307].

Charge dissipation, which can occur by various **Mechanisms**, and is affected by many **Dependencies**, such as humidity as discussed in section 4.7.4 and the environment as in section 4.8.

# 6. Where are we now?

*"Honest disagreement is often a good sign of progress."*

attributed to Mahatma Gandi

Triboelectricity (and contact electrification) are topics for which many people say there is little agreement in the literature, sometimes to promote interest by stating that triboelectricity is unresolved science. However, much of what we have discussed above is established science which has been cross verified in other scientific sub-disciplines beyond being supported by extensive experimental triboelectric results. We believe that in many publications the analysis has only focused on one segment of the problem without considering the wider picture. Much is known, although many details still need additional exploration both experimentally and theoretically.

Is triboelectricity confusing, confused or complex?

- **Confusing:** perhaps, because it spans many different topics in chemistry, mechanics, physics and surface/interfacial science.
- **Confused:** only in the sense that different names have been used for the same fundamental science, and at times concepts rediscovered or reclassified.
- **Complex:** certainly, but there is extensive independent evidence for many of the factors. All that is needed is combining them, which is not always simple.

All the evidence points towards triboelectricity being a combination of conventional tribology with asperities and transfer layers and capacitor and interface physics which follows the extended Ampère's law:

$$\nabla \cdot \mathbf{D} = \varepsilon_0 \nabla \cdot \mathbf{E} + \nabla \cdot \mathbf{P} = \rho_f \tag{21}$$

Here the electric field term includes material and environment dependent effective work functions that account for interface dipoles, band bending and Fermi level pinning, and the polarization includes electromechanical flexoelectric and piezoelectric contributions. In most cases, the charges following the boundary condition at the interface to a relevant conductor or second material:

$$\mathbf{D} \cdot \mathbf{n} = \varepsilon_0 \mathbf{E} \cdot \mathbf{n} + \mathbf{P} \cdot \mathbf{n} = \rho_f^S \tag{22}$$

Some contributing factors are clear, for instance the role of the contact potential (or built-in potential), as well as interface dipoles and adsorbates in changing these. While the early evidence for these was more empirical, with modern *ab-initio* techniques, this is well established science. Interface dipole and band bending have been discussed often qualitatively in the triboelectric literature for Schottky contacts, although we feel that the full consequences of doping, dipoles and Fermi-level pinning needs more attention and careful study than it has had to date. Similar to the semiconductor literature, we feel that "effective work function" or comparable terms such as bult-in potential should be used to account for the documented large changes with Fermi level pinning and impurities. Since most real surfaces have some roughness, a more extensive effective medium theory may well be more representative of the physics. As mentioned earlier, the standard semiconductor physics numerical solutions for these problems seems to have rarely been used [356, 357]. It is worth repeating that there is also extensive evidence for these in contacts involving polymers [248, 259-263]. We suspect that the oft-reported irreproducibility of triboelectric experiments is associated with changes in the contact or built-in potentials, i.e., due to insufficient control of the dependencies.

While not so well defined, a leading role for trap states has been known in the literature for a long time. It is encouraging to see the start of measurements of the relevant states for a range of polymers such as the work by Murata and Hiyoshi [311] and more recently Xu *et al*. [314].

Both mechanochemistry and tribochemistry are also established science. We personally prefer to use the more physical approach with electromechanical contributions and deformation potentials as these can be independently measured and/or calculated. We feel this gives the physical approach more predictive power; this is only our view, and with all terms included the approaches should give identical results, albeit less so with pragmatic approximations in the calculations.

One new advance has been the inclusion of flexoelectric terms. We suspect that if these had been known a century earlier there would now be much more consensus on the underlying science. A wide range of evidence points to it explaining many features which are otherwise considered mysteries, such as same-material tribocharging and the dependence on macroscopic bending. Additionally, inclusion of flexoelectricity in models produces estimates of the charge transfer that are in quite good agreement with experimental results both in terms of trends and absolute values. One advantage of the flexoelectric contributions is that, for selected materials, fairly accurate values of both the relevant strain gradients and the coefficients are available, so quantitative calculations can be done which are accurate to within a factor of 2-3. While this is not as good as in many other fields of science, we believe it is an advance for triboelectricity that can be improved upon by careful experimental measurements, not only of triboelectric charge transfer but simultaneously the many relevant materials properties.

Flexoelectric (and piezoelectric, when they exist) terms add to (or subtract from, depending upon their sign) the contact potential contributions and are more important when there is a significant local force. Therefore, they will not be particularly important for very soft contacts. Variations in the results for different types of contacts for elastic spheres are well established, for instance those described recently by Ireland [96, 549] which appear to be consistent with the contact force argument above, although quantitative analysis would be more definitive – this provides an opportunity for future work. (The flexoelectric contribution scales as the ratio of the flexoelectric coefficient divided by the Youngs modulus to some power [194], Table 1 and 4.) While there is still limited information on flexoelectric coefficients for polymers (see the Supplemental material of reference [19] as well as [156, 157, 550, 551]) the indications are that they can be quite large. Additionally, their elastic moduli are smaller by a factor of 10 to 1000 or more. Hence, the electromechanical terms may be more important with polymers than for harder materials – see Table 4. There are also continuing developments in both the understanding and manipulation of flexoelectricity, for instance [157, 202, 293, 344, 492, 500, 551-555] (a very incomplete list) which almost certainly are relevant for triboelectricity and *vice-versa*.

Where much less appears to be settled is in disengagement since, as mentioned earlier, this does not appear to map so directly to existing experimental and theoretical calculations in other scientific subfields. There are some tantalizing papers which indicate that there may be more interesting science. For instance, Kaponig *et al*. [556] indicate that in their small sphere bouncing experiments, charge transfer occurred on a microsecond timescale. It would be interesting to know whether, with higher time resolution, there are more features to observe.

The question of timescale also applies to sliding charge transfer. It is known that in tribology there are often very rapid events, typically connected to stick-slip processes. There is data that stick-slip can be stochastic and noisy [557-559], for instance effects due to thermal fluctuations which show up with high resolution, tips jumping both forward and backwards with different stick-slip lengths as described by Maier *et al*. [557]. We will speculate that there are similar current reversals during sliding. It may require some effort to accurately incorporate screened electromechanical and mean-inner potential terms into the type of molecular dynamics or other approaches often used in tribology [129, 560-562], similar to work that has been performed to model flexoelectricity [563-565]. One challenge, is that the dielectric permittivity, along with electromechanical contributions [566], can have a strong frequency dependence. This is likely relevant for some triboelectric contacts, especially fast sliding and short-time contacts.

An important topic open to further discussion is exactly what boundary conditions should be used, as we have pointed out recently [194]. In related work Abdollahi *et al.* [567] did not include tribocharge and used as a boundary condition that the normal component of the displacement field at the surface is zero; there is also work such as that of Sidhardh *et al.* [199] who indicate that the electric field can vary by orders of magnitude and sign for beams with a flexoelectric response for different electric boundary conditions. It is common in the triboelectric energy harvesting literature to apply the boundary condition $\mathbf{D} \cdot \mathbf{n} = \rho_f^S$ (sometimes with an inappropriate extra term [185-187]). It may be that piecewise boundary conditions are more appropriate as we have recently considered [194]. In addition, our analyses of triboelectricity neglect many coupling terms, whereas others have used a more complete approach to related problems including higher-order terms [199, 201, 202] or couple stress approaches [203]. We believe that more experimental data is needed to test the correct formulation and determine in which situations these couplings are significant, providing one direction for future work. As this review was being written, one paper for self-consistent contact electrification between a flat plate dielectric and an asperity using a Lennard-Jones potential with electrostatic terms has appeared [471]; we hope that this is just one of many more to come.

Connected to the boundary conditions are also more subtle issues related to polarization accommodation at surfaces or interfaces. It is known that while some surface terminations in materials such as oxides have unbalanced polarizations [568-570], in general oxide surfaces rearrange to remove this [241]. There are constraints on the electrostatic stability of insulating surfaces, for instance the sum of the Born charges is zero at a surface [571, 572]. How this will couple to deformations at surfaces and interfaces is an open question. For instance, to what extent do atomic relaxations at or near the surface away from the contact compensate for the bound surface charge? There is evidence for compensation at surfaces or interfaces involving ferroelectric materials [573-575]; we strongly suspect that there will also be contributions for flexoelectricity and also band bending.

Another area where more could be done is on the tribology/triboelectricity of sliding contacts; we will not attempt a comprehensive analysis here, rather a few illustrative topics. Our recent work has provided a reason why sliding can play a major role: the shear introduces asymmetry which automatically leads to a charge imbalance between the leading and trailing edges of an asperity contact, and hence current flow [132]. Much more is understood about the details of sliding, and there is a large information gap where more advanced tribology methods can be applied to sliding triboelectricity. For instance, *in-situ* studies of tribology inside transmission electron microscopes are now well established [576-580]; in principle these could be extended to include simultaneous electrical measurements.

There are other areas where more could be done using existing tribology science expanded to include triboelectricity. For instance, does the breakdown of continuum models for nanoscale contacts analyzed by Luan and Robbins [581] require a modification for nanoscale triboelectric contacts? If sliding occurs via dislocations [143, 582, 583], what are the consequences of this for the surface polarization? Alternatively if the sliding is more by traveling cracks [584, 585], the known high flexoelectric polarization around crack tips [401-403, 508] may be worth adding to the existing tribology models. There is already starting to be some interesting work in such directions, for instance the addition of contributions from adhesion associated with charge transfer by Lu *et al*. in two recent papers [332, 586]. Another recent paper that is worth mentioning in this context is that of Sobarzo *et al*. [587]. These authors found that the initial charge exchange between nominally identical materials was somewhat random but became reproducible after repeated contacts. They attribute this to nanoscale morphology changes, which strongly implies a connection to classic running-in (aka break-in or wear-in) in tribology [588], which is believed to be due to micro-topological changes of contact surfaces [589]. That changes of surface in the initial stages of triboelectric experiments has been mentioned as early as the work of Shaw in 1917 [68], as well as cursory mentions in many other articles. To our knowledge, this is the first definitive demonstration of triboelectric running-in, a phenomenon which, based upon established tribology, we suspect is general.

There is also the question of collective effects as mentioned earlier. There has been recent work on statistical contacts [120], and also other recent work imaging the real contacts in a triboelectric system [148], but we think far more could be done. In tribology there has been considerable progress in simulation methods to model phenomena associated with sliding, see for instance the reviews by Vanossi *et al*. [561], Cammarata *et al*. [590], and also the overview from a Lorentz Center workshop [129]. As illustrated in figure 32, more can be taking place than just single asperity contacts. There has also been recent work on rough surfaces [127, 128]. It would be relevant to triboelectric studies to add the flexoelectric (and/or piezoelectric) contributions, at the very least analyzing the surface bound charge (polarization) which plays a major role [149].

We think that in the future much more needs to be done in terms of careful, theoretical modelling and analysis as well determination of experimental factors. This is not new, and in 1959 Montgomery [33] made a statement that we believe remains true to this day:

*"Great complexity is inherent in the problem, and painstaking experimentation and patient analysis must be expected to precede understanding."*

While there is now a small database of trap states for some polymers, for each of these the dielectric, flexoelectric and piezoelectric (if they have one) parameters need to be determined, as well as better details about the band bending. Coming at this from the theoretical side will involve some of the existing multiphysics approaches which combine all the relevant terms, hopefully including contributions from Fermi-Dirac occupancy. Here the scale has to extend from the continuum level of several times the Debye length in any dielectrics, down to the subnanometer scale of interface dipoles and metal induced gap states. Many of the relevant theoretical parameters can be calculated *ab-initio* with modern DFT codes, and with such work it will be informative to see how well experiment and theory mesh. It appears that with oxides which do not have strong correlation terms the flexoelectric terms are not very sensitive to the choice of functional [490], although as Hong and Vanderbilt showed pseudopotential calculations need a correction term in order to reproduce higher-level all-electron calculations [159, 221]. The question of how high one has to climb the Jacob's ladder of functionals [591] for more strongly correlated materials (including work function and related contributions) we pose for future work.

It may be possible to model some contacts using density functional methods, but we will caution that such simulations will have to involve significant numbers of atoms, much larger than published calculations to date as otherwise there can be artifacts introduced by truncation of the long-range screening by electrons and also atom displacements. While for metals the screening lengths are small, of the order of interatomic spacings, for dielectrics the screening length is of the order of the depletion width which will range from ~10nm for a dielectric with many free carriers to a micron or larger. There are also subtle issues such as the use of the Mermin functional [592] to include the electronic temperature which will otherwise lead to wrong dopant state occupancies, how to include low concentrations of electronic dopants, as well as accuracy issues with charged cells and dipoles [593] and numerical issues in the limit of very small occupancies of states even with double precision numerics. These are not "black-box" calculations, and may also be hard to converge or be unstable unless care is taken particularly with the fixed-point iterations [594]. (Note the comments above about the limitations of pseudopotential methods without corrections.)

Turning to experimental work, there is already a fair amount of information in the literature on optimizing triboelectric energy harvesting devices (TENG), for instance the review by Zhou *et al.* [595]. Some of the approaches taken such as modifications of the work functions, dielectric constants, changes of the elastic moduli and surfaces circumstantially connect to various aspects discussed herein. We will challenge the experimental TENG community to increasingly focus on quantitative connections, which in our view will drive better informed, and therefore faster-improving, design of triboelectric devices. In many cases, missing knowledge of key materials properties makes quantitative analysis impossible. Therefore, in all triboelectric experiments we suggest determining the following (all are already known for a few select materials, and some are known for many other materials):

- The electronic structure and properties of the materials, including work functions, band energies, doping levels, and trap characteristics.
- For dielectrics, the band bending and any pinning in different cases.
- Mechanical properties, including plastic and size-dependent effects when appropriate.
- Electromechanical properties, including flexoelectric and piezoelectric coefficients, as well as deformation potentials and strain-induced changes of the mean inner potential or work function. Note that these are not just simple material properties, but will depend upon doping, defects and for polymers additives.
- Surface characteristics, including the roughness, crystallographic orientation, surface terminations, and chemical species present.
- Dynamic properties of the system, including the friction coefficients, dielectric behavior, and the mobility of charged species.

Given the role of dopants, surface adsorbates and surface crystallography mentioned earlier, ideal experiments will have tight control of these. This will not simply involve the use of a vacuum gauge measuring a static UHV background pressure. Moving parts always lead to desorption of surface bound species, and detection of submonolayer levels is not trivial; as reviewed earlier, it only takes a fraction of a monolayer to shift work functions and therefore contact potentials. For instance it is unclear how many experiments to date have been completely free of chemisorbed water, since detection of hydrogen or hydroxyl groups on surfaces is tricky, though possible with careful STM [596-598] or grazing-incidence XPS [599, 600]. Various procedures for different materials can be found in the surface science literature to which we will refer the interested reader.

We will propose here some specific experimental tests which we believe will move the field forward (in addition to improved models and other points mentioned above). Our suggestions are:

1. It is known that temperature changes the flexoelectric coefficients of strontium titanate [601]; does the triboelectric charge transfer vary quantitatively in the same way? (Similar coefficient variations have been documented for a relaxor PIN-PMN-PT [495] and $BaTi_{1-x}Zr_xO_3$ [602].) Such experiments would need to track the electromechanical changes as well as elastic constants and tribocharge. We note that there is circumstantial evidence that there could be interesting science in the 1978 work of Ohara [603], who observed peaks in tribocharge between polymer films near the glass transition temperatures.
2. There is recent data indicating that doping alters apparent flexoelectric coefficients quite dramatically [342-344] which are believed to be due to the barrier-layer effect [345-347] as mentioned earlier. It is reasonable that barrier-layer and doping contributions will similarly change tribocharging. There are already hints toward this [604], but without companion measurements of flexoelectric coefficients the evidence is circumstantial at best. Here the work function and band bending would need to be tracked together with charge transfer and electromechanical coefficients. Similar to standard practice in semiconductor preparation, careful control of material surfaces, as well as quantification of trap states, will also be helpful for detailed analysis of the effects of the electronic structure.
3. It is known that there can be strong crystallographic dependence of flexoelectric coefficients, for instance a sign reversal for strontium titanate between [100] and [110] directions [160]; do these quantitatively map to tribocharge changes?
4. Going further afield, what are the flexoelectric (and piezoelectric) coefficients, work functions and similar of different types of pharmaceuticals, powders such as grain or flour, desert sand (from different parts of the world or Mars) or volcanic plume material? Do these correlate with tribocharging measurements already present in the literature [11, 605, 606]? We note that measurements indicate that different volcanic ash has different tribocharging characteristics [607], which is an alternative set of experiments.
5. A mechanism for sliding DC electrification due to sliding depletion regions in Schottky contacts has been proposed, but not yet verified in detail [326]. We have argued that the timescales of sliding and depletion formation are typically incongruous [132]. Is there clear, quantitative evidence for contributions of this mechanism, which would be especially strong in experiments where sliding is fast or depletion formation is unusually slow?
6. Friction of asperities involves a stick-slip process with periods of force built up followed by sudden and rapid movement. Can the microsecond-scale or even nanosecond dynamics of the tribocharging that is associated with these events be resolved? Note that the frequency dependence of dielectric and electromechanical parameters may need to be considered for accurate work with short time scales.

As a final comment, we believe that in many ways there is an analogy between triboelectricity and the early cat's whisker diodes [608] such as the one for which Jagadis Chandra Bose received the first patent in 1904 [609]. In these, a metal wire such as a phosphor bronze was pressed into a semiconductor such as a piece of lead sulfide. These predate our modern understanding of band structure and interfaces by many decades, so they evolved by trial and error, sometimes never working. In hindsight the reason is clear: uncontrolled impurities and contacts. What is needed in future triboelectric experiments is to take the same path that the semiconductor physics community took about 100 years ago; care and attention to detail about the materials with better than one part per million level . This is not impossible.

## Acknowledgements

This work would not have been possible without contributions from many scientists. The work started with Pratik Koirala, who saw a strange bending in an electron microscope and decided not to ignore it. Together with Chris Mizzi it was worked out that this was due to flexoelectricity [494]. Then Chris together with Alex Lin was able to combine flexoelectricity and tribology [19]. This line of work was then pushed further by Chris and then involved Karl Olson [56, 101], and Karl has pushed it much further [131, 132, 149, 194].

We would also like to thank Martin Castell, Lincoln Lauhon and Peter Voorhees for comments as this work has developed, and also Julian Gale, Jun Liu, Chris Mizzi, Martin Mueser, Scott Waitukaitis and Yang Xu for comments on a draft of this manuscript.

KPO acknowledges funding from the McCormick School of Engineering at Northwestern University.

## Figure Captions

Figure 1. Amber bear figurine from the Meolithic or Proto-Neolithic period (ca. 9600-4100 BCE) from the Muzeum Naradows in Szcezecin. Public domain image from Wikimedia.

Figure 2. Triboelectric decision tree. Starting from the top, **Drivers** are shown in green boxes, while **Mechanisms** are shown in orange rounded boxes. Environmental conditions such as the surrounding gas, humidity, and temperature are **Dependencies** and shown by the yellow circles. Finally, points related to charge transfer remaining after disengagement are shown in dashed boxes at the bottom.

Figure 3. Schematic of triboelectricity at different physical scales. What appears as smooth contact of flat faces (left) is actually contact between many asperities (middle), each of which involve nanoscale contacts between the two bodies (right).

Figure 4. Unitless Hertzian (a) stress and (b) strain for a contact of force $\pi$ between a rigid sphere of radius 4/3$\pi$ and an elastic half-space with effective modulus 1 and Poisson's ratio 0.24. These values are chosen so that the contact radius and mean contact pressure are both 1. The $\rho$ and $z$ coordinates are the in-plane distance and depth from the center of the contact, respectively. The contour lines are shown on the color bars.

Figure 5. Tribocurrent vs. load for a nickel needle on a boron doped diamond single crystal, (001) orientation showing the tribocurrent for the as-annealed track (empty squares) and for the oxidized one (filled squares) as a function of load. The solid lines are the best fits as force to the one-third power. Adapted with permission from Escobar, Chakravarty and Putterman [90].

Figure 6. Representations of the radial stress field for a contact with shear. Shown is both half the top surface, and a vertical section. Units are normalized by the contact radius and mean contact pressure. In a) there is only a normal force, in b) purely tangential whereas in c) they are combined which is the case for sliding.

Figure 7. (a) Contact stiffness $K_c$ and (b) charge transfer $Q_s$ measurements for fractal surfaces with differing roughness amplitudes $R_t$ contacting metal electrodes. The samples were 3D printed VisiJet M3 crystal material. The contact stiffness and force $F_c$ applied are normalized with the modulus $E$ of the printed material and apparent contact area $A$. Reprinted with permission from Zhai *et al*. [120].

Figure 8. A two-dimensional sketch showing that strain leads to polarization only in non-centrosymmetric materials, whereas in general a strain gradient breaks the inversion center and leads to polarization and in any material.

Figure 9. a) The Belincourt method for piezoelectric coefficient measurement, and the (b) three point bending and (c) pyramid compression methods for flexoelectric measurements all involve applying an oscillating strain to a dielectric and measuring the charge on the metal electrodes, effectively using the tribocharge.

Figure 10. (a) Polarization in the dielectric of a capacitor is compensated by surface charges on the metal plates. (b) Strain-induced polarization at asperity contacts is compensated by surface charges that are involved in charge transfer. The physics convention for the dipole arrows is used with the arrow tip at the positive pole of the bulk polarization.

Figure 11. (a) Charge transfer is reduced by charge of the same sign initially on the contacting body. (b) Charge on the particles reaches a saturation values. Reproduced with permission from Matsusaka, Ghardiri and Masuda [225].

Figure 12. Work function of MgO (111) surface (y) in volts as a function of fractional coverage by chloride ions (x) replacing hydroxide ions. Above 1/3 monolayer coverage the chloride is unstable due to interionic repulsions, and the work function saturates.

Figure 13. Dipoles formed due to work functions steps at metal-oxide surfaces may point either (a) toward the metal, as at Ni/NiO interfaces, or (b) away from the metal, as at Al/NiO interfaces.

Figure 14. (a) PN junction, (b) heavily doped N-type/lightly doped n-type junction, (c) Schottky junction, (d) metal/insulator/semiconductor junction, and (e) Ohmic junctions due to (i) $\phi_m < \phi_s$ and (ii) highly-doped semiconductors leading to easy tunneling through the barrier. In each case, the left shows the band structure before contact and the right after contact. $\phi_m$ is the metal work function, $\phi_s$ and $\chi_s$ the semiconductor work function and electron affinity, $\phi_b$ the barrier height, $\phi_i$ the built-in potential, and $W$ the depletion width.

Figure 15. Mechanical stress can drive “mechanochemical” or “tribochemical” reactions by changing reaction energy landscapes. For example, species (blue spheres) may preferentially adsorb on one body (a), as against on the other (b) illustrated with no force, the black line and markers. When force is applied, they can move from (d) to (f), the green line. Additionally, the energy barriers to form intermediate species (c) versus (e) can be modified.

Figure 16. Effect of combined electric force and field to break the S-S bond of a high-symmetry linear disulphide at the bond indicated in red. The field is indicated by the blue arrow, and the dipole moment in Debye is in red. Adapted with permission from Scheele and Neudecker [306].

Figure 17. Two examples of trap state possibilities. In a) for a metal to p-type junction states in the gap which are empty (open circles) may be filled (arrow) due to band bending in the depletion zone or electromechanical terms. Similarly in b) for an n-type the trap state can donate to the metal. As drawn the trap states are inside the surface, but they could also be across the surface away from the contact.

Figure 18. (a) General distribution of trap states in the band structure of a polymer contacting a metal. (b) De-trapping and other processes that change the population of shallow and deep traps. (c) Measured trap state energy and density distribution for four polymers contacting steel. Each are a sum of two peaks, corresponding to deep and shallow traps at higher and lower energies, respectively. Reprinted with permission from Xu *et al.* [314].

Figure 19. Cyclic triboelectric series where each material will charge positive with respect to that clockwise from it.

Figure 20. Illustration of bridging electrons at a contact. For two metals with different work functions (blue and orange here) there will be electron transfer indicated as a red arrow which will lead to an interface dipole, and other charges away from the contact depending upon the geometry.

Figure 21. Work functions for both a metal $\phi_M$ and a semiconductor or insulator $\phi_S$ where there is both band bending and Fermi level pinning (the red states). Here the contact potential $V_c$ includes the dipoles at the interface as well. Both the Fermi level $E_F$ and the Debye length $\lambda_D$ are indicated, as well as the two terms $V_D$ and $V_{CB}$ for the band bending with Fermi level pinning.

Figure 22. Abrupt junction band bending in a) and in b) with a vacuum gap. Shown in the diagrams is both the potential V and the electric displacement D (normal to the junction), as well as the depletion widths $x_p$ and $x_n$, the built-in potential $V_c$ (contact potential) and the separation d.

Figure 23. Pressures causes changes in the reaction energy of hydrogen transfer and the potential barrier at a hydrogen terminated silica/gold contact. (a) The contact is modeled ab initio simulating the initial and the final state at a fixed interface distance, and then progressively reducing this distance to increase the pressure. (b),(d) Reaction energy at every indentation step, respectively, for hydrogen transfer of fused silica and quartz. (c),(e) Potential barrier before (unfilled circles) and after (filled circles) hydrogen transfer, respectively, for fused silica and quartz. Reproduced with permission from Fatti *et al.* [320]

Figure 24. Schematic of the types of carbon forming at the contact during a sliding experiment leading to what is called a tribofilm as well as wear of material. Adapted from Wang *et al.* [23] with permission.

Figure 25. (a) Work functions or contact potentials can be modified by adsorbed species such as water. (b) Adsorbed water can passivate or introduce trap states. (c) Humidity affects various conduction mechanisms, for example through water bridges between bodies or through diffusion of adsorbed and/or dissociated water molecules (particularly the protons) over relatively small times $t$ and temperatures $T$. (d) Humidity reduces the charge density needed for atmospheric breakdown.

Figure 26. The population of trap states for different humidities, reproduced with permission from Amiour, Ziari, and Sahli [444]. Traps energy distribution in Kapton HN polyimide films through surface potential decay model under humidity conditions.

Figure 27. Gold lead electrostat, showing the charges (in red) if a positively charged rod is brought near to the top plate. Original version: Sylvanus P. Thompson [457], Derived version: Chetvorno, Public domain, image via Wikimedia Commons.

Figure 28. Sphere-on-flat contact band diagrams for a Si sphere (orange) on a $SrTiO_3$ flat (blue). (a)-(c) show the conduction and valence band edges as a function of depth (normalized by the indenter radius R) at different radial distances (in units of the contact radius a) from the contact point as defined in Figure 2(a) with a contact pressure of 6 GPa. The unstrained Fermi level of each material is assumed to be at its band gap center and zero energy is taken to be the unstrained $SrTiO_3$ Fermi level.

Figure 29. At the top, a cross-section of the polarization field for a pure normal, tangential and combined sliding contact with the asperity shown schematically as a grid and black arrows for the force directions; below the bound surface charge density. The polarization and charge are normalized, using a mean contact pressure $p_m$, contact radius $a$, flexoelectric coefficient $\mu$, and Young's modulus $Y$ all equal to 1. The polarization and charge scale as $\mu p_m/aY$. The white arrows indicate the direction of the polarization vector, the color scale the magnitude.

Figure 30. Hysteresis loops for a TENG contact showing distinct irreversibility in the backwards open-circuit voltage. Reproduced with permission from Seol *et al.* [527].

Figure 31. Reproduction of the results of John *et al.* [226] for NaCl particles on a Titanium plate. Positive and negative precharges have a different effect on the charge transfer in the next collision, suggesting (as the authors state) that there is a rectifying pn junction being formed.

Figure 32. Schematic of the many processes that can take place at contacts. While they are not going to all be relevant to every triboelectric experiment, the diversity of processes needs to be considered. Reproduced from Vakis *et al.* [129] with permission.

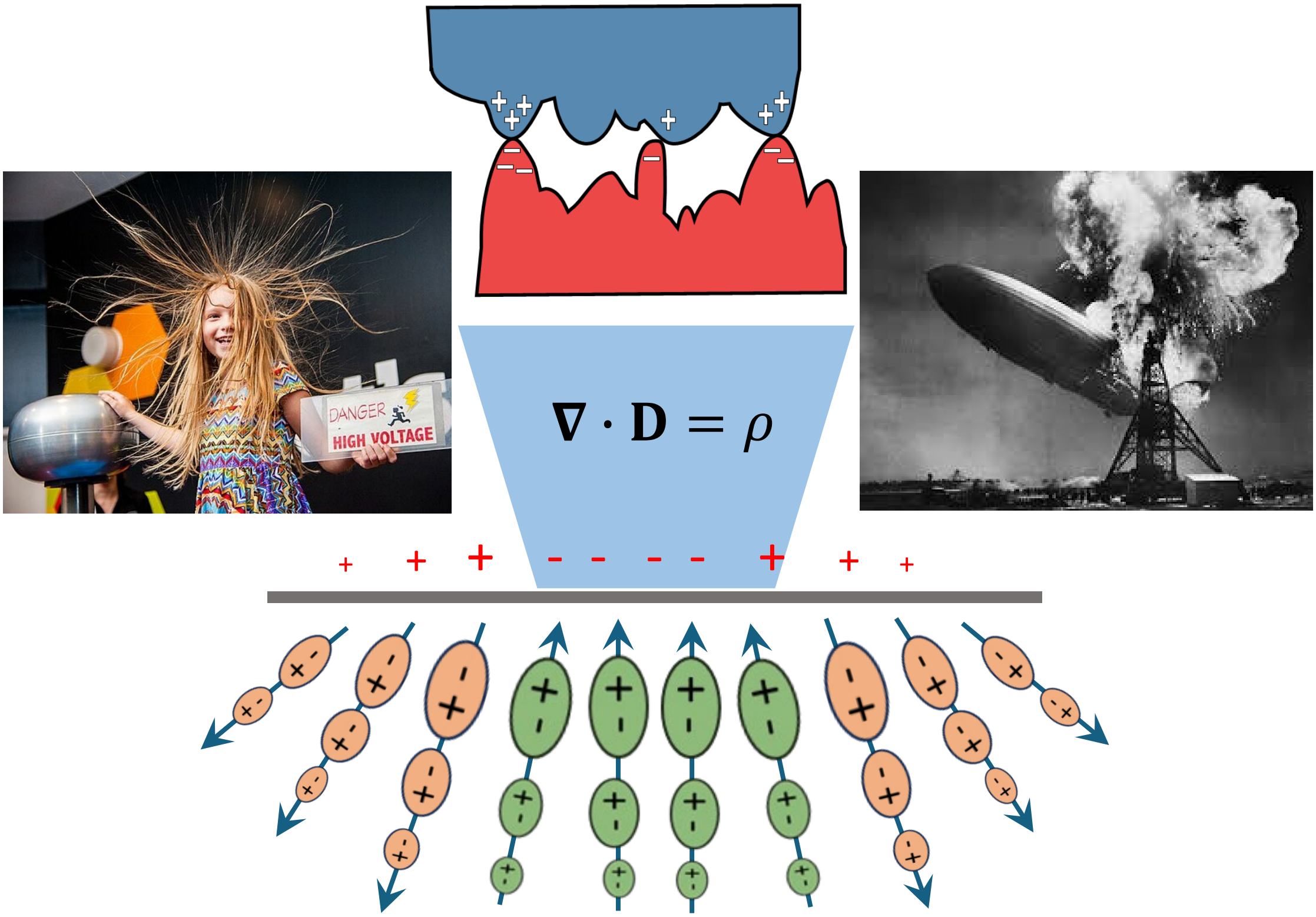

TOC FIGURE. What we see at the macroscale and leads to accidents such as the Hindenburg fire is due to asperity contacts and the general Ampère law with the electric displacement field and polarization.

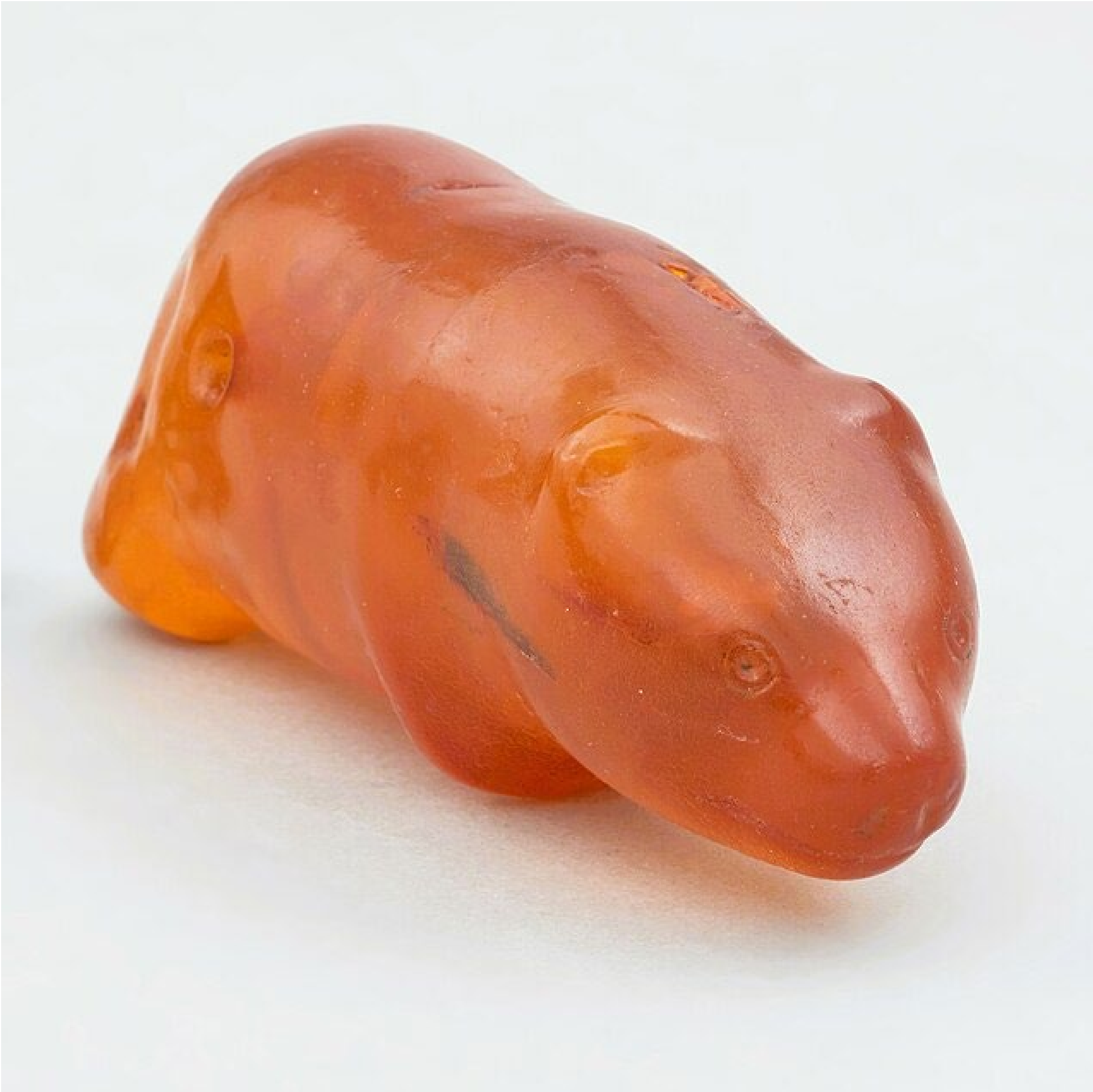

Figure 1. Amber bear figurine from the Meolithic or Proto-Neolithic period (ca. 9600-4100 BCE) from the Muzeum Naradows in Szcezecin. Public domain image from Wikimedia.

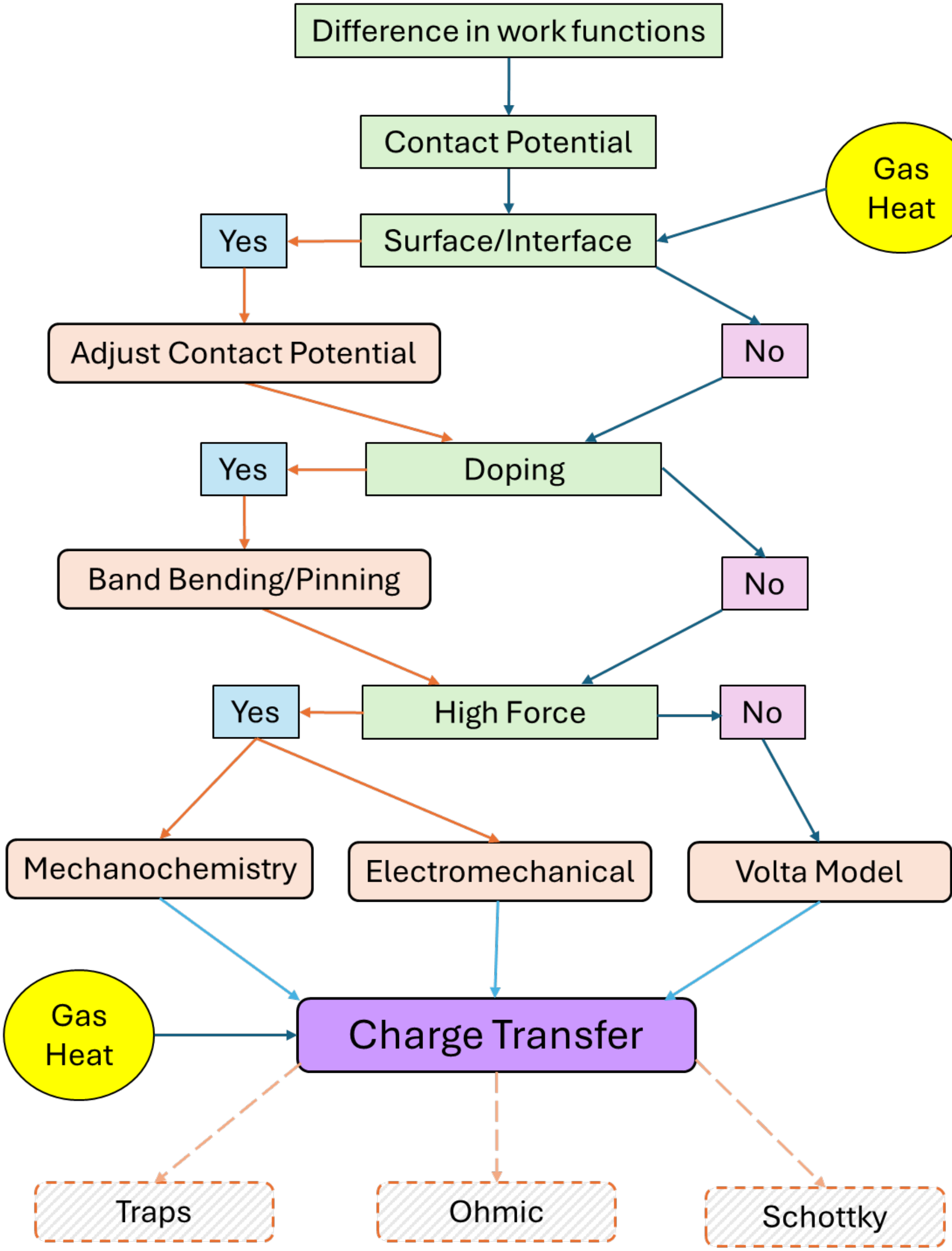

Figure 2. Triboelectric decision tree. Starting from the top, **Drivers** are shown in green boxes, while **Mechanisms** are shown in orange rounded boxes. Environmental conditions such as the surrounding gas, humidity, and temperature are **Dependencies** and shown by the yellow circles. Finally, points related to charge transfer remaining after disengagement are shown in dashed boxes at the bottom.

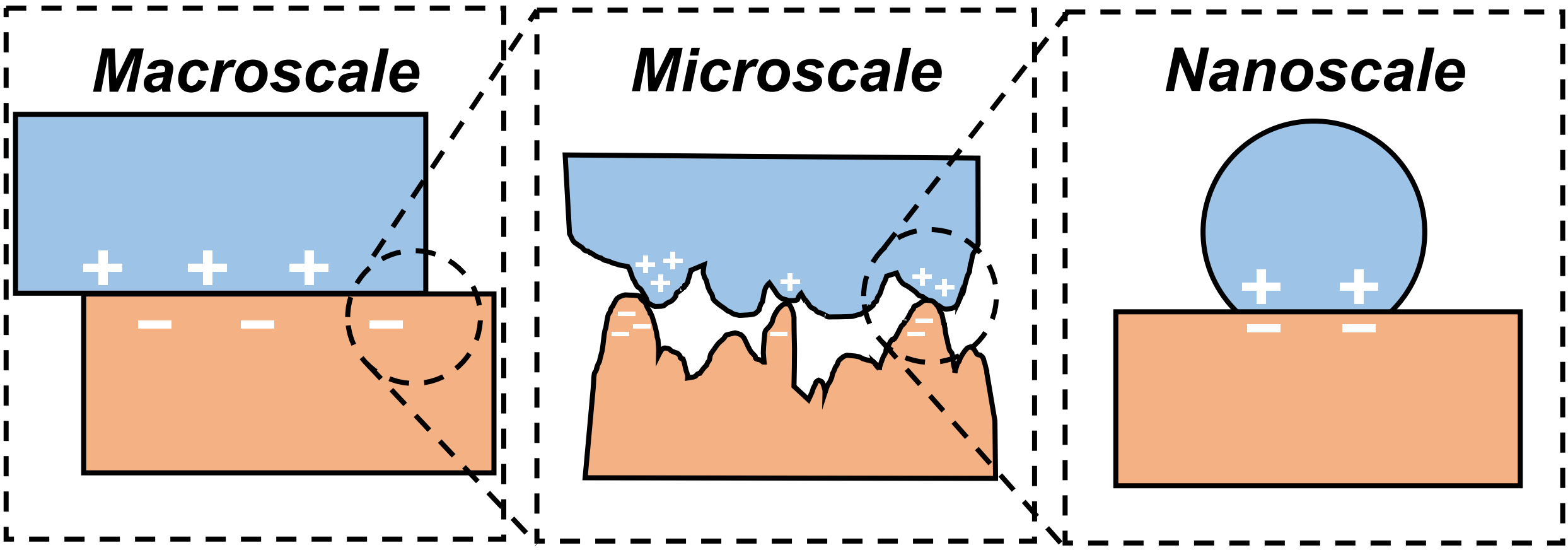

Figure 3. Schematic of triboelectricity at different physical scales. What appears as smooth contact of flat faces (left) is actually contact between many asperities (middle), each of which involve nanoscale contacts between the two bodies (right).

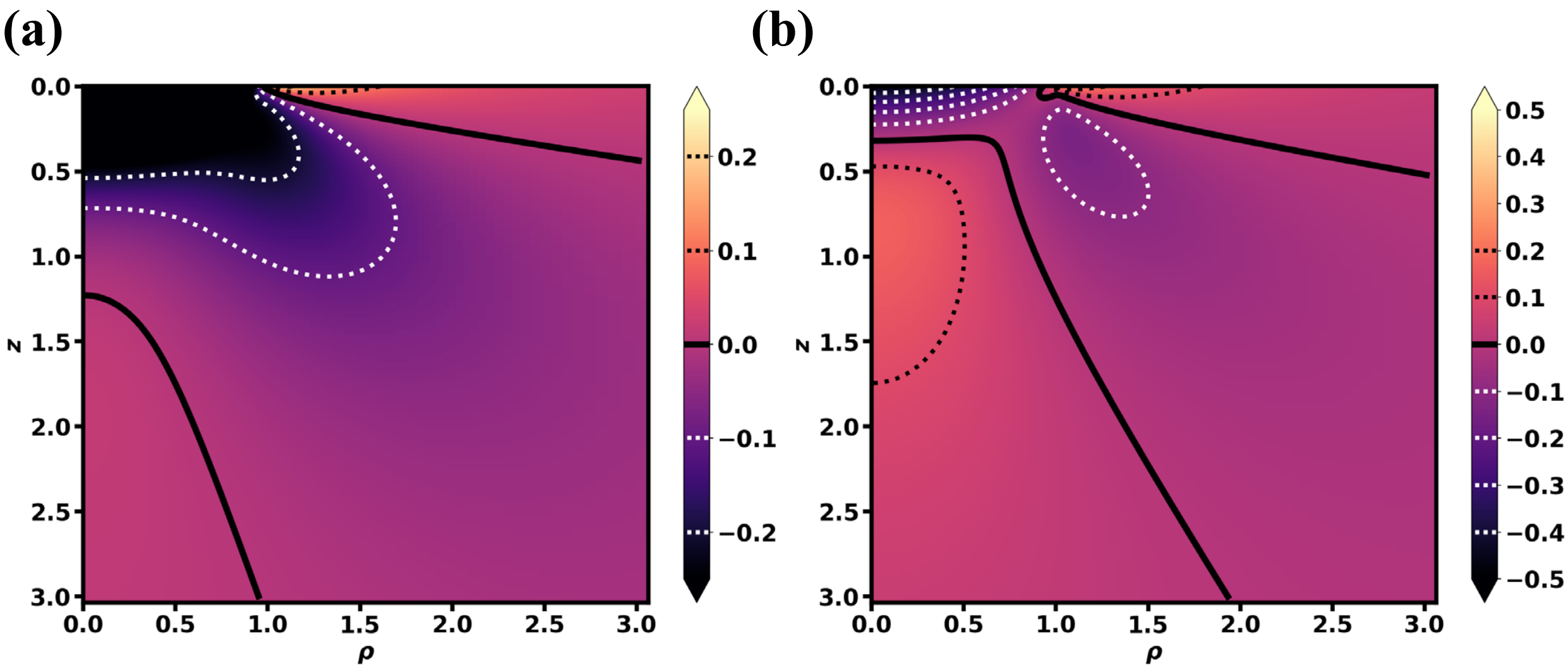

Figure 4. Unitless Hertzian (a) stress and (b) strain for a contact of force $\pi$ between a rigid sphere of radius $4/3\pi$ and an elastic half-space with effective modulus $1$ and Poisson's ratio 0.24. These values are chosen so that the contact radius and mean contact pressure are both $1$. The $\rho$ and $z$ coordinates are the in-plane distance and depth from the center of the contact, respectively. The contour lines are shown on the color bars.

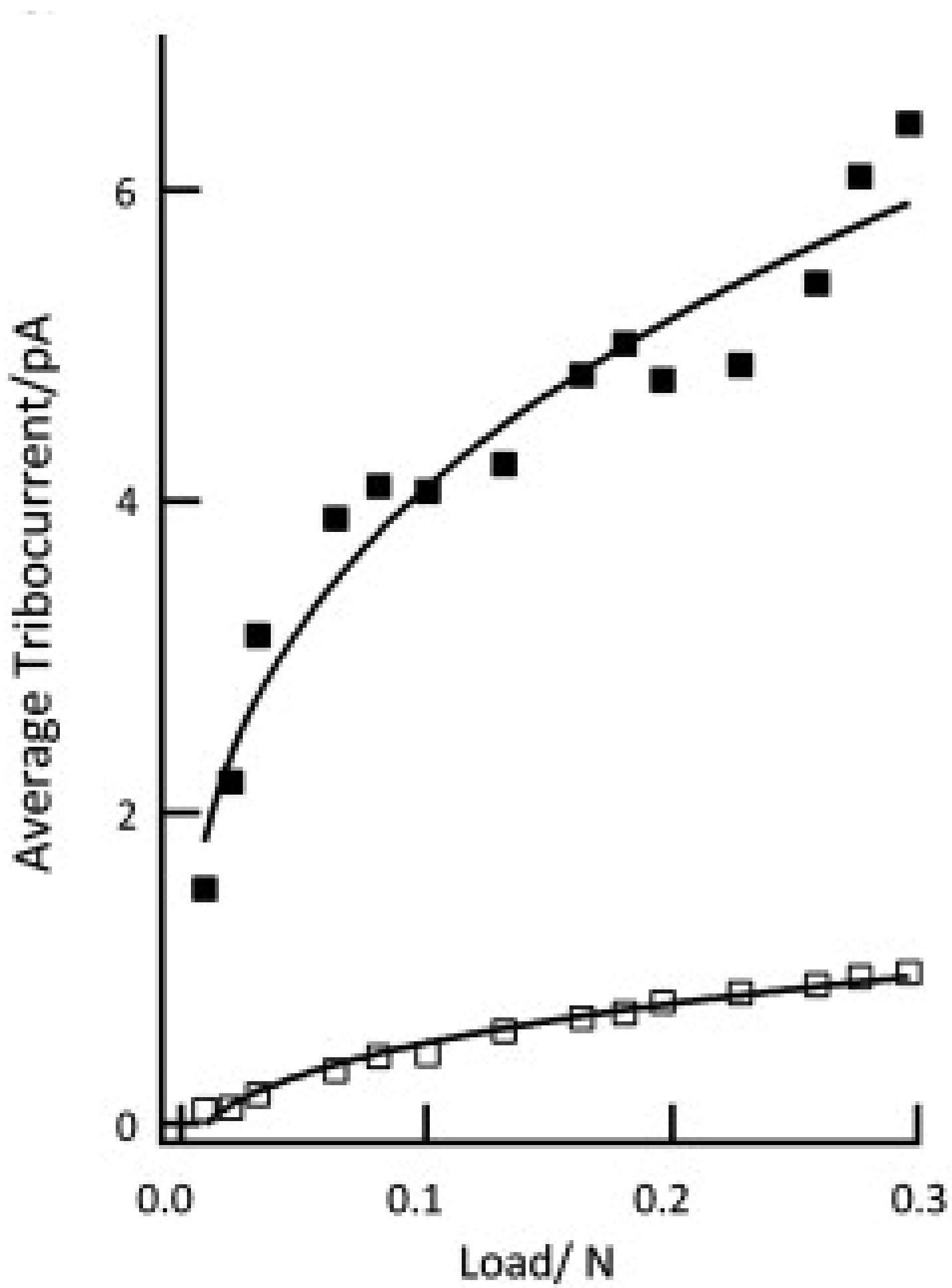

Figure 5 Tribocurrent vs. load for a nickel needle on a boron doped diamond single crystal, (001) orientation showing the tribocurrent for the as-annealed track (empty squares) and for the oxidized one (filled squares) as a function of load. The solid lines are the best fits as force to the one-third power. Adapted from Escobar, Chakravary and Putterman

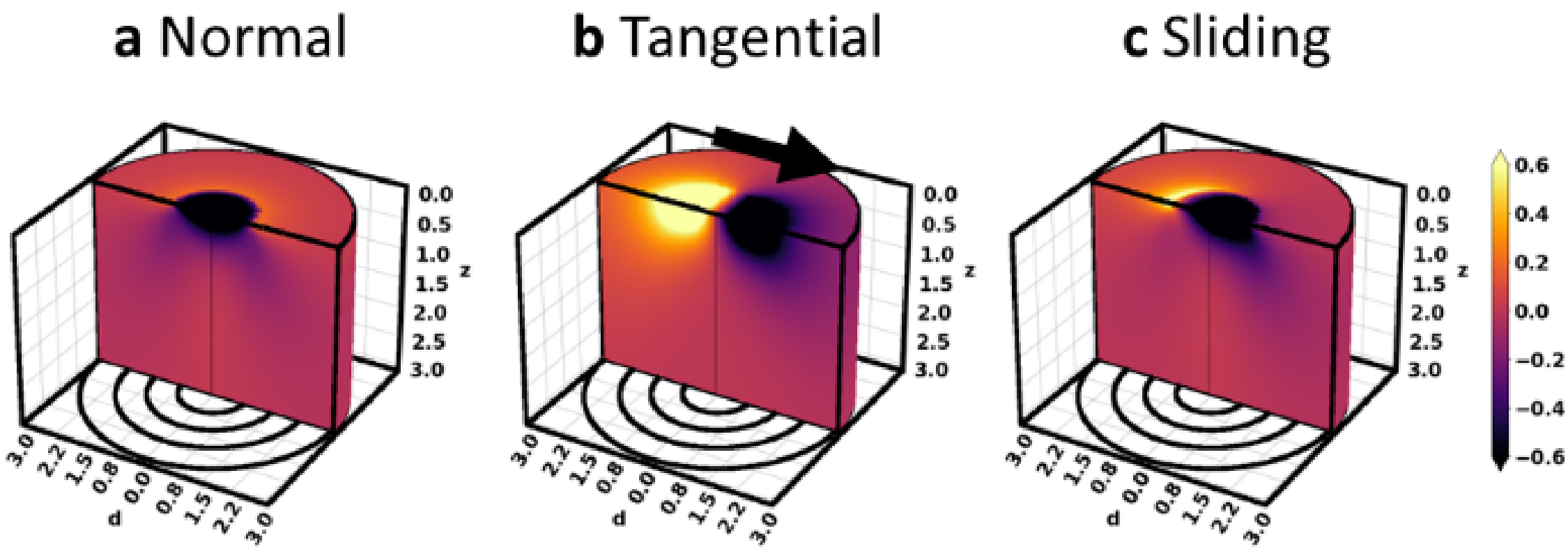

Figure 6. Representations of the radial stress field for a contact with shear. Shown is both half the top surface, and a vertical section. In both cases normalized units are used. In a) there is only a normal force, in b) purely tangential whereas in c) they are combined which is the case for sliding.

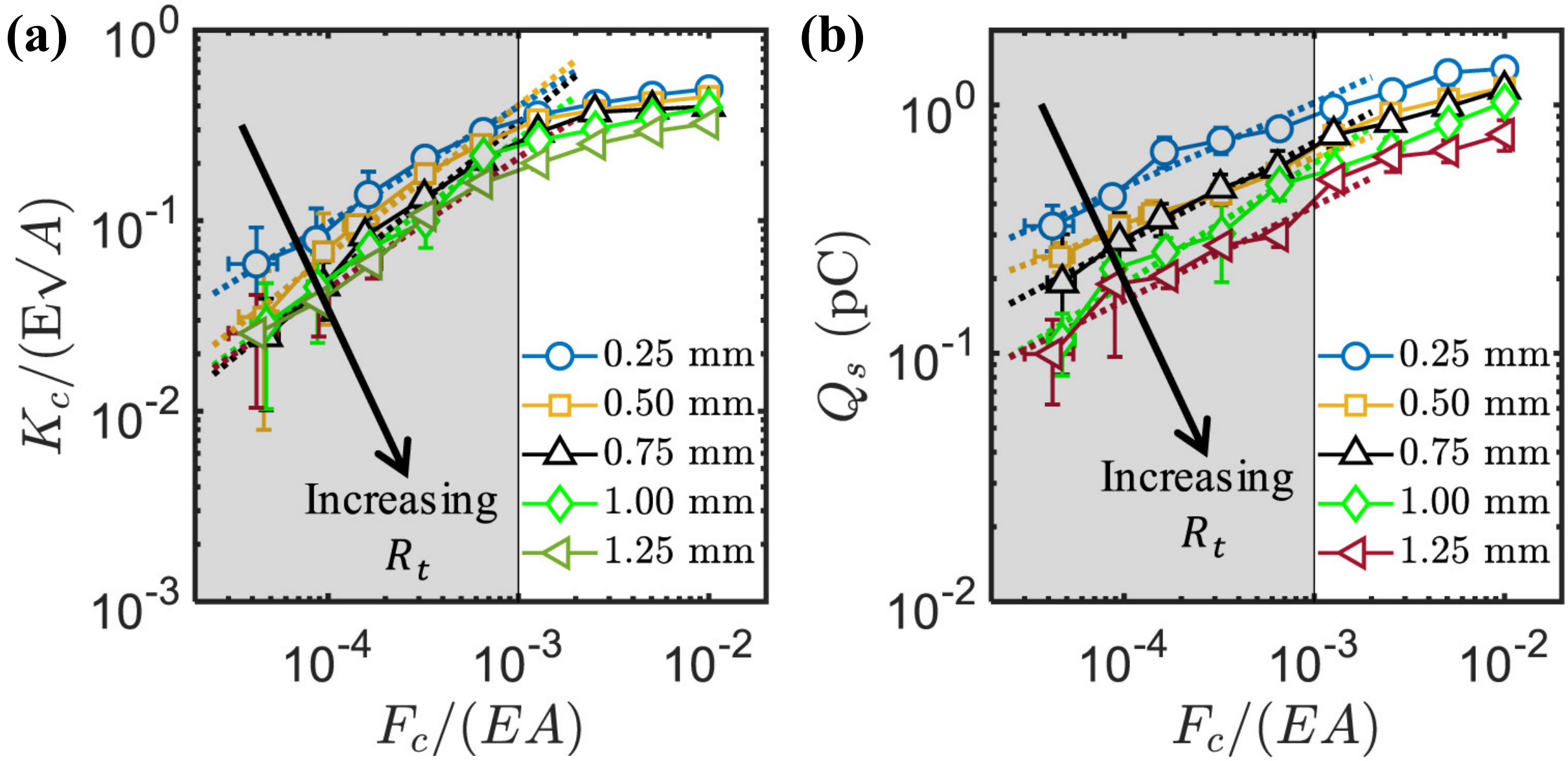

Figure 7. (a) Contact stiffness $K_c$ and (b) charge transfer $Q_s$ measurements for fractal surfaces with differing roughness amplitudes $R_t$ contacting metal electrodes. The samples were 3D printed VisiJet M3 crystal material. The contact stiffness and force $F_c$ applied are normalized with the modulus $E$ of the printed material and apparent contact area $A$. [Reprinted with permission from Zai et al.

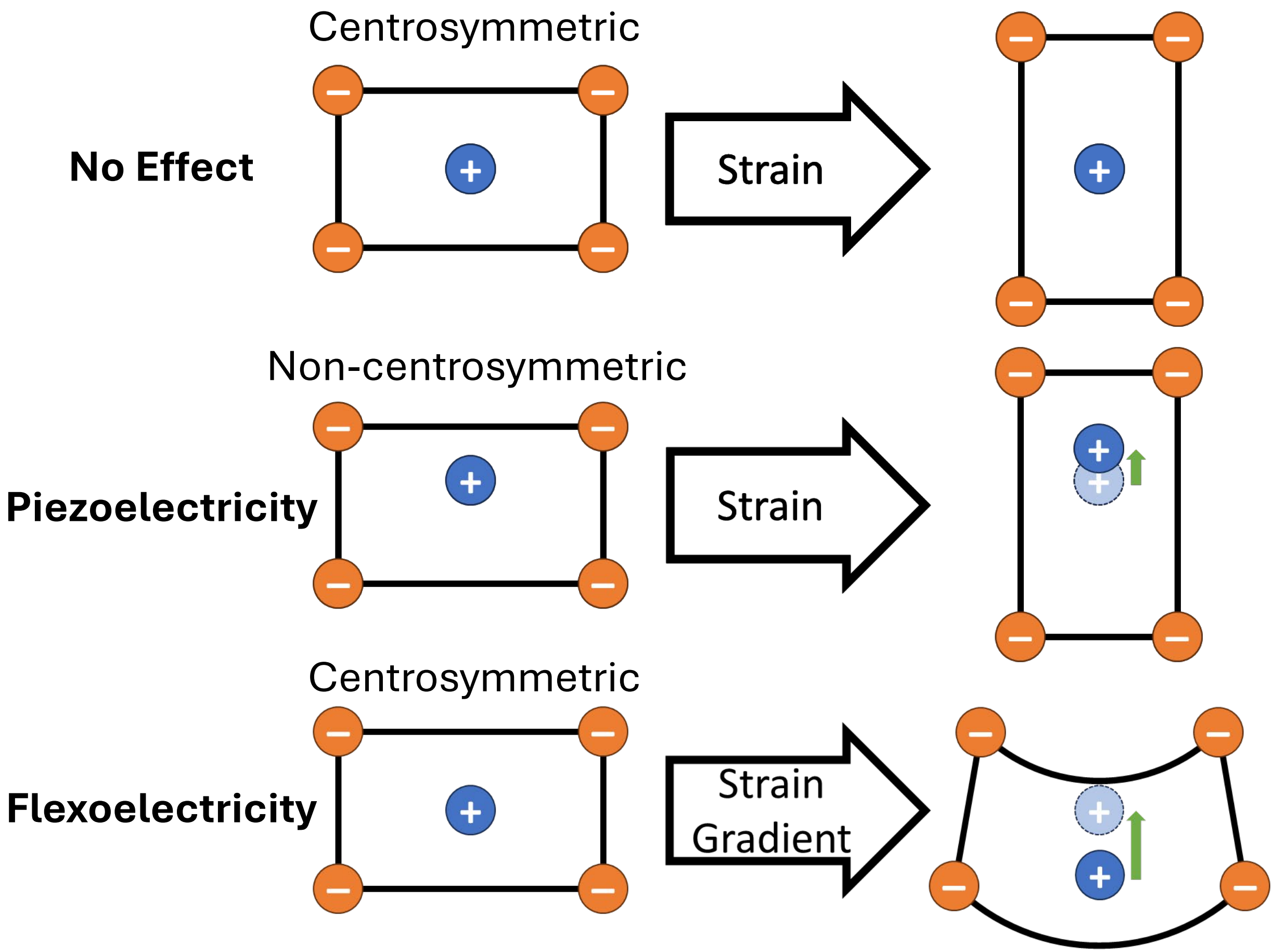

Figure 8. A two-dimensional sketch showing that strain leads to polarization only in non-centrosymmetric materials, whereas in general a strain gradient breaks the inversion center and leads to polarization and in any material.

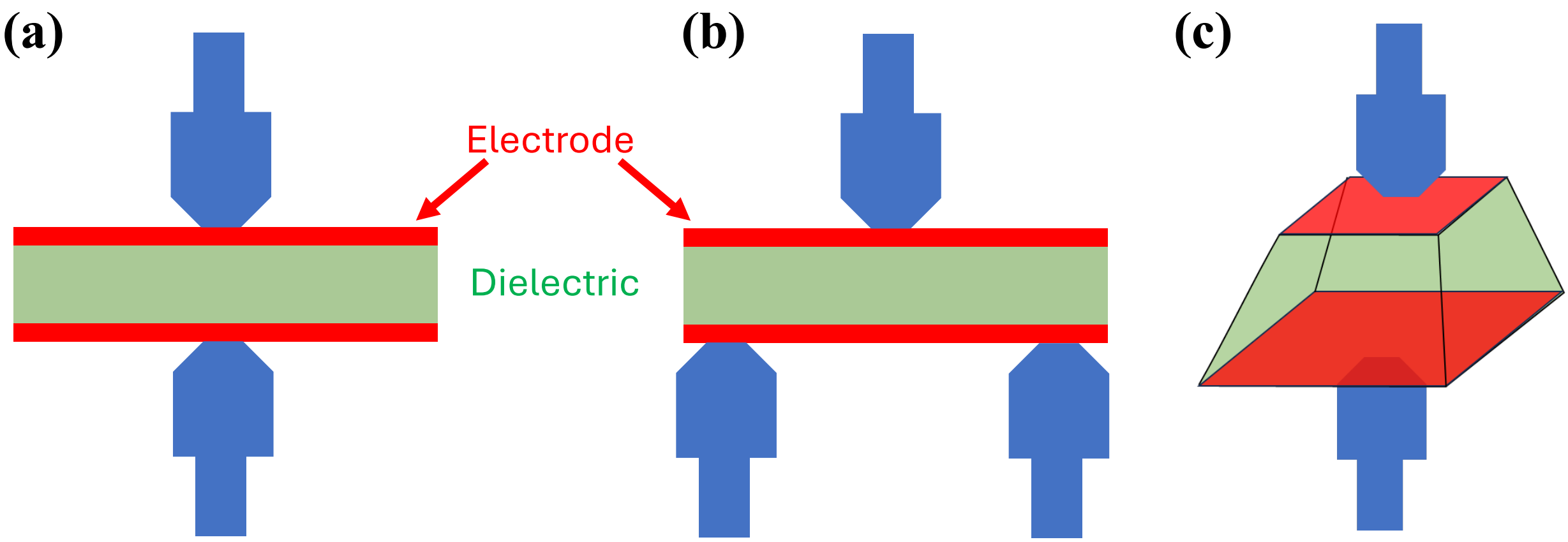

Figure 9. (a) The Belincourt method for  piezoelectric coefficient measurement, and the (b) three point bending and (c) pyramid compression methods for flexoelectric measurements all involve applying an oscillating to a dielectric and measuring the charge on the metal electrodes, effectively using the tribocharge.

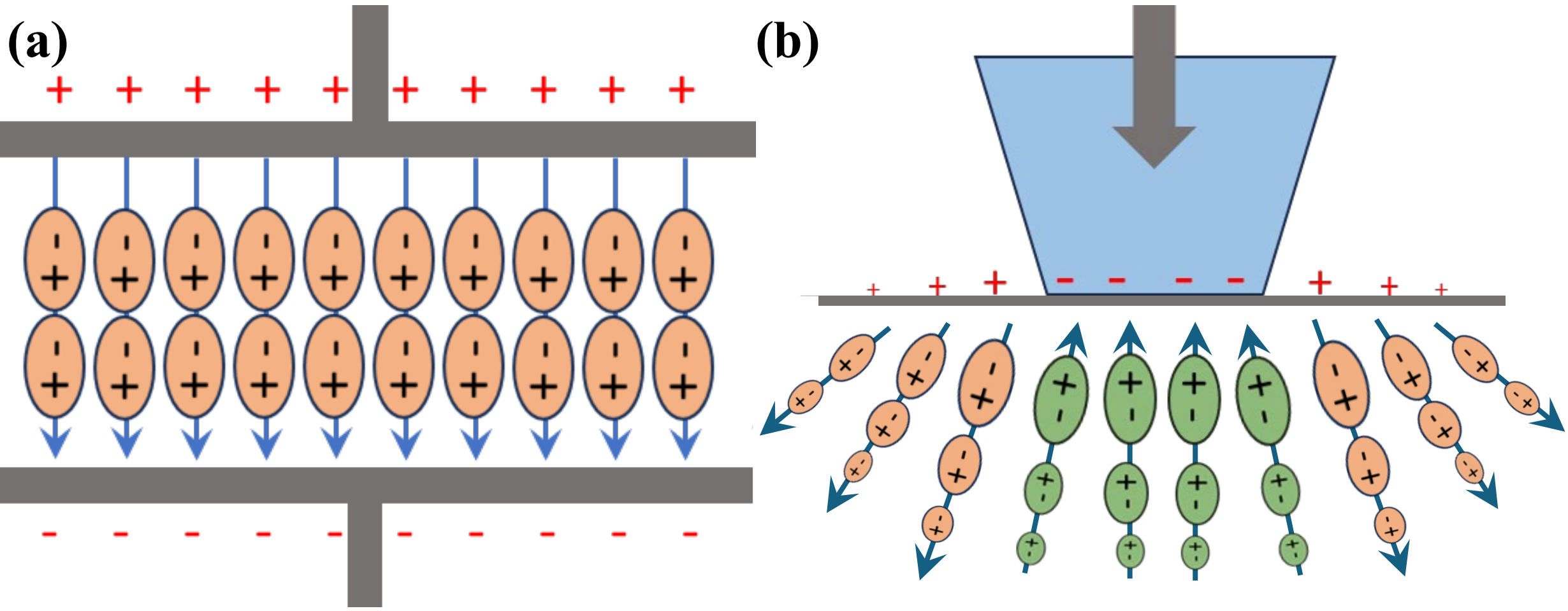

Figure 10. (a) Polarization in the dielectric of a capacitor is compensated by surface charges on the metal plates. (b) Strain-induced polarization at asperity contacts is compensated by surface charges that are involved in charge transfer. The physics convention for the dipole arrows is used with the arrow tip at the positive pole of the bulk polarization.

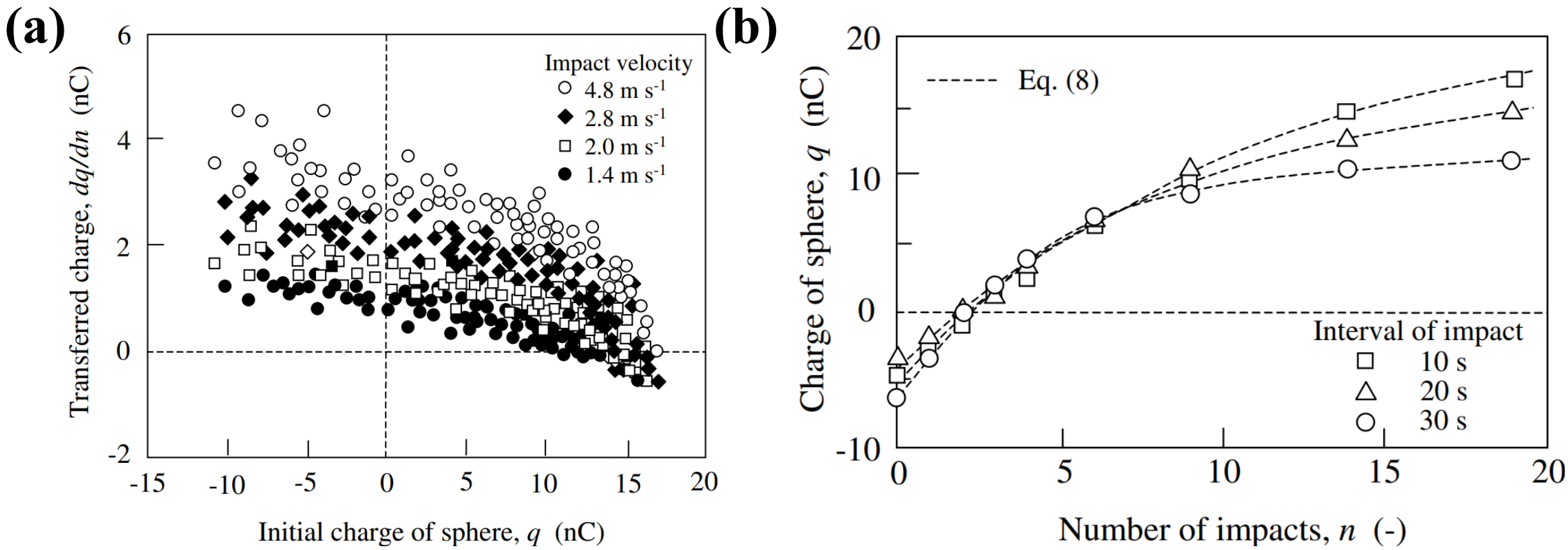

Figure 11. (a) Charge transfer is reduced by charge of the same sign initially on the contacting body. (b) Charge on the particles reaches a saturation values. Reproduced with permission from Matsusaka, Ghardiri and Masuda.

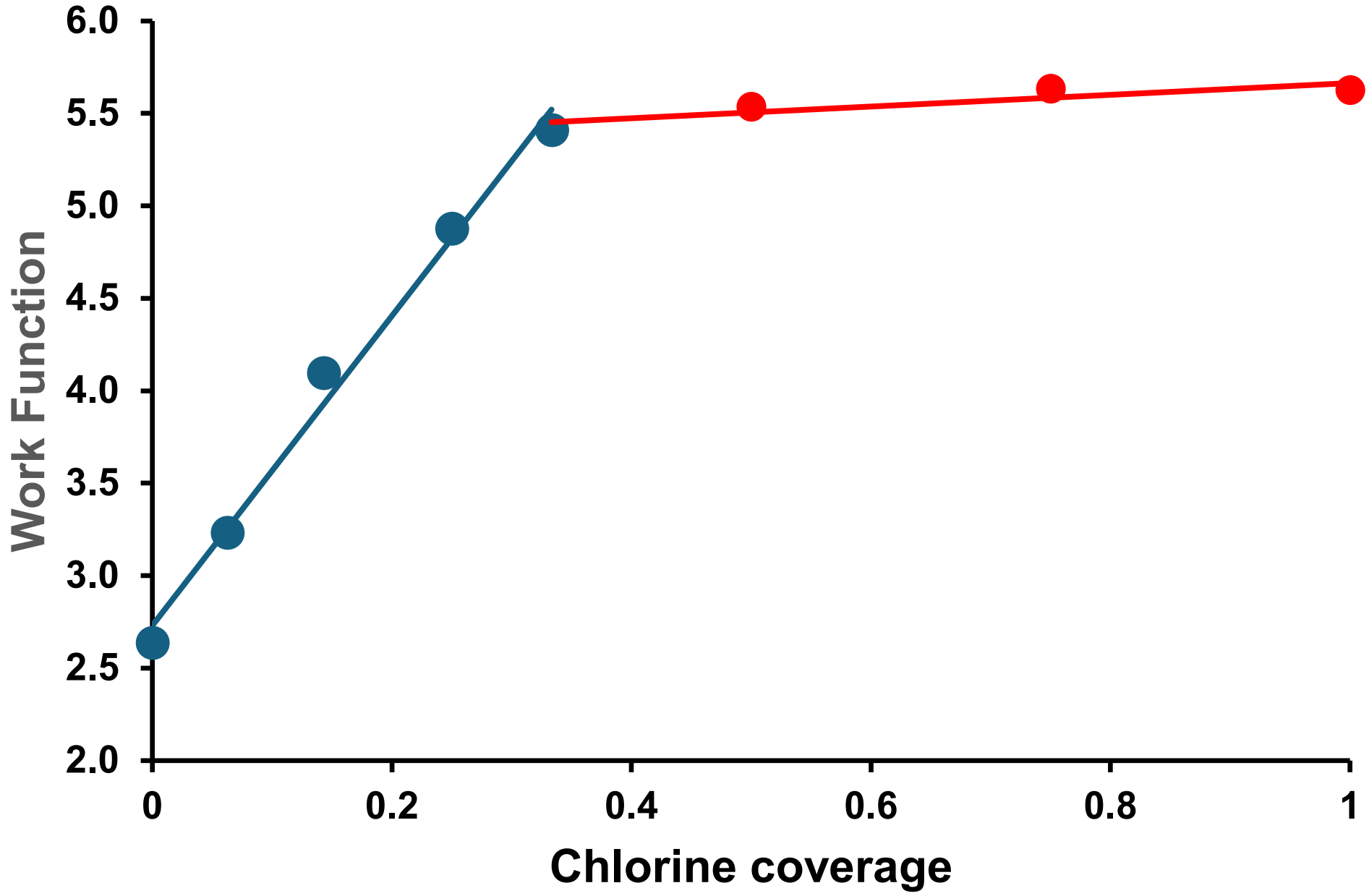

Figure 12. Work function of MgO (111) surface (y) in volts as a function of fractional coverage by chloride ions (x) replacing hydroxide ions. Above 1/3 monolayer coverage the chloride is unstable due to interionic repulsions, and the workfunction saturates.

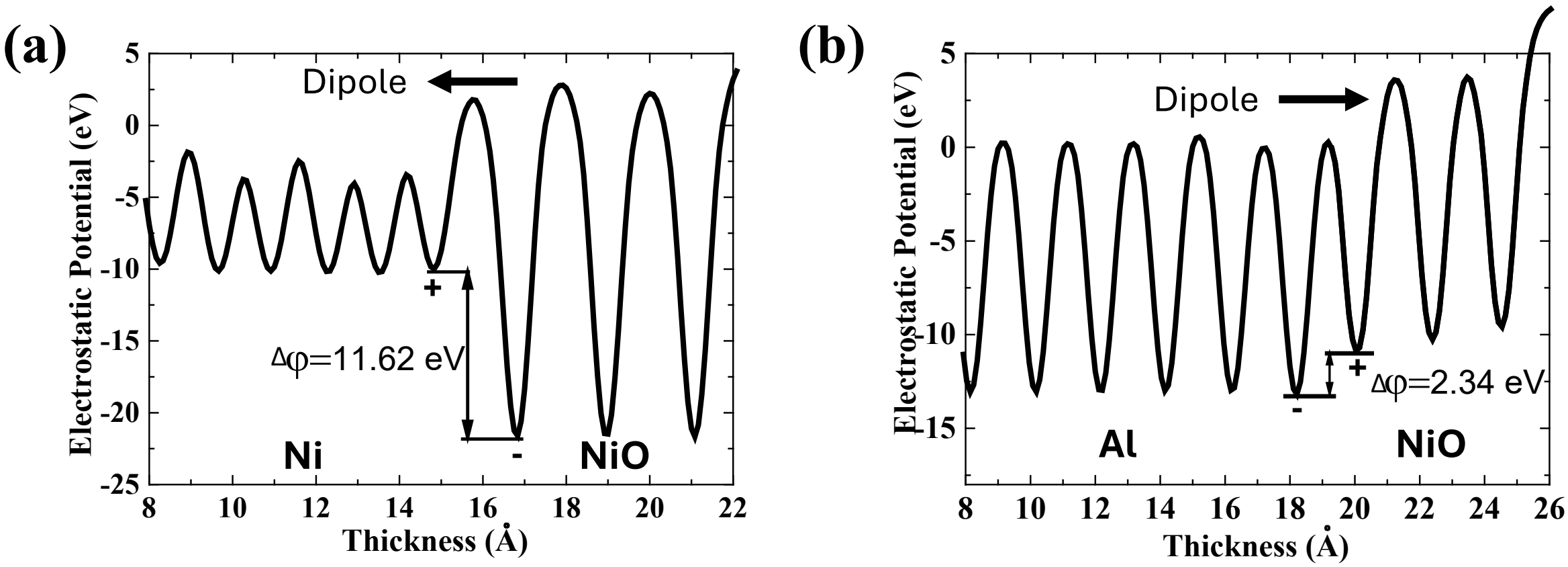

Figure 13. Dipoles formed due to work functions steps at metal-oxide surfaces may point either (a) toward the metal, as at Ni/NiO interfaces, or (b) away from the metal, as at Al/NiO interfaces.

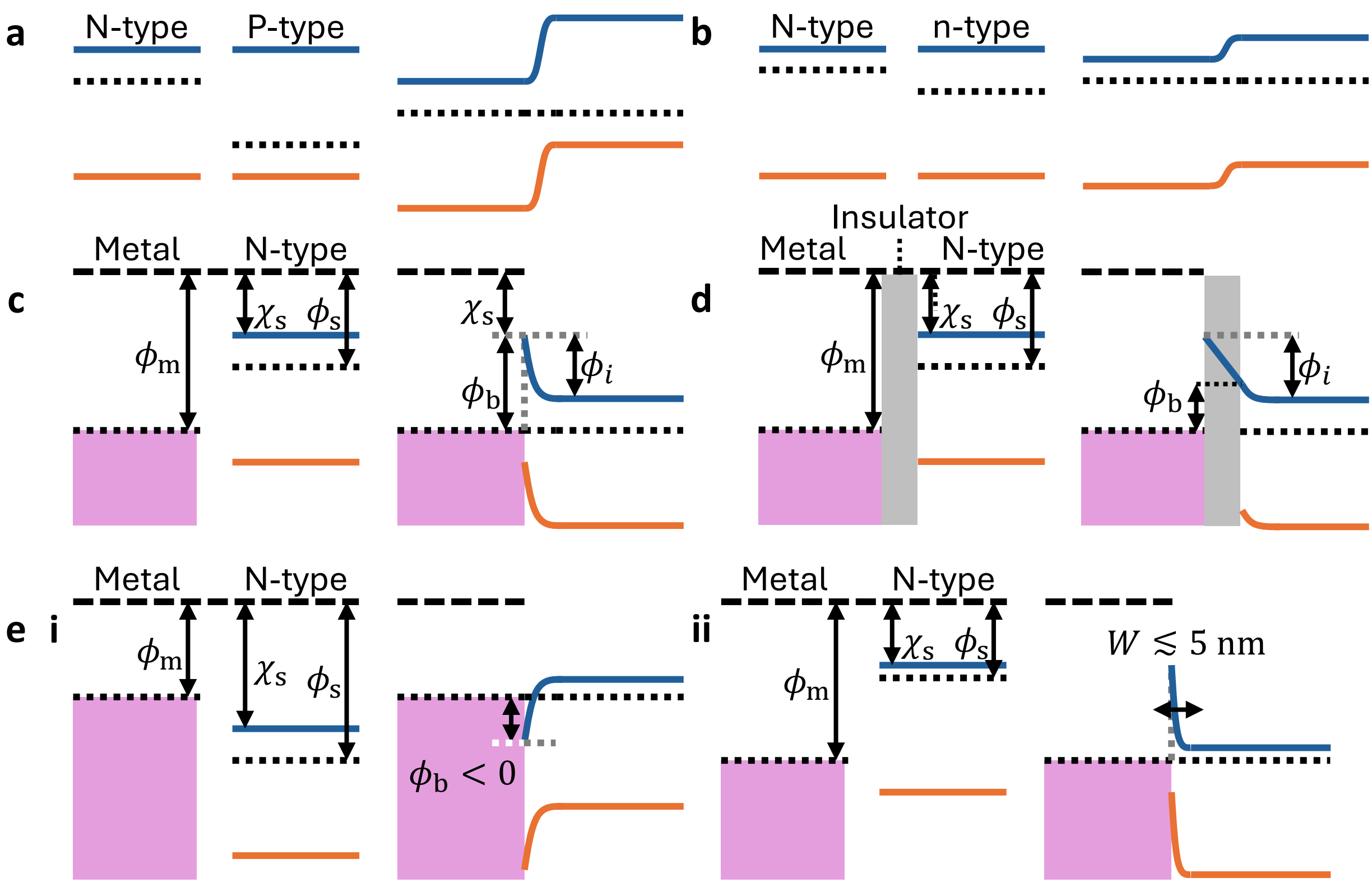

Figure 14. (a) PN junction, (b) heavily doped N-type/lightly doped n-type junction, (c) Schottky junction, (d) metal/insulator/semiconductor junction, and (e) Ohmic junctions due to (i) $\phi_m < \phi_s$ and (ii) highly-doped semiconductors leading to easy tunneling through the barrier. In each case, the left shows the band structure before contact and the right after contact. $\phi_{\mathrm{m}}$ is the metal work function, $\phi_{\mathrm{s}}$ and $\chi_{\mathrm{s}}$ the semiconductor work function and electron affinity, $\phi_{\mathrm{b}}$ the barrier height, $\phi_i$ the built-in potential, and $W$ the depletion width.

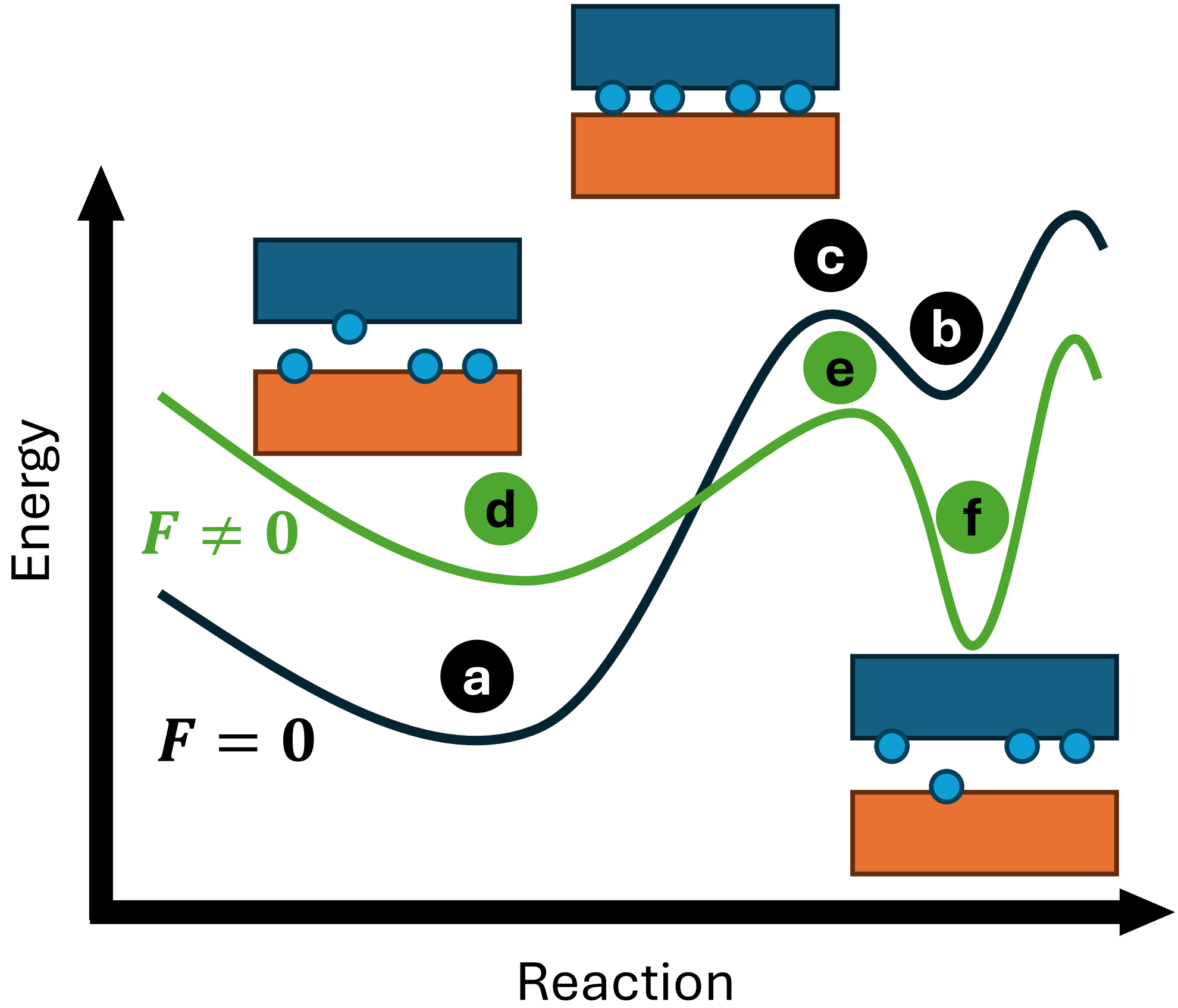

Figure 15. Mechanical stress can drive “mechanochemical” or “tribochemical” reactions by changing reaction energy landscapes. For example, species adsorbed on one body (a) may be favored with no force, while others (b) may be favored when force is applied. Additionally, the energy barriers to form intermediate species (c) can be modified.

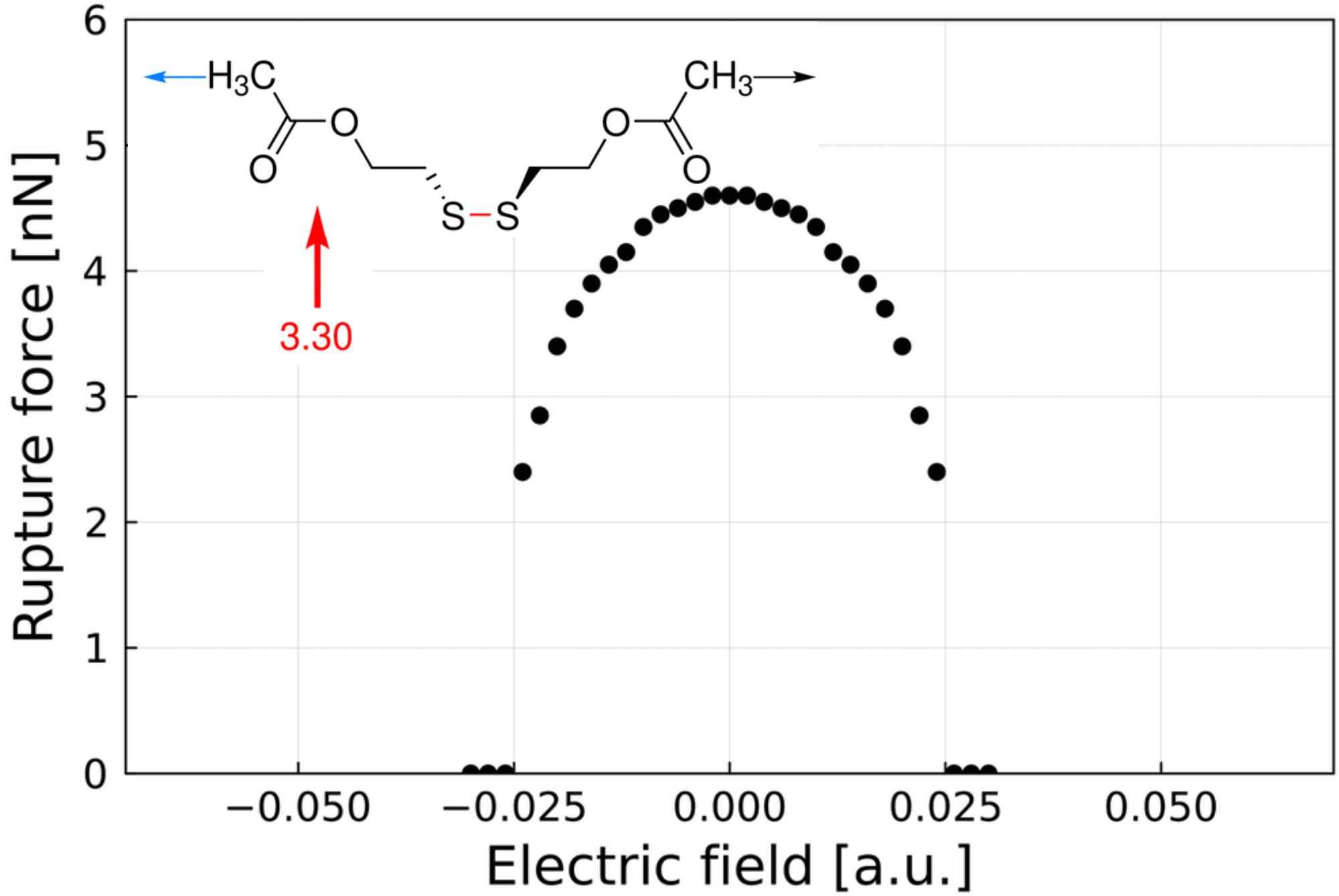

Figure 16. Effect of combined electric force and field to break the S-S bond of a high-symmetry linear disulpide at the bond indicated in red. The field is indicated by the blue arrow, and the dipole moment in Debye is in red. Adapted with permission from T. Scheele and T. Neudecker

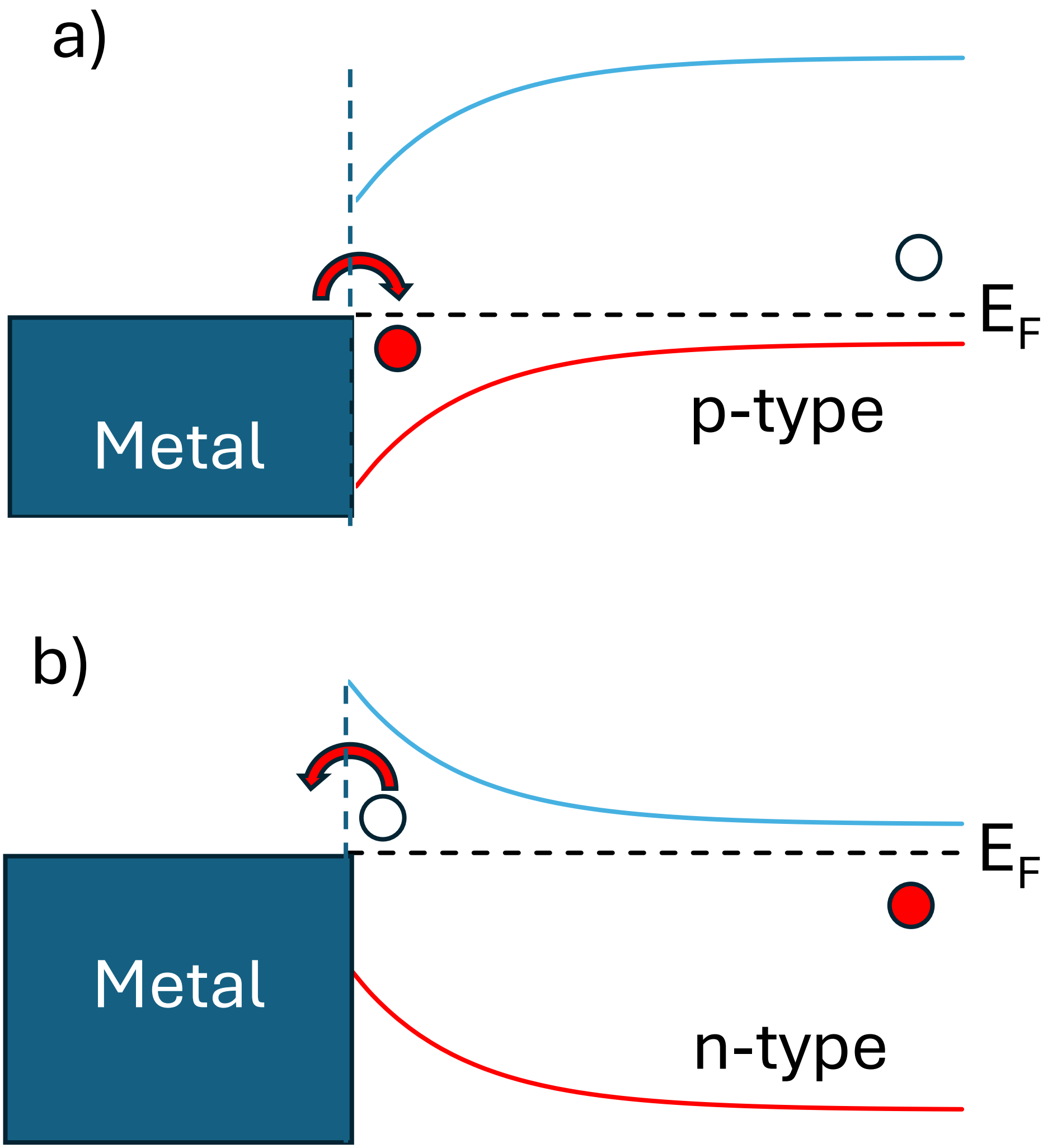

Figure 17. Two examples of trap state possibilities. In a) for a metal to p-type junction states in the gap which are empty (open circles) may be filled (arrow) due to band bending in the depletion zone or electromechanical terms. Similarly in b) for an n-type the trap state can donate to the metal. As drawn the trap states are inside the surface, but they could also be across the surface away from the contact.

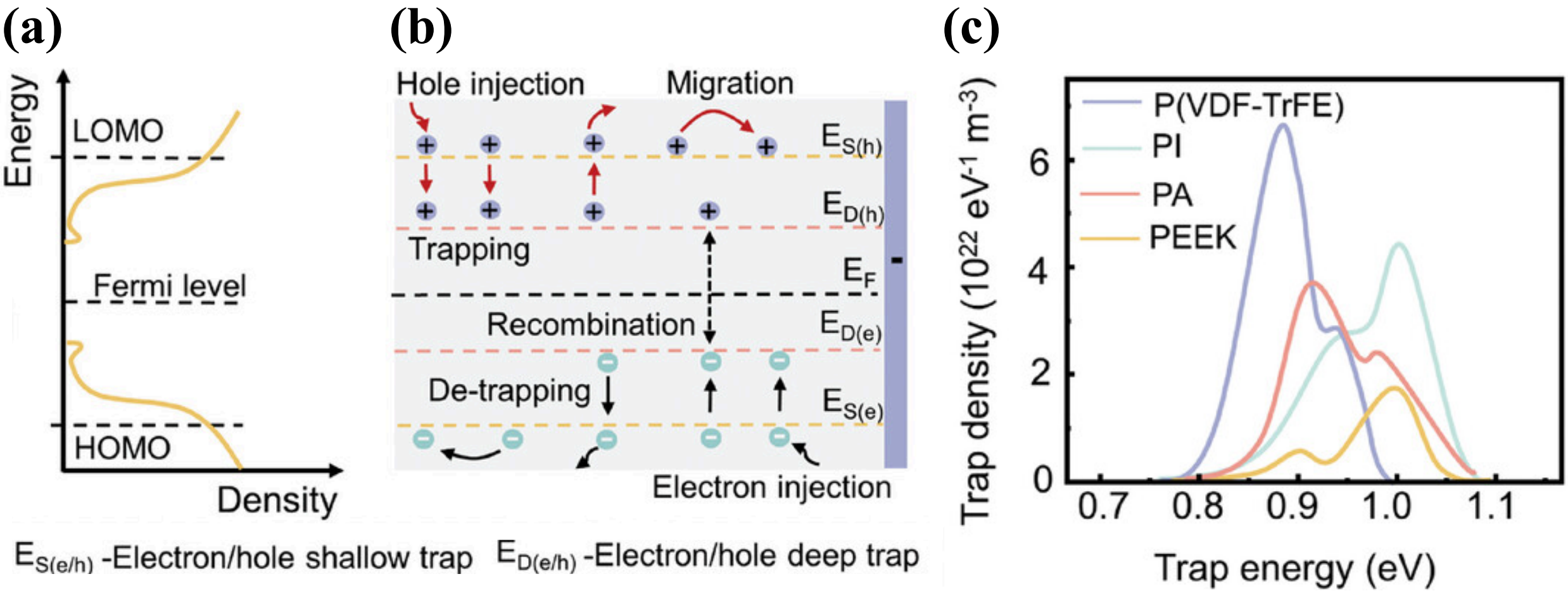

Figure 18. (a) General distribution of trap states in the band structure of a polymer contacting a metal. (b) De-trapping and other processes that change the population of shallow and deep traps. (c) Measured trap state energy and density distribution for four polymers contacting steel. Each are a sum of two peaks, corresponding to deep and shallow traps at higher and lower energies, respectively. Reprinted with permission from Xu *et al*.

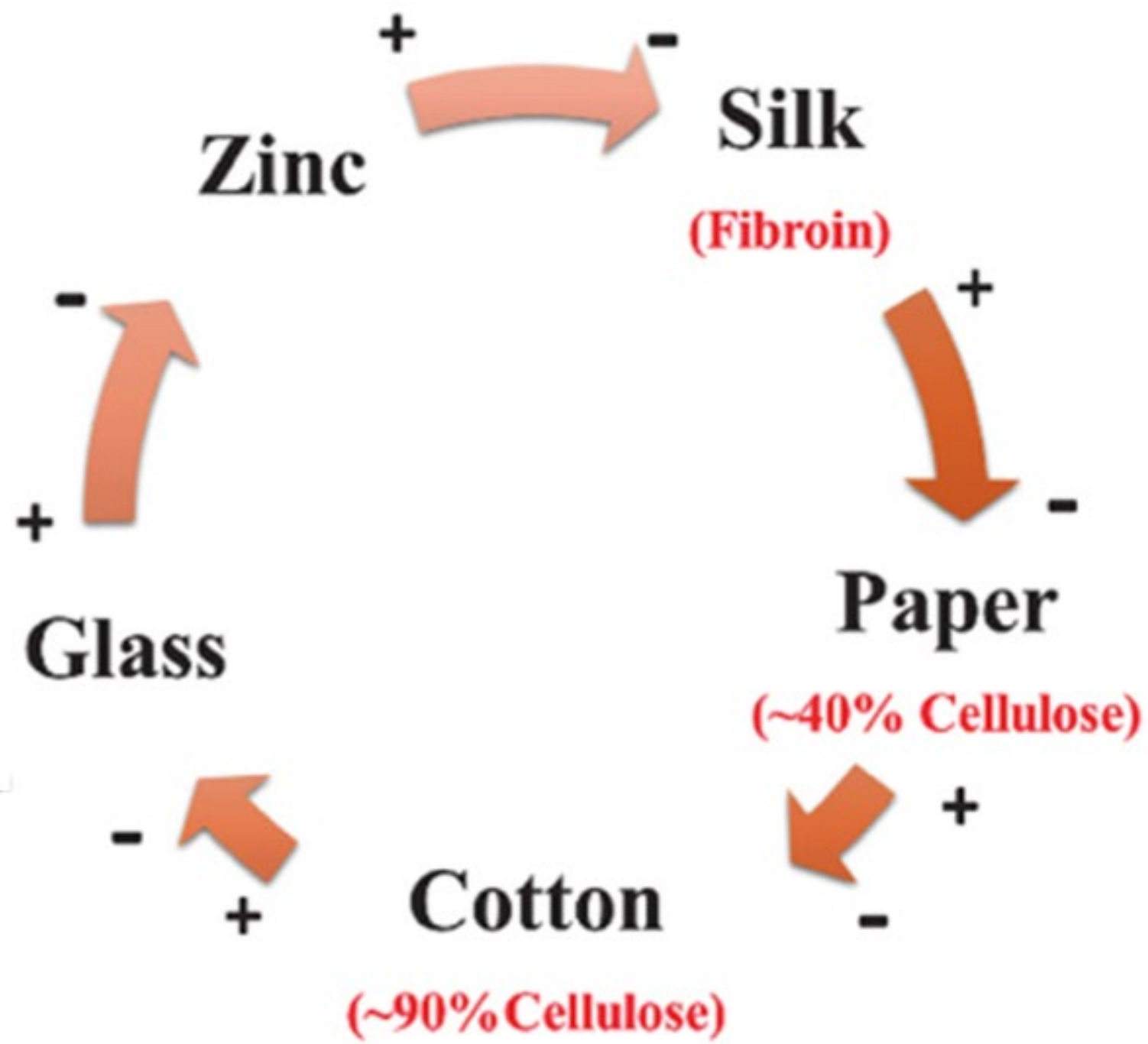

Figure 19. Cyclic triboelectric series where each material will charge positive with respect to that clockwise from it.

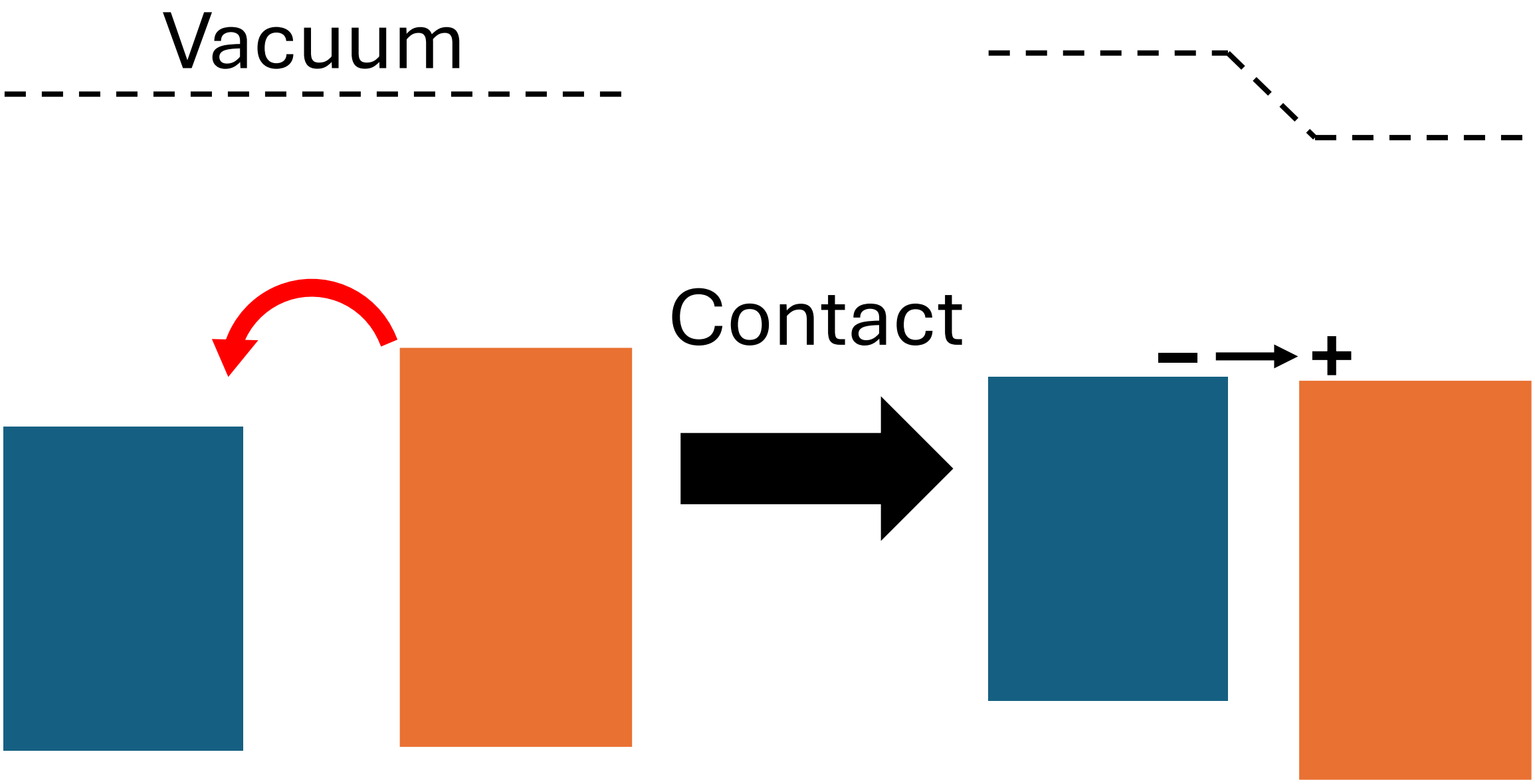

Figure 20. Illustration of bridging electrons at a contact. For two metals with different work functions (blue and orange here) there will be electron transfer indicated as a red arrow which will lead to an interface dipole, and perhaps other charges away from the contact depending upon the geometry.

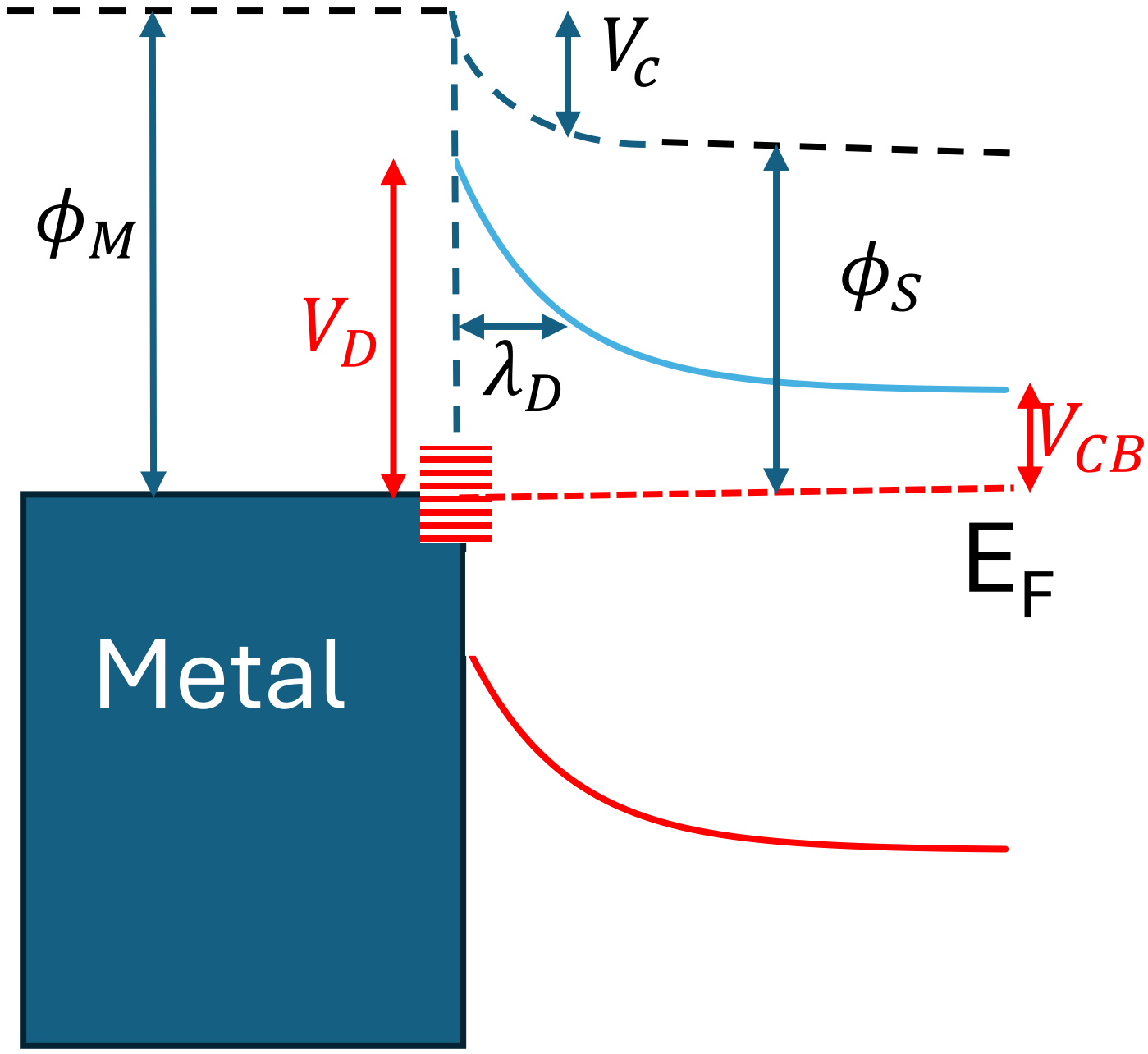

Figure 21. Work functions for both a metal $\phi_M$ and a semiconductor or insulator $\phi_S$ where there is both band bending and Fermi level pinning (the red states). Here the built-in potential (contact potential) $V_c$ includes the dipoles at the interface as well. Both the Fermi energy $E_F$ and the Debye length $\lambda_D$ are indicated, as well as the two terms $V_D$ and $V_{CB}$ for the band bending with Fermi level pinning.

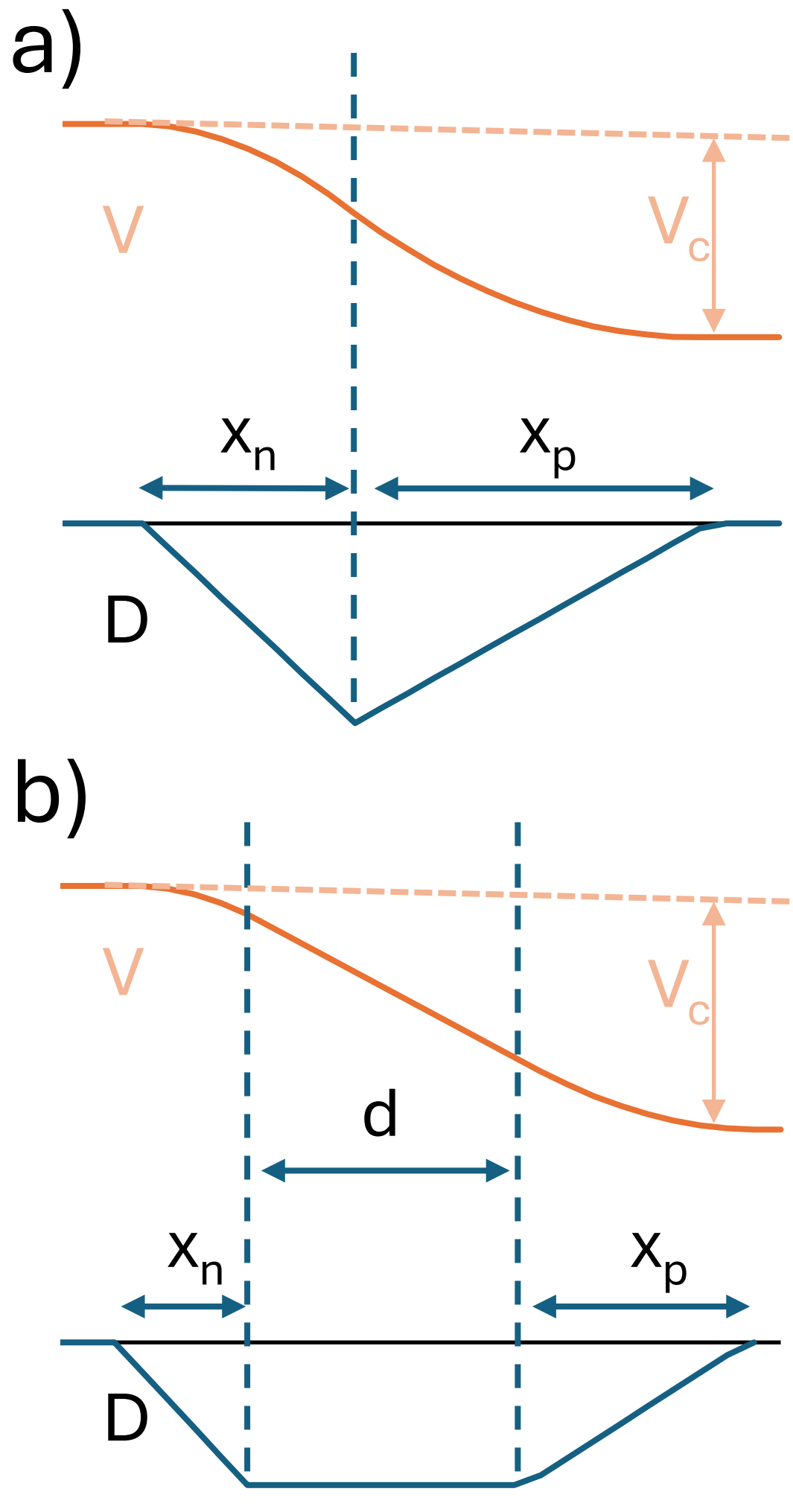

Figure 22. Abrupt junction band bending a) in contact, and b) with a vacuum gap. Shown in the diagrams is both the potential V and the electric displacement D (normal to the junction), as well as the depletion widths $x_p$ and $x_n$, the built-in potential $V_c$ (contact potential) and the separation d.

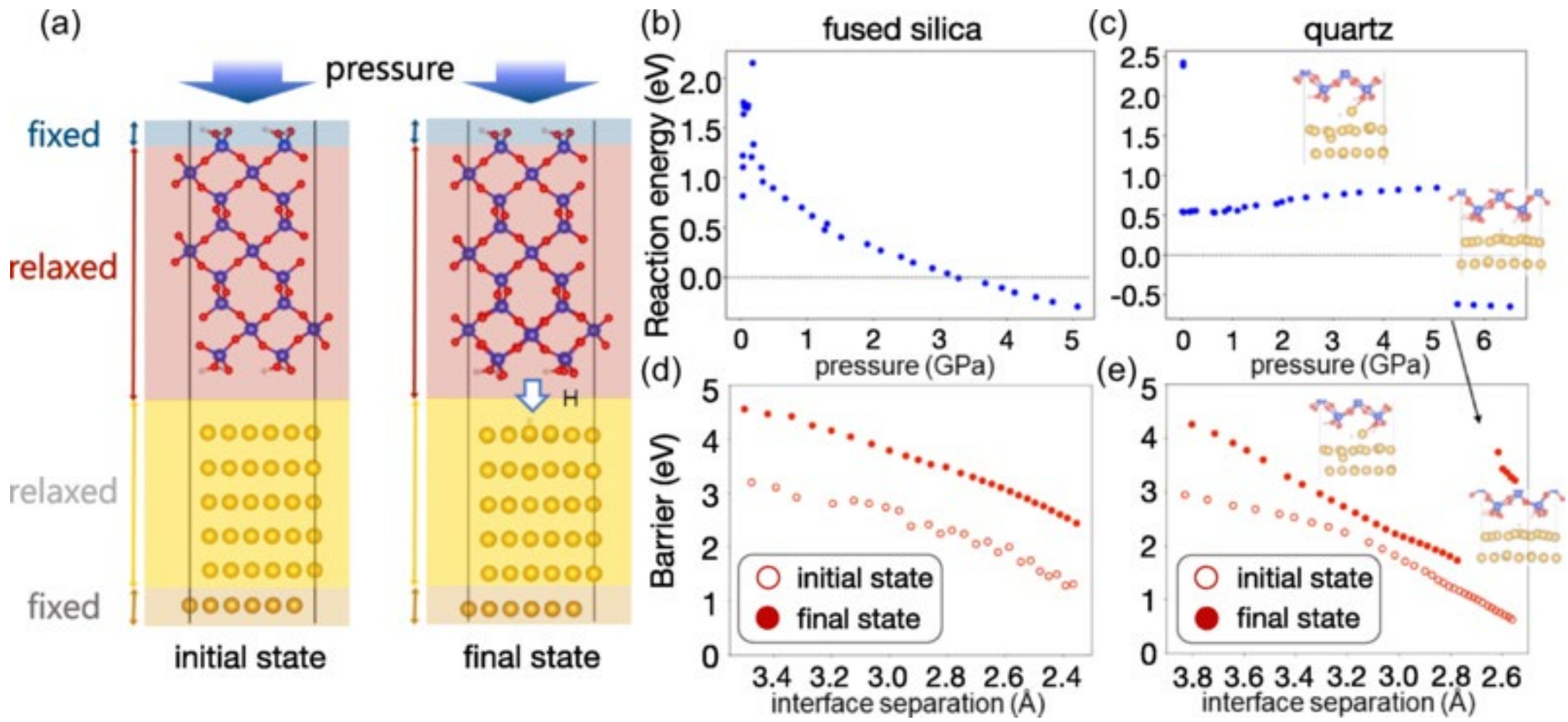

Figure 23. Pressures causes changes in the reaction energy of hydrogen transfer and the potential barrier at a hydrogen terminated silica/gold contact. (a) The contact is modeled *ab initio* simulating the initial and the final state at a fixed interface distance, and then progressively reducing this distance to increase the pressure. (b),(d) Reaction energy at every indentation step, respectively, for hydrogen transfer of fused silica and quartz. (c),(e) Potential barrier before (unfilled circles) and after (filled circles) hydrogen transfer, respectively, for fused silica and quartz. Reproduced with permission from Fatti *et al.*

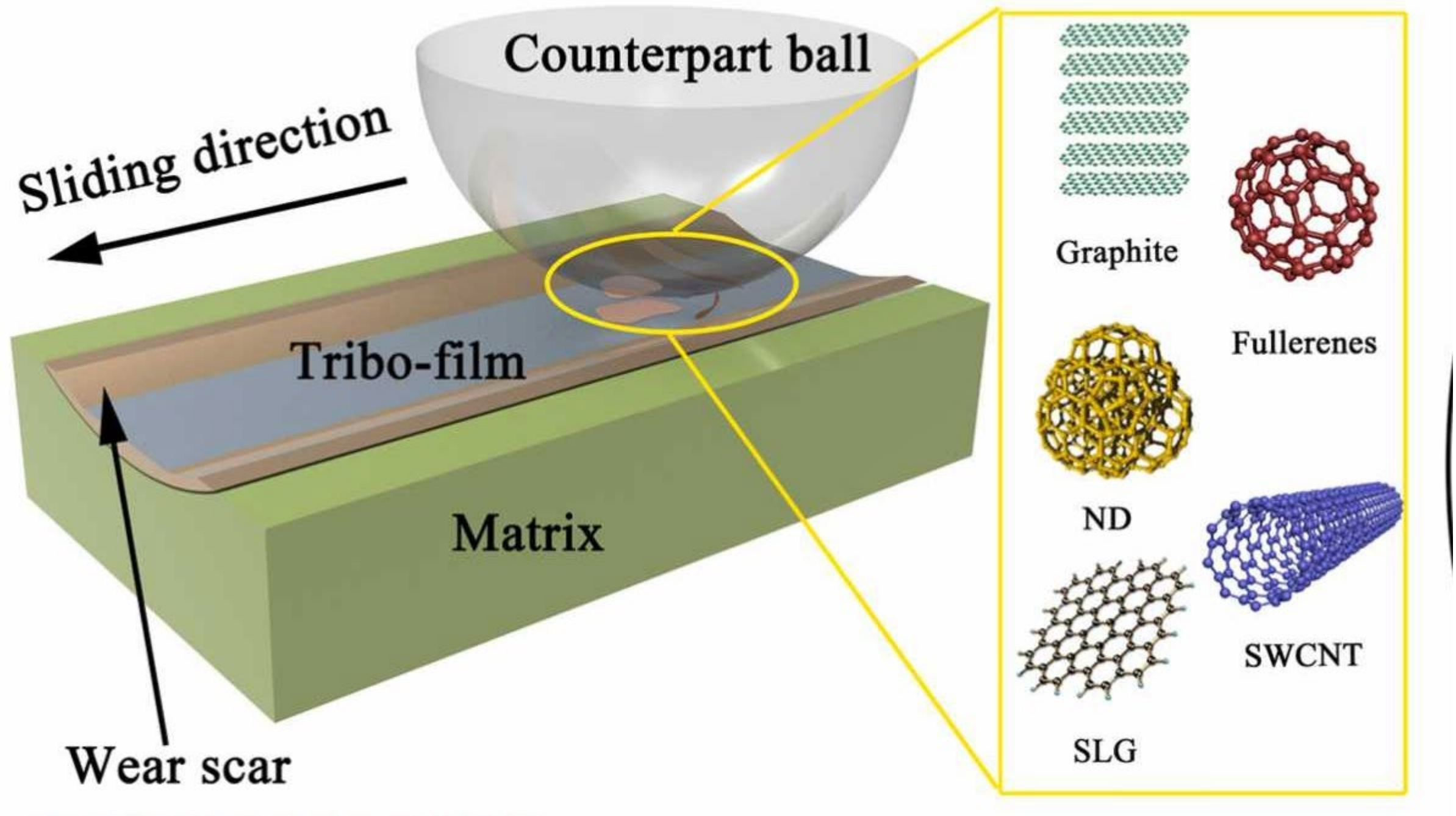

Figure 24. schematic of the types of carbon forming at the contact during a sliding experiment leading to what is called a tribofilm as well as wear of material. Adapted from Wang *et al* [21] with permission.

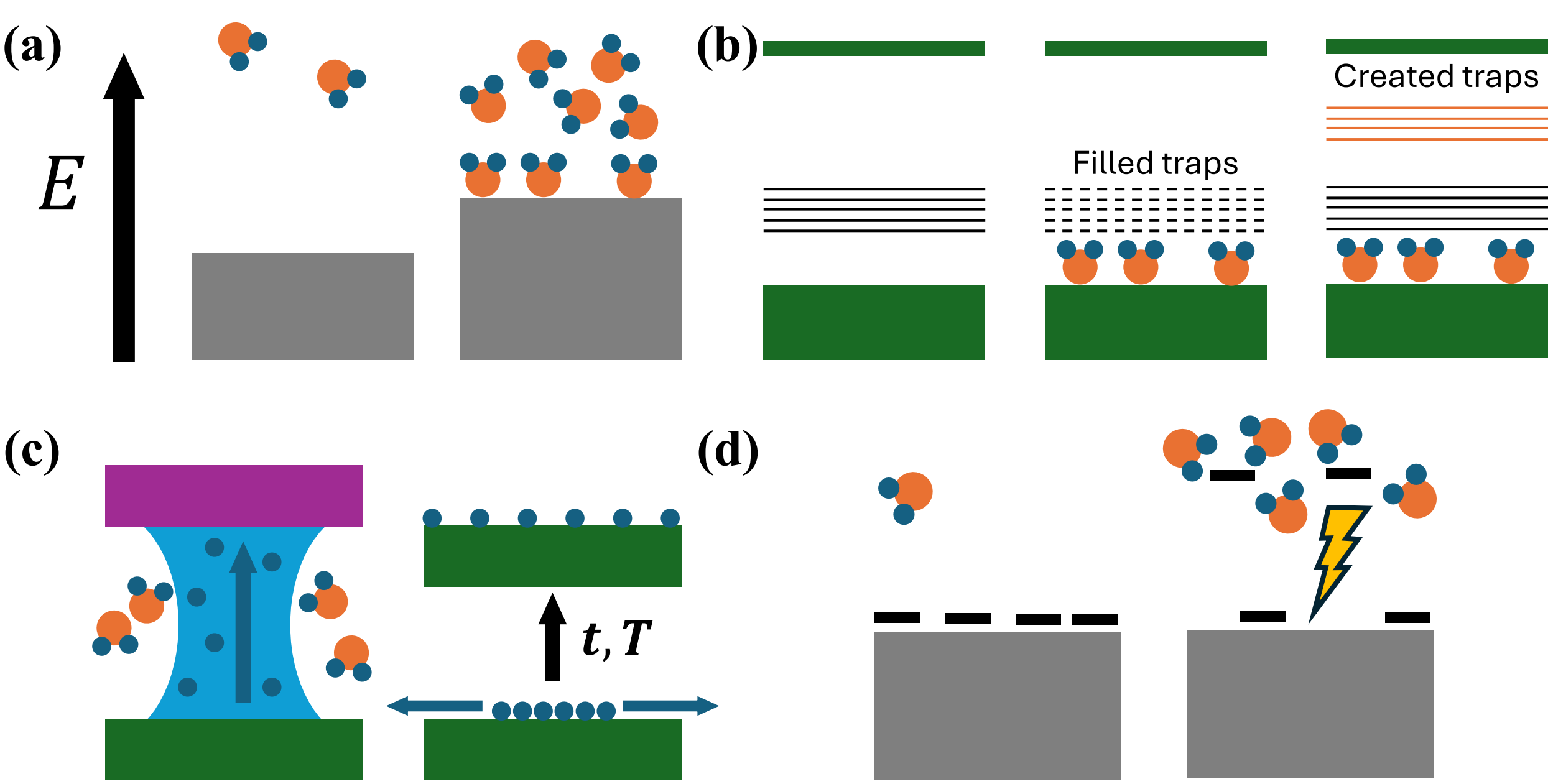

Figure 25. (a) Work functions or contact potentials can be modified by adsorbed species such as water. (b) Adsorbed water can passivate or introduce trap states. (c) Humidity affects various conduction mechanisms, for example through water bridges between bodies or through diffusion of adsorbed and/or dissociated water molecules (particularly the protons) over relatively small times $t$ and temperatures $T$. (d) Humidity reduces the charge density needed for atmospheric breakdown.

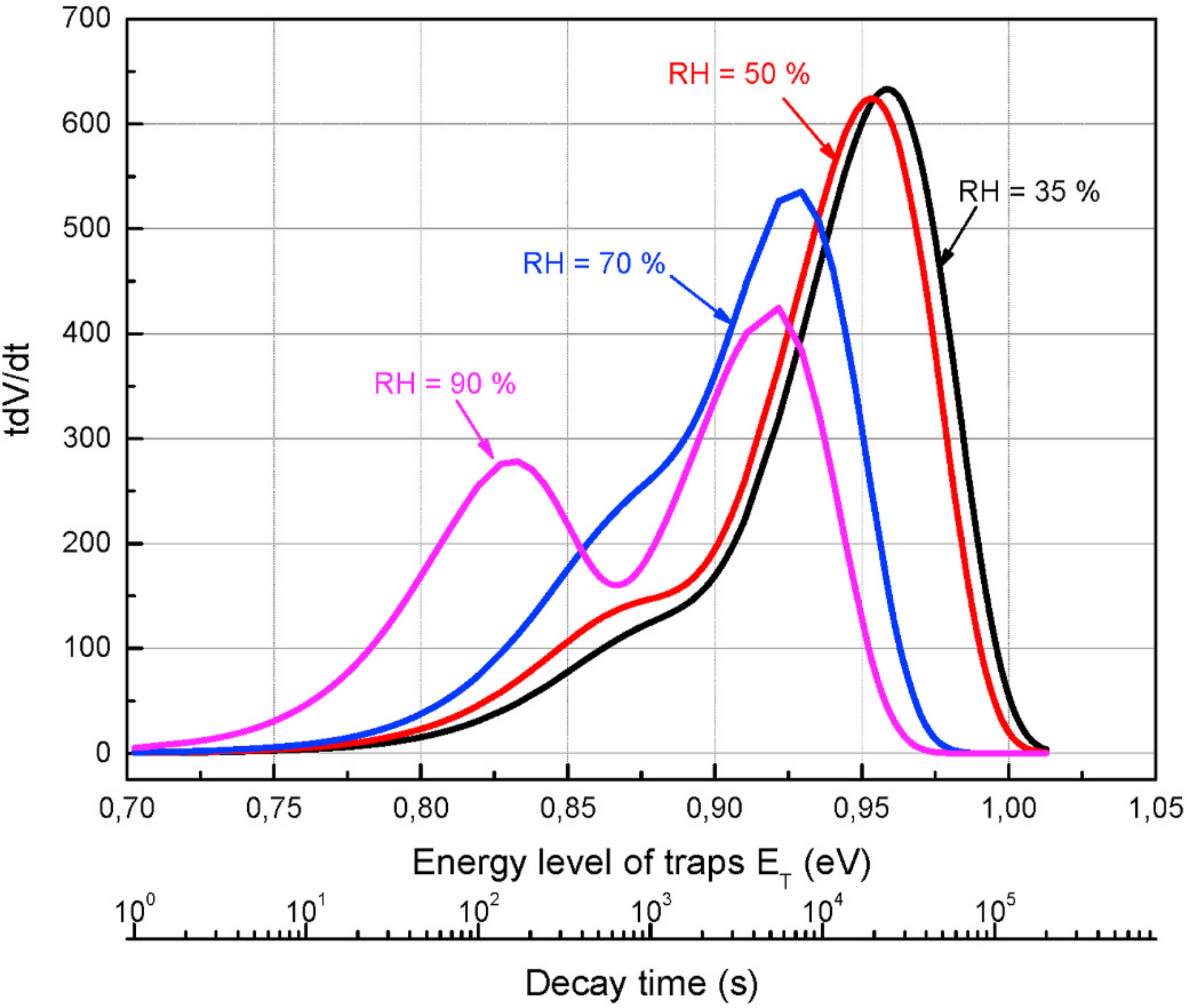

Figure 26. the population of trap states for different humidities, reproduced from ...Amiour, N., Z. Ziari, and S. Sahli, Traps energy distribution in Kapton HN polyimide films through surface potential decay model under humidity conditions.

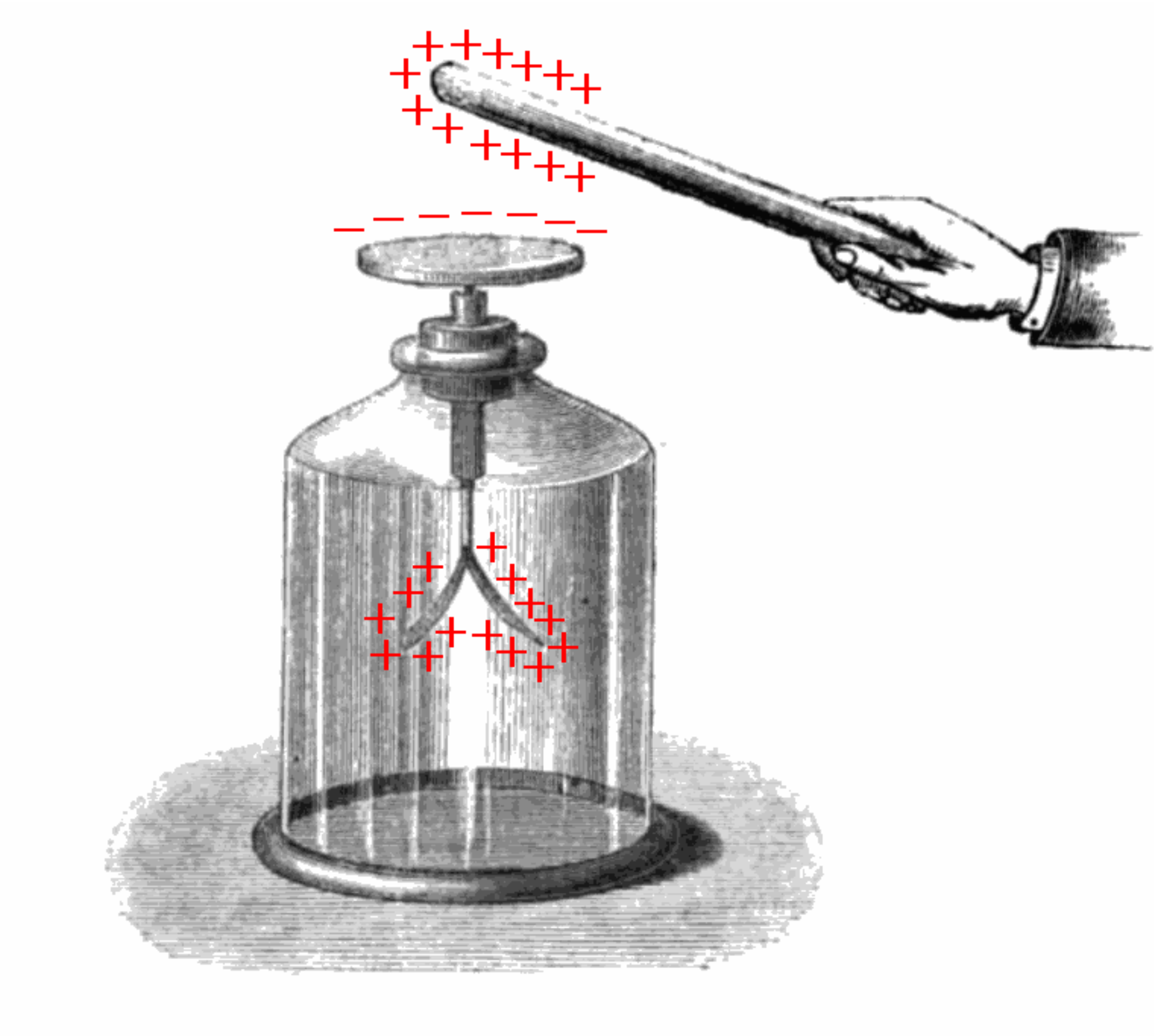

Figure 27. Gold lead electrostat, showing the charges (in red) if a positively charged rod is brought near to the top plate. Original version: Sylvanus P. Thompson, Derived version: Chetvorno, Public domain, image via Wikimedia Commons.

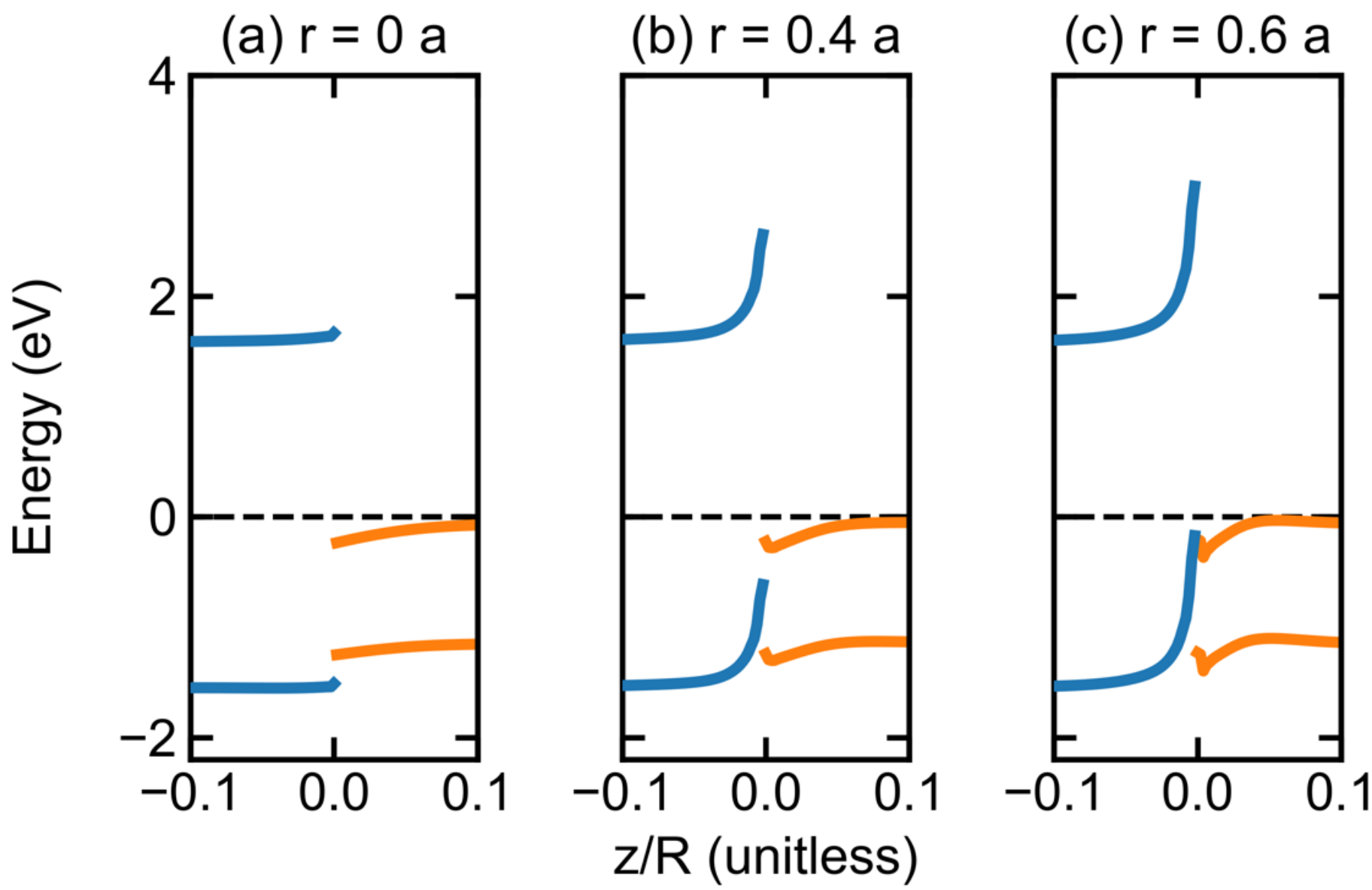

Figure 28. Sphere-on-flat contact band diagrams for a Si sphere (orange) on a $SrTiO_3$ flat (blue). (a)-(c) show the conduction and valence band edges as a function of depth (normalized by the indenter radius R) at different radial distances (in units of the contact radius a) from the contact point as defined in Figure 2(a) with a contact pressure of 6 GPa. The unstrained Fermi level of each material is assumed to be at its band gap center and zero energy is taken to be the unstrained $SrTiO_3$ Fermi level.

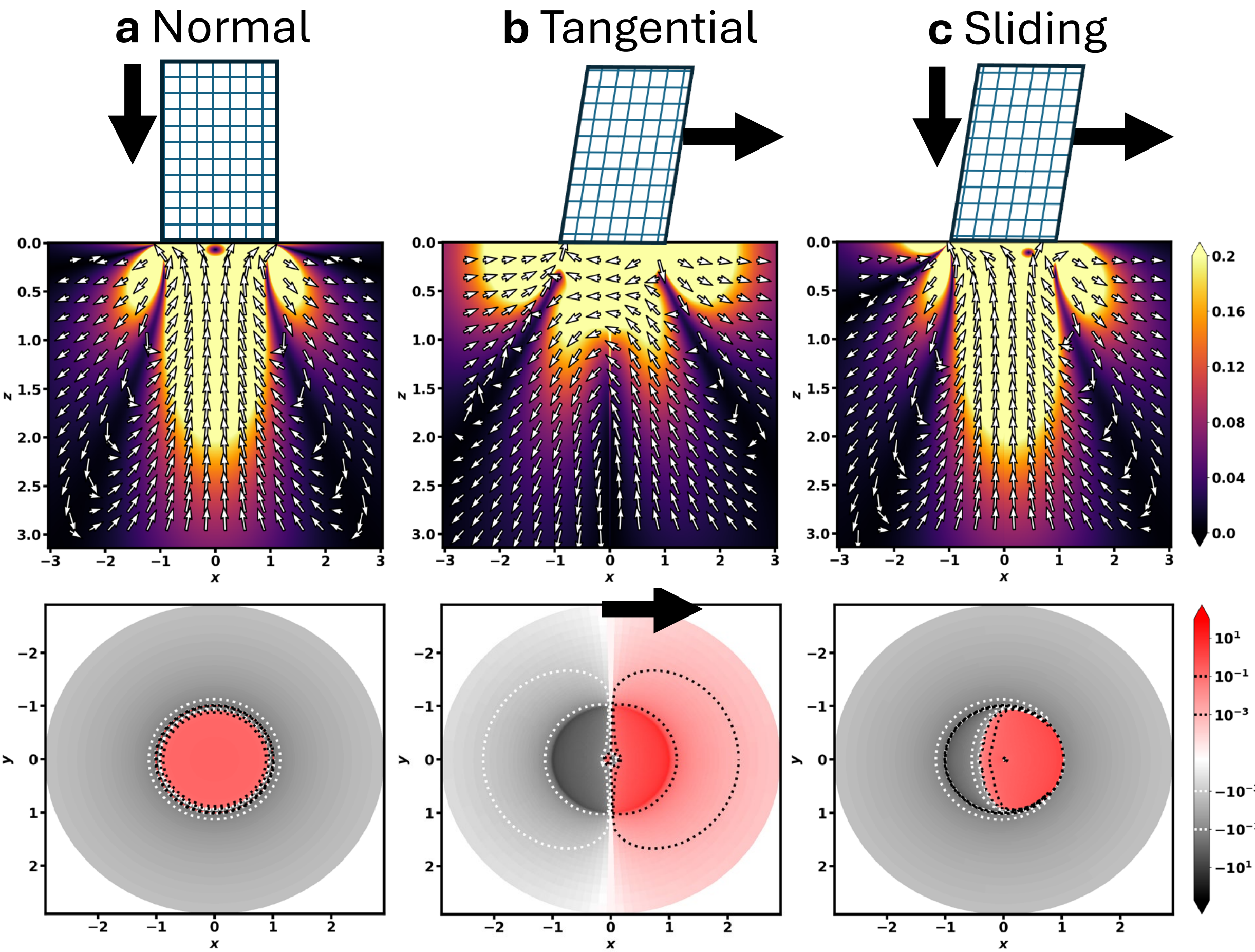

Figure 29. At the top a cross-section of the polarization field for a pure normal, tangential and combined sliding contact with the asperity shown schematically as a grid and black arrows for the force directions; below the bound surface charge density. The polarization and charge are normalized, using a mean contact pressure $p_m$, contact radius $a$, flexoelectric coefficient $\mu$, and Young's modulus $Y$ all equal to 1. The polarization and charge scale as $\mu p_m / aY$. The white arrows indicate the direction of the polarization vector, the color scale the magnitude.

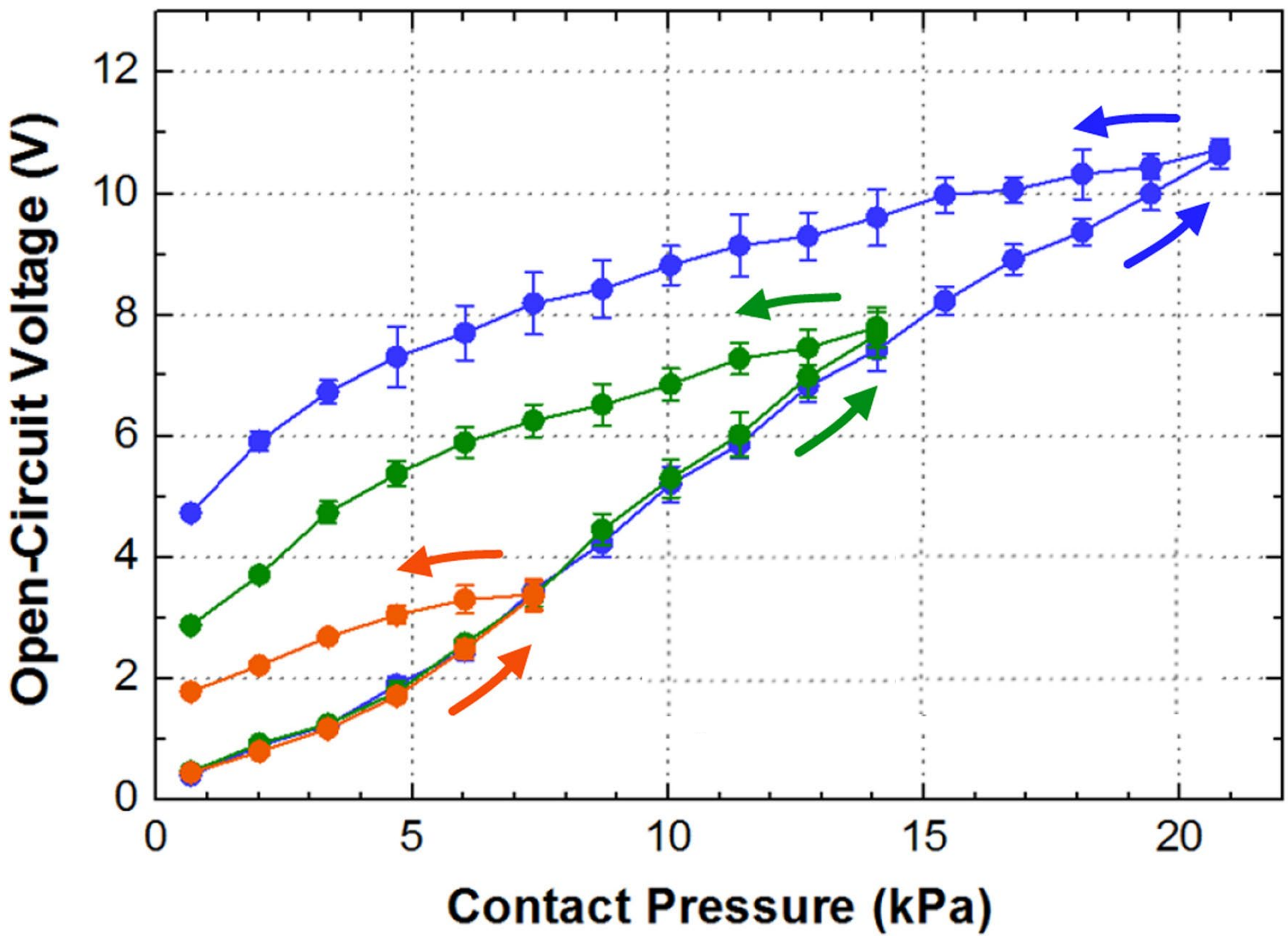

Figure 30. Hysterisis loops for a TENG contact showing distinct irreversibility in the backwards open-circuit voltage. Reproduced with permission from Seol et al.

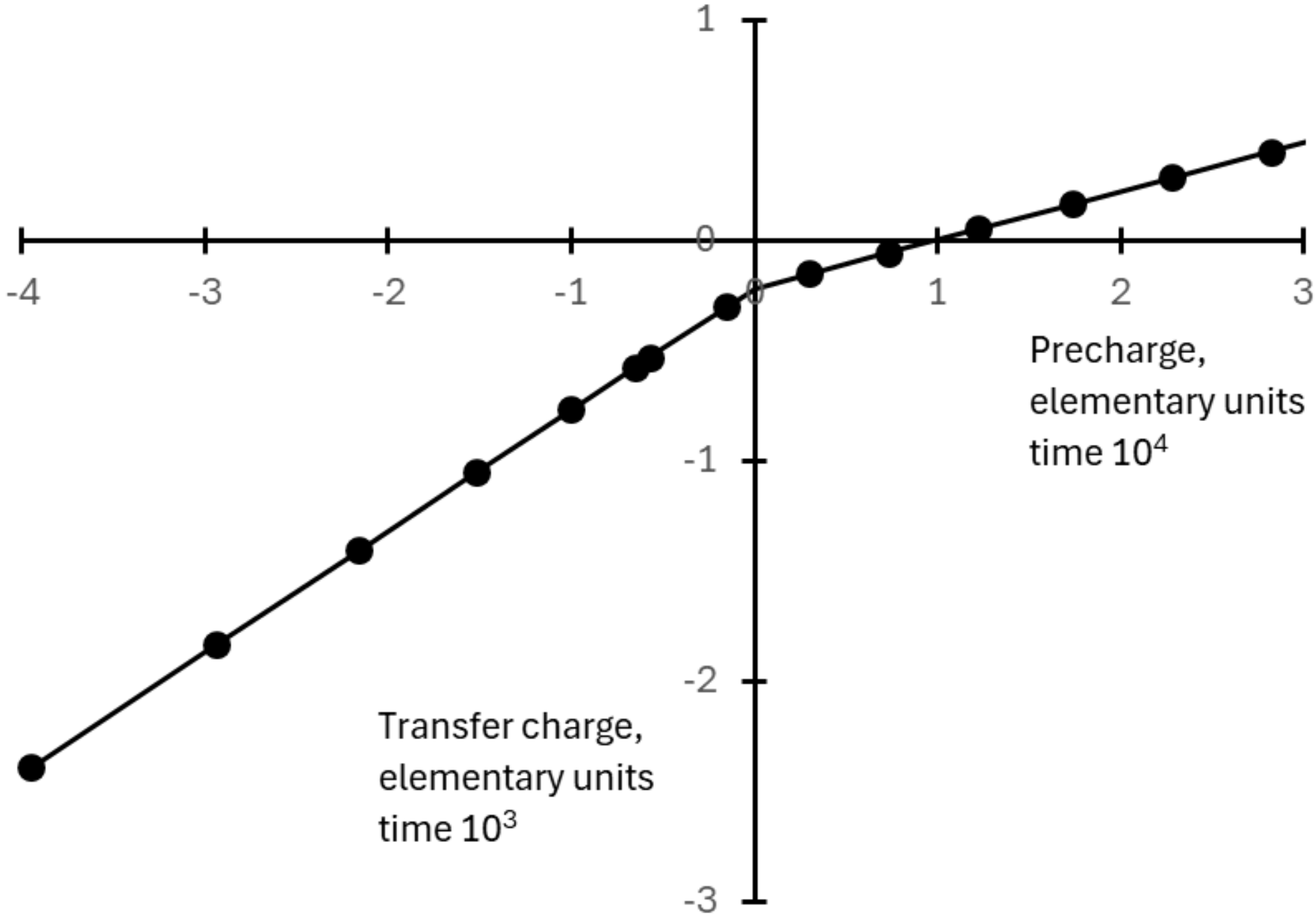

Figure 31. Reproduction of the results of John *et al.* for NaCl particles on a Titanium plate. Positive and negative precharges have a different effect on the charge transfer in the next collision, suggesting (as the authors state) that there is a rectifying pn junction being formed.

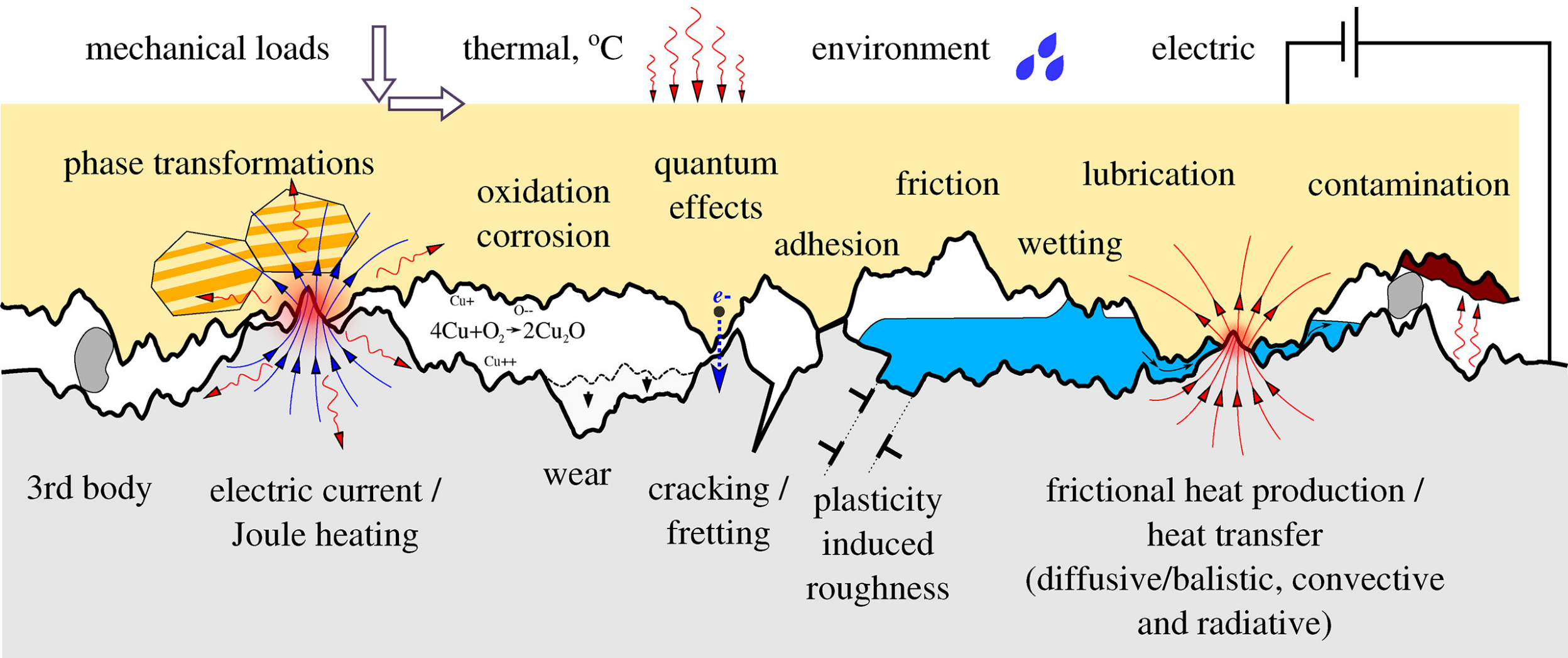

Figure 32. Schematic of the many processes that can take place at contacts. While they are not going to all be relevant to every triboelectric experiment, the diversity of processes needs to be considered. Reproduced from Varkis et al with permission.